\documentclass[preprint,12pt]{elsarticle}

\usepackage{amssymb}
\usepackage{amsmath}
\usepackage{geometry}

\usepackage{lineno}

\usepackage{hyperref}
\usepackage{comment}
\usepackage{float}
\usepackage{xcolor}

\newsavebox\CBox

\usepackage{caption}
\usepackage{subcaption}

\usepackage{booktabs}
\usepackage{array, makecell}

\journal{Chemometrics and Intelligent Laboratory Systems}  

\begin{document}

\begin{frontmatter}



\title{Modeling cyclostationarity in time series using ASCA} 

\author[1]{Daniel Vallejo-España\corref{cor1}}
\ead{dvalesp@ugr.es}
\author[1]{Jesús García Sánchez}
\ead{gsus@ugr.es}
\author[2]{Manuel Villar-Argaiz}
\ead{mvillar@ugr.es}
\author[3]{Concepción De Linares}
\ead{delinare@ugr.es}
\author[1]{José Camacho}
\ead{josecamacho@ugr.es}
\affiliation[1]{organization={Department of Signal Theory, Telematics and Communications, University of Granada},
            addressline={Calle Periodista Daniel Saucedo Aranda},
            city={Granada},
            postcode={18014},
            state={Andalucía},
            country={Spain}}
\affiliation[2]{organization={Department of Ecology, University of Granada},
            addressline={Avenida de Fuentenueva},
            city={Granada},
            postcode={18071},
            state={Andalucía},
            country={Spain}}
\affiliation[3]{organization={Department of Botany, University of Granada},
            addressline={Avenida de Fuentenueva},
            city={Granada},
            postcode={18071},
            state={Andalucía},
            country={Spain}}
\cortext[cor1]{Corresponding author.}



\begin{abstract}
Modern data analysis across diverse disciplines increasingly relies on time series. Many of these datasets exhibit cyclostationarity, where patterns approximately repeat in a regular manner, often across multiple time scales, such as daily, weekly or yearly cycles. In this context, statistical inference is essential to distinguish genuine underlying effects from random variability. While tools like Analysis of Variance (ANOVA) provide such inference, they often lack interpretability and struggle with the complexities of multivariate data.
To address these limitations, we propose a unified pipeline for the exploratory analysis of cyclostationary time series using ANOVA Simultaneous Component Analysis (ASCA). ASCA is an extension of ANOVA that is able to work in both univariate and multivariate cases. Combining inference with the visualization capabilities of Principal Component Analysis (PCA), ASCA provides powerful options for interpretability. The capabilities of ASCA have been well-established in the analysis of experimental data, but they remain largely unexplored for observational data like time series. Our workflow introduces an algorithmic approach to modeling time-dependent data using ASCA, enabling control over multiple cyclostationary time scales while also accounting for the specific challenges of this type of data, such as autocorrelation. We demonstrate the efficacy of this methodology through two real-world case studies: water temperature trends in mountain lakes in Sierra Nevada, Spain, and airborne pollen trends over 30 years recorded in the city of Granada, Spain.

\end{abstract}




\end{frontmatter}


\section{Introduction}
\label{sec:intro}

Time series have become increasingly important in modern data analysis due to the growing availability of temporally resolved data across a wide range of disciplines. In established areas such as finance and economics, time series have long played a central role. More recently, advances in sensing technologies and digital infrastructures have propelled the prominence of time series in fields such as environmental research \citep{Shenoy2022ExtremeUSWeather, Botrel2023aquatic}, urban planning \citep{Lin2024}, and eHealth \citep{Klonoff2025, Serhani2020}. In disciplines such as these, data are collected over time to capture evolving processes, often resulting in complex, high-dimensional time series. As these applications increasingly inform high-stakes decisions, the analysis and interpretation of time series data have become essential, enabling domain experts to understand temporal patterns and draw accurate conclusions from the data.

A key aspect of time series analysis is recognizing that time series often exhibit \textit{cyclostationarity}, where patterns approximately repeat in a regular manner over time. These cycles may be linked to human activity, such as weekly patterns in electricity consumption, or to natural processes, such as daily variations in temperature or variations of airborne pollen concentrations over the seasons. In many applications, several cycles coexist at different temporal scales. This is the case for temperature data, which typically display a daily cycle driven by the alternation of day and night, as well as a yearly cycle associated with seasonal changes, with warmer periods in summer and cooler periods in winter. In the case airborne pollen, their presence in the air indicates the flowering period of anemophilous plants, which, although seasonal, is influenced by meteorological factors, pollution, etc. Taking \textit{multiple temporal scales} into account is therefore essential when interpreting time-dependent data with cyclostationary behavior, as it allows recurring patterns to be analyzed separately. Although time series data are usually observational rather than experimental, separating temporal scales closely resembles the structure used in statistical experimental design, where data are organized in terms of factors and levels \citep{montgomery2017design}. For example, for temperatures measured hourly over multiple years, one can define an `hour of the day' factor with 24 levels and a `day of the year' factor with 365 levels, as well as a `year of record' factor with one level per year. Similarly, to define daily airborne pollen concentrations over several years, one could create a `day of the year' factor and a `year of record' factor. This representation makes it possible to analyze cyclic behavior on multiple scales using statistical models, such as Analysis of Variance (ANOVA).

ANOVA enables statistical inference by testing whether observed differences between groups are likely to reflect genuine underlying effects rather than random variability. In the context of temperature or airborne pollen concentration time series, this allows analysts to formally assess questions such as whether average temperatures or concentrations differ across years. Such inference is essential for decision-making, as it provides quantitative evidence to support conclusions. However, classic ANOVA has important limitations in the context of time series analysis, mainly due to its assumptions (normality, homoscedasticity and independence of residuals) and limited visualization support.

An alternative to classic ANOVA for time-dependent data is Functional Data Analysis (FDA). FDA treats observations that vary over a continuum, such as time, as functions rather than discrete points. For example, both temperatures and airborne pollen measured throughout a day or year could be considered functions and modeled using FDA methods. Within that framework, functional ANOVA methods like \citet{Cuevas2004ANOVAFunctional} and \citet{Zhang2014ANOVAFunctional} extend classical ANOVA to functional responses, providing a single, global test for differences between groups across the entire temporal scale. A multivariate functional ANOVA is also available \citep{Gorecki2016MANOVAFunctional}, enabling the simultaneous comparison of multiple functional responses. Other approaches use pointwise tests, with corrections for multiple testing \citep{Pataky2015twowayanova, Niiler2020ttests}, which makes it possible to identify the intervals where groups differ.

While FDA methods offer advantages over standard ANOVA, they still lack intuitive, visually interpretable results, which can limit their usefulness for the exploration of time series data. To address this limitation, we propose the use of ANOVA Simultaneous Component Analysis (ASCA; \citealp{smilde2005asca}), a multivariate extension of ANOVA designed to separate the individual effects of multiple factors and their interactions in settings with many response variables. ASCA provides visualizations similar to those of Principal Component Analysis (PCA), such as score and loading plots, which act as graphical post-hoc tools that facilitate the interpretation of group differences and help identify the variables driving those differences. Although ASCA was originally developed for data arising from statistical experimental design, it has only recently begun to be applied to observational data \citep{villarargaiz2022heat, camacho2024netmob} and experimental time series data \citep{ezenarro_characterisation_2024}.

Although datasets are often multidimensional, ASCA requires a two-dimensional matrix as input. To address this mismatch, we propose an unfolding approach that converts a multidimensional array or tensor into a matrix suitable for ASCA. Unfolding techniques have been widely used to transform three-dimensional tensors into two-dimensional matrices for methods such as PCA and Partial Least Squares (PLS), for example in batch process data \citep{camacho_bilinear_2008} and hyperspectral imaging \citep{prats-montalban_multivariate_2011}. Similarly, unfolding has been applied in Multivariate Curve Resolution (MCR), where it is used to extract component contributions in chemical mixtures \citep{de_juan_multivariate_2021}. When unfolding, the tensor is rearranged so that selected dimensions form the rows, interpreted as factors and their levels, while the remaining dimensions form the columns, interpreted as response variables. In our context, this provides explicit control over the multiple time scales present in a time series, allowing these scales to be treated either as factors or as responses, depending on the hypothesis under study. Building on this idea, we present a unified pipeline that combines unfolding techniques with ASCA to analyze and interpret cyclostationary behavior in time series data.

In this work, we make the following contributions:
\begin{itemize}
\item Propose the use of ASCA as a tool for exploring cyclostationarity in time series, combining statistical inference with highly interpretable, visual results.
\item Introduce an algorithmic unfolding approach that transforms cyclostationary time-dependent data into a matrix representation suitable for ASCA.
\item Demonstrate the proposed methodology on two real-world time-dependent datasets.
\end{itemize}

The rest of this paper is structured as follows: Section~\ref{sec:asca} summarizes the basics of ASCA, Section~\ref{sec:methodology} introduces a method for modeling time series with cycles using ASCA, Section~\ref{sec:autocorr} elaborates on the choices of permutation strategies and model diagnostics for ASCA, Section~\ref{sec:casestudies} describes two case studies, and Section~\ref{sec:conclusion} concludes the work.

\section{ANOVA Simultaneous Component Analysis (ASCA)} \label{sec:asca}

ASCA is a multivariate extension of the classic ANOVA, designed for analyzing datasets arising from experimental design. The ASCA pipeline is composed of three steps: (1) factorization of the data matrix according to the factors and interactions of the experimental design; (2) significant testing for factors and interactions; (3) visualization of the factors and interactions’ effects via Principal Component Analysis (PCA) or alternative techniques, such as PARAFAC \citep{jansen2008parafasca}.

\subsection{Data Factorization}

Let $\mathbf{X} \in \mathbb{R}^{N \times M}$ be a data matrix with $N$ observations and $M$ response variables. ASCA decomposes $\mathbf{X}$ into the contributions of the experimental factors and their interactions. The decomposition for a model with two factors \textit{A} and \textit{B} would be:
\begin{equation}\label{eq:asca}
\mathbf{X}=\mathbf{1}\mathbf{m}^{\text{T}} + \mathbf{X}_A + \mathbf{X}_B + \mathbf{X}_{AB} + \mathbf{E},
\end{equation}
where $^{\text{T}}$ indicates matrix transposition and:
\begin{itemize}
\item $\mathbf{1}$ is a column vector of ones of length $N$,
\item $\mathbf{m}$ is a column vector of variable means,
\item $\mathbf{X}_A$, $\mathbf{X}_B$ and $\mathbf{X}_{AB}$ are matrices capturing the effects of factors $A$, $B$, and their interaction $AB$,
\item $\mathbf{E}$ is the residual matrix.
\end{itemize}

The technique used in this paper to compute the factorization is ASCA+ \citep{thiel2017asca}, that allows for mild unbalancedness in the data. In ASCA+, the decomposition is obtained as the least squares solution of a regression problem, where $\mathbf{X}$ is regressed onto a coding matrix $\mathbf{D}$:
\begin{equation}
\mathbf{X} = \mathbf{D}\Theta + \mathbf{E} = \mathbf{1}\theta_m + \mathbf{D}_A \Theta_A + \mathbf{D}_B \Theta_B + \mathbf{D}_{AB} \Theta_{AB} + \mathbf{E},
\end{equation}
and $\mathbf{D}$ is constructed using sum coding \citep{thiel2017asca}.
$\mathbf{E}$ is the residual matrix, $\mathbf{D}_A$, $\mathbf{D}_B$ and $\mathbf{D}_{AB}$ represent the coding for factors
$A$ and $B$ and their interaction $AB$, respectively, and
$\Theta = [\theta_m;\Theta_A;\Theta_B;\Theta_{AB}]$, is composed of the coefficients of the factorization estimated by least squares:
\begin{equation} \label{eq:asca_coef}
\Theta = (\mathbf{D}^{\text{T}}\mathbf{D})^{-1}\mathbf{D}^{\text{T}}\mathbf{X},
\end{equation}
\begin{equation}
\mathbf{E} = \mathbf{X} - \mathbf{D}\Theta.
\end{equation}

\subsection{Statistical Significance Testing} \label{sec:ascainf}
Following decomposition, ASCA tests the statistical significance for factors and interactions using permutation testing. Permutation testing can be conducted by randomly permuting the rows of $\mathbf{X}$ in Equation~\eqref{eq:asca_coef}, resulting in a new set of regression coefficients and, consequently, a new factorization:
\begin{equation}
\Theta^* = (\mathbf{D}^{\text{T}}\mathbf{D})^{-1}\mathbf{D}^{\text{T}}\mathbf{X}^*,
\end{equation}
\begin{equation}
\mathbf{E}^* = \mathbf{X}^* - \mathbf{D}\Theta^*.
\end{equation}
The p-value is then calculated as:
\begin{equation} \label{eq:pvalue}
p = \frac{\#\{F^*_k \geq F; \; k=1,\dots,K\} + 1}{K + 1},
\end{equation}
where:
\begin{itemize}
  \item $F$ is the F-ratio, calculated from the original factorization as the ratio between the mean squares (MS) of the factor or interaction (i.e., the sum of squares divided by its degrees of freedom) and the MS of a reference term. Often, this reference term are the residuals \citep{Camacho2026tutorial}.
  \item $F^*_k$ denotes the value of the F-ratio obtained in the $k$-th permutation of $\mathbf{X}$.
  \item $K$ is the total number of permutations performed.
\end{itemize}
The p-value provides an empirical estimate of the probability of observing a result as extreme or more extreme than the one obtained, assuming the null hypothesis is true.

Multiple strategies for permutation testing exist. A more detailed discussion focused on permutation schemes for time series analysis is provided in Section~\ref{sec:autocorr}. 

\subsection{``Post-Hoc'' Visualization}

Significance testing in ASCA assesses whether the effects of factors and their interactions are statistically significant, but it does not determine the significance of individual levels or specific combinations of levels. To evaluate differences between levels in ANOVA, post-hoc tests are commonly used. In ASCA, a visual counterpart to post-hoc analysis is the application of PCA \citep{zwanenburg2011anova}. It is worth noting that, while these ``post-hoc visualizations'' help highlight pattern differences between levels, they do not provide a formal statistical test. In this approach, PCA loadings are computed from the factor or interaction matrix, and score plots are generated from the sum of this matrix and the residuals. These score plots allow visual comparison of each of the main and interaction effects against the variability captured in a reference \citep{montgomery2017design}. Often---and in this study as well---residuals are used as the reference, serving as the denominator of the F-ratio \citep{Camacho2026tutorial}.

\section{Structuring cyclostationary time series data for ASCA analysis}\label{sec:methodology}

\begin{figure}[ht]
\centering
\includegraphics[width=0.95\textwidth]{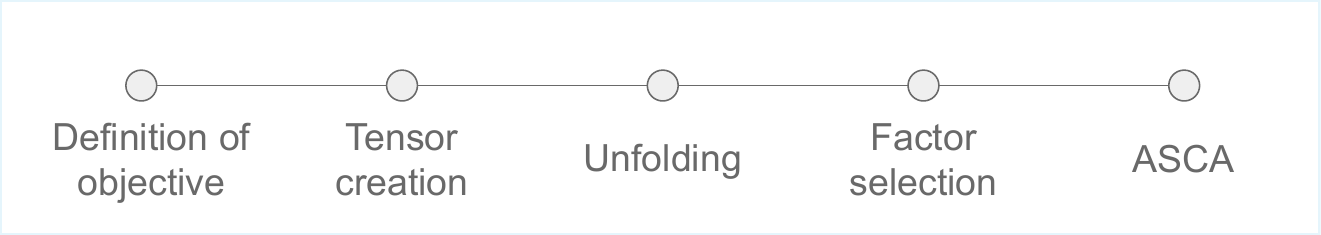}
\caption{Pipeline for modeling cyclostationarity with ASCA.}
\label{fig:pipeline}
\end{figure}

ASCA requires input data to be structured as a matrix, where rows represent factors for significance testing and columns correspond to response variables. To obtain this structure from the data, it is helpful at first to conceptualize the data as a multidimensional tensor, where each dimension is referred to as a ``mode''. Since time series data often spans multiple temporal scales, each scale can be represented as a separate tensor mode. For instance, measurement times may be indexed by hour of the day, day of the year, and year of the total span, with each serving as a mode. A time series usually allows more than one tensor representation, so the choice is guided by the objective of the study. In addition to these time scales, the tensor may encompass non-temporal modes. The resulting tensor is subsequently unfolded into a matrix, in a way that accounts for potential autocorrelation and leverages the cyclostationary nature of some time scales. This matrix serves as the input for ASCA following the selection of relevant factors. This section details the workflow from raw data to the analysis with ASCA, divided into four main steps, as seen in Figure~\ref{fig:pipeline}: definition of objective, tensor creation, unfolding, and factor selection.

\begin{figure}[ht]
    \centering
    \small
    
    \begin{minipage}{\linewidth}
        \textbf{A} \par \vspace{1pt}
        \centering
        \includegraphics[width=0.95\textwidth, height=6cm, keepaspectratio]{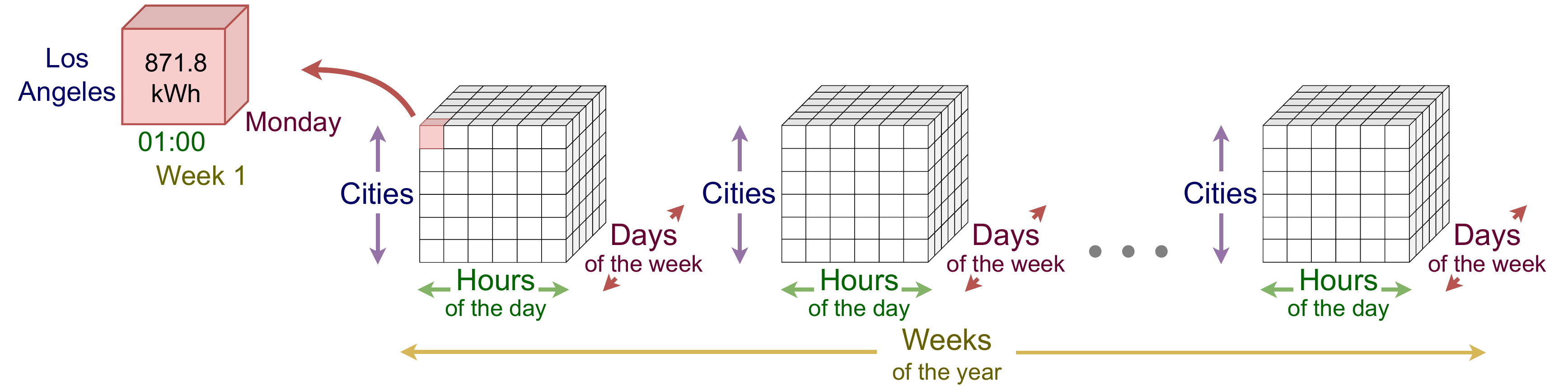}
    \end{minipage}

    \vspace{0.25cm}

    \begin{minipage}{0.54\linewidth}
        \textbf{B} \par \vspace{1pt}
        \centering
        \includegraphics[width=\linewidth, height=6cm, keepaspectratio]{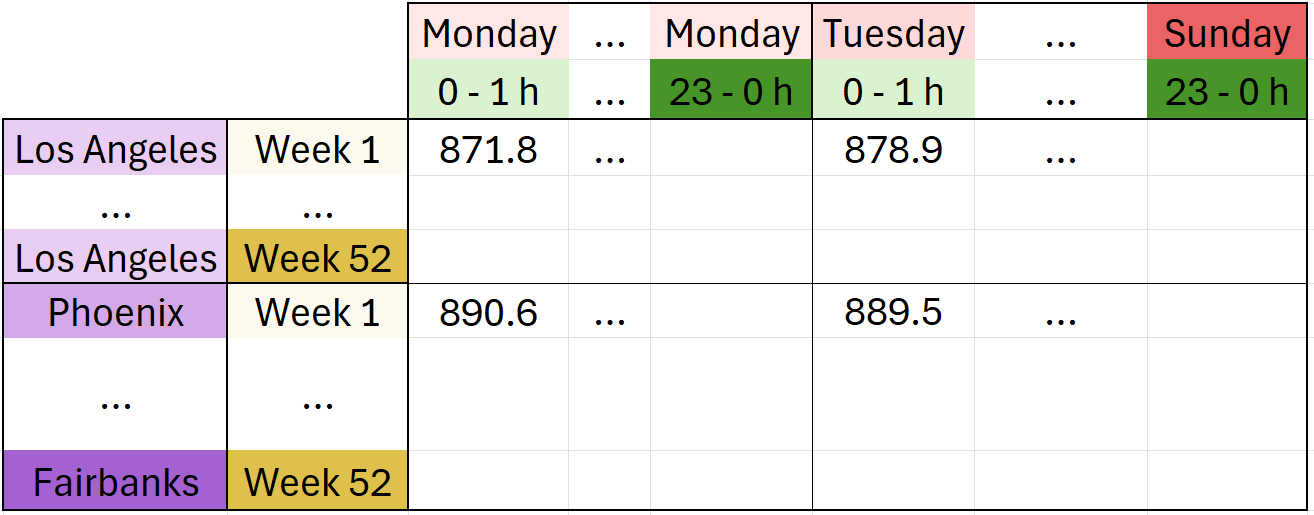}
    \end{minipage}
    \hspace{0.02\linewidth}
    \begin{minipage}{0.42\linewidth}
        \textbf{C} \par \vspace{1pt}
        \centering
        \includegraphics[width=\linewidth, height=6cm, keepaspectratio]{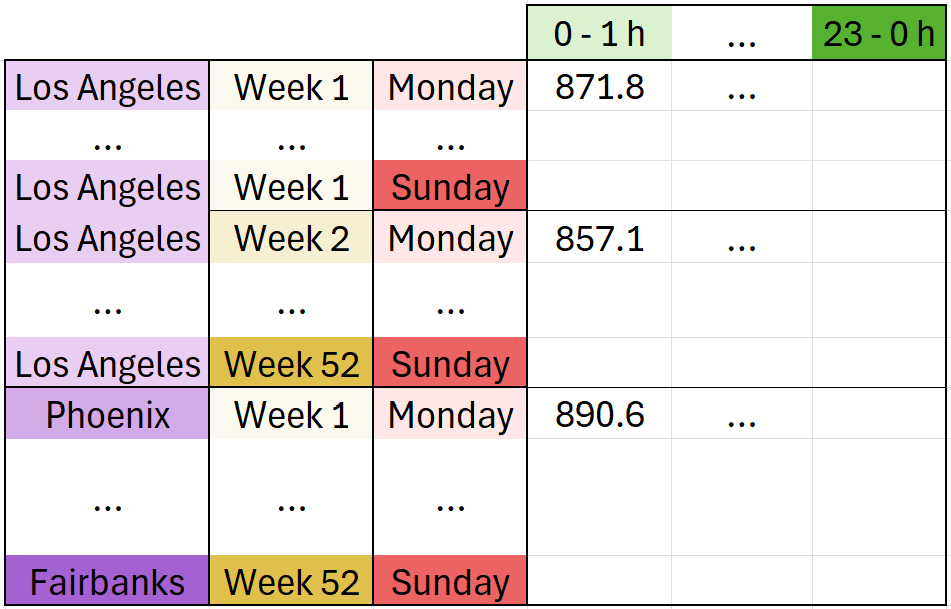}
    \end{minipage}

    \vspace{0.1cm}

    \caption{\textbf{Energy consumption dataset representation.} \textbf{A} Tensor representation where the four modes represent city, hour of the day, day of the week, and week of the year. \textbf{B} Matrix unfolding with modes `city' and `week of the year' as rows, and `hour of the day' and `day of the week' as columns. \textbf{C} Matrix unfolding with modes `city', `week of the year' and `day of the week' as rows, and `hour of the day' as columns.}
    \label{fig:tensor_unfolding}
\end{figure}

Throughout the explanation of the procedure, we will use as an example a subset of simulated energy consumption data, available on the \href{https://data.openei.org/submissions/153}{Open Energy Data Initiative website} \citep{ong2014loadprofiles}. This subset includes load profiles for one building type across eight different cities in the United States. Each load profile represents the hourly electricity consumption in kilowatt-hours (kWh) over the course of a typical year. These data exhibit cyclostationarity, driven by the periodic nature of human activity and environmental conditions. Specifically, in one year, two distinct cycles are expected: a daily cycle, characterized by fluctuations between peak active hours and nighttime inactivity; and a weekly cycle, reflecting the differences in demand between workdays and weekends. These cycles define the tensor featured in Figure~\ref{fig:tensor_unfolding}A. Figure~\ref{fig:tensor_unfolding}B--C shows two unfolding possibilities for said tensor, each representing a data matrix used as input for an ASCA model.

\subsection{Definition of the objective}

The proposed pipeline begins by defining the analysis objective or question to be addressed. For instance, in the context of energy consumption data, the technicians in charge of the electrical grid may need to determine how to split the power between different spaces at a given time. To make an informed decision, it is useful to know the weekly consumption patterns of the buildings, as well as their evolution over the course of the year. The question to be answered dictates the construction of the tensor and its subsequent unfolding. Should multiple objectives be defined, multiple unfoldings might be necessary.

\subsection{Tensor creation}

Building the data tensor requires the inclusion of every factor relevant to the analysis. Specifically, these factors fall into three categories: non-temporal factors, cyclostationary temporal factors, and an additional time scale representing the temporal evolution of the entire time series.

Non-temporal factors are those independent of time, such as the city in the energy consumption dataset. Conversely, temporal factors are derived directly from the time series. Constructing the tensor requires considering all possible cyclostationary behaviors among these temporal factors. Once all known or potentially relevant cycles have been established, the time series is split into several modes that become part of the data tensor, each representing a cycle. In the example dataset, two modes arise from the daily and weekly cycles. This methodology proposes a structured approach to decomposing the time series into these modes. We adopt the naming convention \textit{frequency of the period} for these cyclostationary modes (e.g., `hour of the day', `day of the week'). These two concepts, \textit{frequency} and \textit{period} are defined as follows:
\begin{itemize}
\item Frequency unit: The granularity at which the data is measured.
\item Period unit: The duration of time over which the pattern repeats.
\end{itemize}
For instance, the `hour of the day' mode isolates a time scale with cyclostationary behavior by grouping all data points that occur within the a day---the period unit---at the resolution of one hour ---the frequency unit---. When multiple cyclic patterns are present, we create one mode per pattern, beginning with the most fine-grained time scale (the daily cycle in the example). To ensure the complete isolation of different cyclic patterns, each subsequent mode adopts the period of the previous mode as its new frequency. For example, `day of the week' inherits the `day' period from the previously defined `hour of the day' mode.

Once all cyclostationary modes are defined, a time scale might remain outside the cyclic structure. In the case of the example, the largest cycle is the weekly one, but there are data available for multiple weeks throughout the year. This scale (`week of the year') is more coarse-grained than the ones in the cycles and is also included into the tensor, as it serves as a means to evaluate the evolution of the cyclostationary behavior over time. We call this the \textit{evolution mode}.

Thus, in our case, a tensor (Figure~\ref{fig:tensor_unfolding}A) has been built with one non-temporal mode (`city'), two cyclostationary modes (`hour of the day' and `day of the year'), and one evolution mode (`week of the year').

\subsection{Unfolding} \label{sec:unfolding}

ASCA requires a data matrix (i.e., a bidimensional array) as input. Consequently, the data tensor must be unfolded into a matrix by assigning specific modes to rows and others to columns. The rows constitute the observations; the modes placed along the rows define the factors to be tested during the inference phase of ASCA. When a temporal mode is treated as a factor, its number of levels is determined by its resolution: the `hour of the day' factor would have 24 levels, whereas the `day of the week' factor would have 7. Conversely, the columns represent the variables. Placing a mode along the columns allows for its visualization via loading plots, which identify the variables that contribute most to the differences between groups of observations.

A tensor allows different unfoldings depending on which modes are assigned to rows or columns. Depending on the specific problem, there may be multiple viable unfolding strategies, resulting in different ASCA models that provide insight into the data from distinct perspectives. Two ASCA models using different unfoldings of the energy consumption dataset are presented and discussed in Section~1 of the Supplementary Materials (SM).

To unfold the tensor, we begin with the non-temporal modes. These modes are placed in the rows when they are required to serve as factors for significance testing. This is the case of the `city' mode in our example, to test for spatial differences in consumption, as marked by the objective. Otherwise, they can be placed in the columns. However, several considerations should be made when assigning them to the columns. First, doing so would increase the number of columns several-fold, potentially raising computational costs. More importantly, a high number of modes in the columns can complicate the interpretation of loading plots, as multiple sources of variability are displayed simultaneously.

Regarding temporal modes---cyclostationary modes and the evolution mode---autocorrelation is a key concern. Autocorrelation occurs because nearby measurements in time tend to resemble each other, so each observation carries information from previous ones instead of being fully independent. 
If autocorrelation in observations is high, the factorization of ASCA might not eliminate it entirely, and it may reflect as autocorrelation in the residuals of the model. This is problematic because ASCA assumes exchangeability of relevant units, which inherently means that the residuals in both the data under analysis and the permutations need to be identically distributed~\citep{anderson2003permutation}. For this reason, our methodology avoids using fine-grained temporal modes---like `hour of the day' in our example---directly as observations (rows) in the ASCA model. Section~\ref{sec:autocorr} presents a criterion to assess whether the residuals of an ASCA model exhibit autocorrelation.

To determine the appropriate placement of cyclostationary temporal modes, we evaluate each mode sequentially, from the finest to the coarsest scale. The first step is to assess the presence of autocorrelation (e.g., by computing the sample autocorrelation function). If measurements at the resolution of a given mode exhibit so much autocorrelation that it propagates into the residuals, three alternatives are available: (1) place the mode in the rows and use a permutation scheme that permutes blocks of correlated rows together; (2) place the mode in the rows after averaging consecutive observations, thereby reducing temporal resolution; or (3) place it directly in the columns. The first two options can act as practical solutions when placing the mode in the columns would result in an unmanageable increase in dimensionality. Adapting the permutation scheme of ASCA can minimize the effect of autocorrelation on the inference, albeit at potential the cost of statistical power (i.e., the model will be more likely to miss significant effects) if the number of possible permutations is too small. On the other hand, averaging is a simple solution, but results in loss of resolution, which also translates to a potential decrease in statistical power for the same reason explained before, as possible group differences at finer time scales may be overlooked. Generally speaking, solution (1) is preferable to (2). An extended discussion on the choice of permutation schemes and their relation to autocorrelation is presented in Section~\ref{sec:autocorr}. Returning to our example, we begin with the finest-grained cyclostationary mode: `hour of the day'. Energy consumption is expected to be heavily correlated from one hour to the next. Measurements throughout the day could be averaged to obtain a single daily measure, reducing autocorrelation, or placed along the columns. We chose the latter to avoid losing information. The second temporal factor is `day of the week'. This factor presents some degree of autocorrelation, although less than the previous one. By placing it in the columns (Figure~\ref{fig:tensor_unfolding}B), the intra-week cycle can be viewed in the loadings. Alternatively, it could be situated in the rows alongside a permutation scheme that shuffles blocks of multiple consecutive days at a time.

If a temporal mode does not exhibit autocorrelation, it can be placed either in the rows or in the columns. Placing the mode in the rows makes it a factor that can be statistically tested using ASCA. However, this may result in a factor with many levels, thus consuming a large number of degrees of freedom and limiting the analysis of interactions with other factors. In our dataset, if we determine that measurements from one day to the next are far enough apart that autocorrelation in the residuals is not a concern, factor `day of the week' could be assigned to the rows (Figure~\ref{fig:tensor_unfolding}C). Doing so would allow us to test whether the intra-day cycle differs significantly across days of the week. Alternatively, the mode can be placed in the columns. In this case, statistical significance testing is not possible, but the temporal structure can still be interpreted through the loading plots. Examining score and loading plots together may provide qualitative insight into interactions, even though these cannot be formally tested. Note that, for the unfolding procedure to yield a matrix suitable for analysis, at least one mode must be assigned to the columns.

The evolution mode, however, cannot be placed in the columns because it is required for significance testing, which evaluates how the cyclic pattern evolves across different cycle iterations. Therefore, the evolution mode must be placed in the rows. If it exhibits autocorrelation, it must be averaged before inclusion, or the choice of permutations must be adapted. In our dataset, the `week of the year' mode is the evolution mode. Measurements taken one week apart are sufficiently independent. Since autocorrelation is not an issue, this mode goes in the rows unchanged, to assess the temporal evolution of energy consumptions, given our objective.

\subsection{Factor selection}

Once the data matrix is constructed, the factors of the ASCA model must be defined. These factors typically correspond to the modes associated with the rows. For example, in Figure~\ref{fig:tensor_unfolding}C, there are three factors: `city', with one level per city; `day of the week', with one level per day in a week, and `week of the year', with one level per week in the time series. However, certain considerations must be addressed. When a temporal factor other than the evolution factor is present, as is the case in Figure~\ref{fig:tensor_unfolding}C, it is important to define how they relate to each other. Factors can be either crossed or nested. Two factors are crossed when every level of one factor co-occurs with every level of the other. In contrast, a factor is nested within another when its levels belong to a specific level of the other. When factors are nested, interaction terms are confounded into the nested factor. Nested relationships are useful when you want to study the evolution of a time series at a finer time scale than one of the cyclostationary factors. In this case, the evolution factor can be nested in the cyclostationary factor. This allows for the examination of the evolution of the time series at a finer resolution.

Another key consideration for the evolution factor is the distinction between ordinal and nominal factors. A nominal factor possesses a level for each of its possible values (e.g., ``Monday'', ``Tuesday'', etc., for the `day of the week' factor). When performing inference on this type of factor, ASCA tests for differences between levels. For ordinal factors, however, values are considered ordered, and the test focuses on identifying a significant trend. Although factors are typically treated as nominal, it is beneficial to consider the evolution factor as ordinal if the study aims to detect a sustained trend. In the context of the energy consumption dataset, no trend is expected over the weeks, as consumption is subject to seasonal fluctuations that are not monotonic. Even if a trend were expected, it is recommended to first analyze the data with a nominal evolution factor and check for linear patterns in the visualizations (SM Figure~6) to assess the suitability of an ordinal factor. Here, ordinal factors are entered into the coding matrix $\mathbf{D}$ by mean-centering\footnote{Note that the ASCA model is based on a least squares regression of the response onto $\mathbf{D}$ and, as such, the model does not depend on the scale of the columns in $\mathbf{D}$.} the factor levels directly.

\section{Inference and diagnostics for autocorrelated data} \label{sec:autocorr}

When applying ASCA to time series data, inherent temporal dependence can violate the assumptions of the model. If autocorrelation persists after the initial data unfolding, permutation testing becomes unreliable. This section outlines how to identify residual dependence in the model and adapt the underlying permutation strategy to guarantee statistically valid inference.

\subsection{Model diagnostics}

Once the ASCA model is applied, the residuals must be checked to verify that they comply with the assumption of exchangeability before extracting conclusions. While the unfolding process (Section~\ref{sec:unfolding}) helps handle autocorrelation by rearranging the data, it might not be sufficient on its own to eliminate all dependence without averaging too many consecutive observations together---which could result in the loss of relevant information.

To easily test the residuals for remaining autocorrelation, we can use the Q-statistic of the residual matrix $\mathbf{E}$. This is computed for each row as the sum of squares across all columns. From this, if rows are chronologically ordered, the autocorrelation function (ACF) of the Q-statistics can be calculated and interpreted.

\begin{figure}[htb]
    \centering
    
    \begin{minipage}{0.85\linewidth}
        \textbf{A} \par \vspace{1pt}
        \centering
        \includegraphics[height=6.5cm, width=\linewidth, keepaspectratio]{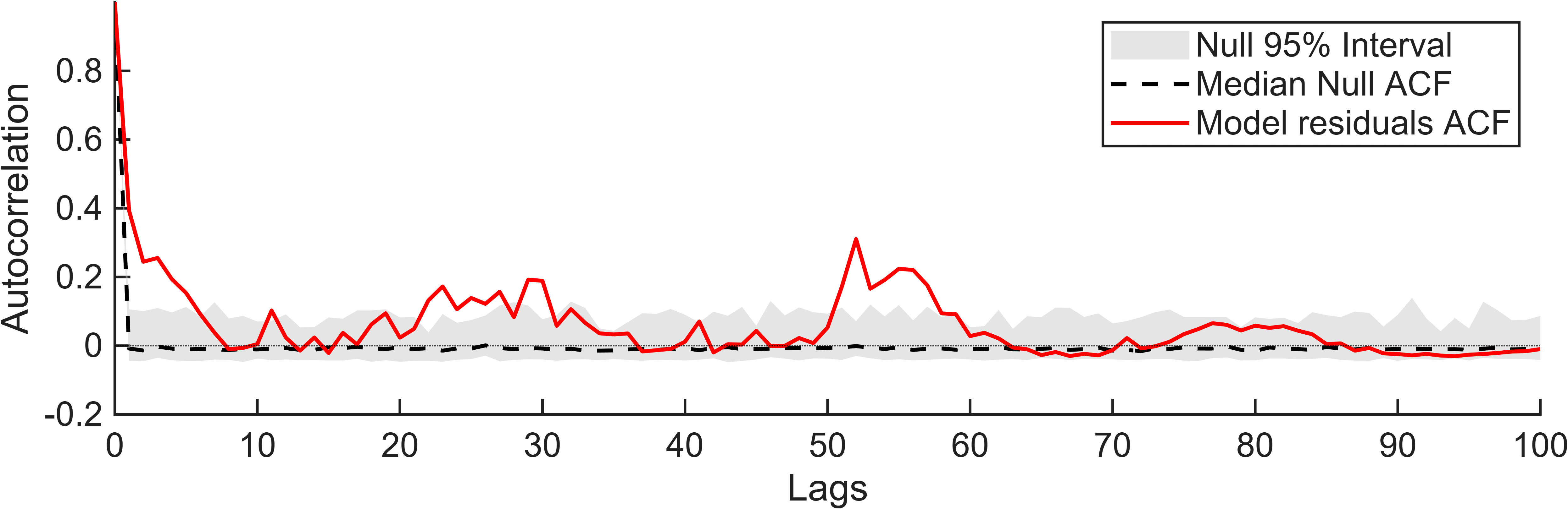}
    \end{minipage}
    
    \begin{minipage}{0.85\linewidth}
        \textbf{B} \par \vspace{1pt}
        \centering
        \includegraphics[height=6.5cm, width=\linewidth, keepaspectratio]{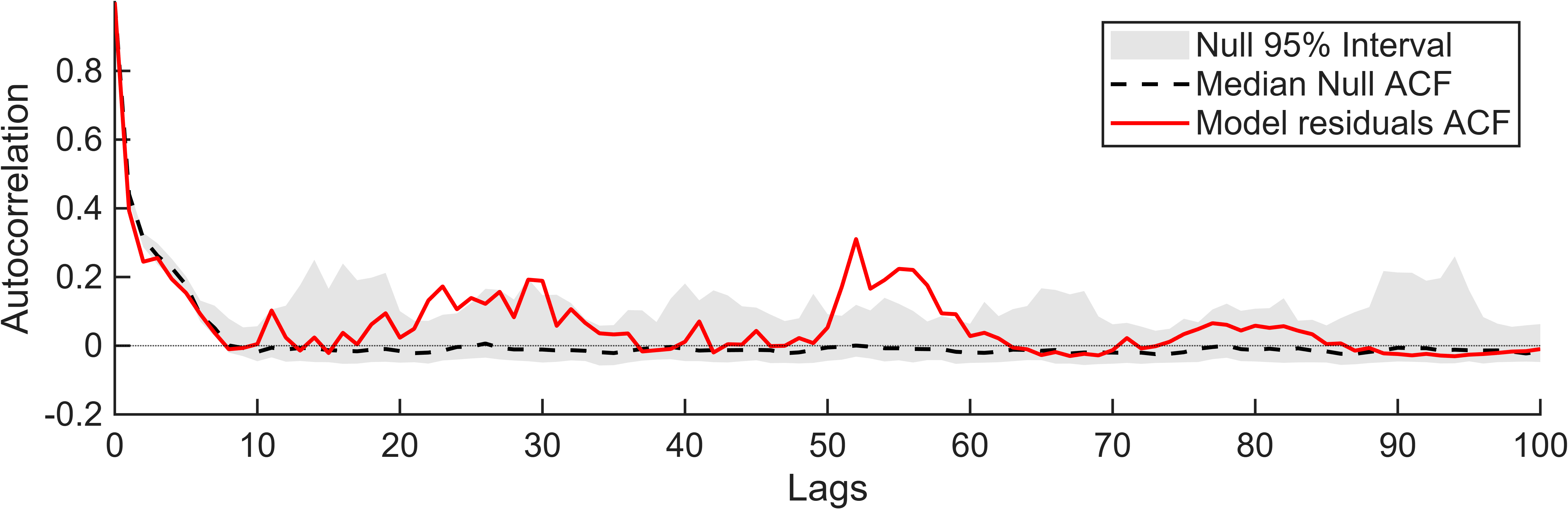}
    \end{minipage}
    

    \caption{\textbf{Sample autocorrelation function (ACF) of the Q-statistic of the residuals of the airborne pollen ASCA model.} The red line represents the proposed model, while the black dotted line shows the median across 100 unconstrained permutations (completely random row permutations in \textbf{A}, 6-month exchangeable units in \textbf{B}). The shaded region denotes the 95\% empirical interval across all permutations.}
    \label{fig:pollen_acf}
\end{figure}

Figure~\ref{fig:pollen_acf} illustrates how comparing the ACF of the residuals of our ASCA model against the ACF of an approximated null distribution can help determine if unhandled autocorrelation is present in the residuals. If the ACF of the residuals of our model deviates from the null ACF (Figure~\ref{fig:pollen_acf}A), the residuals exhibit autocorrelation that is not integrated into the null distribution. In this case, we adjust the permutation scheme (Figure~\ref{fig:pollen_acf}B), as detailed in the following subsection.

\subsection{Permutation schemes}

To address the autocorrelation detected in the model diagnostics, the permutation scheme of ASCA can be adapted. Permutation strategies generally fall into two categories: \textit{unrestricted permutations}, where units are shuffled freely across the dataset, and \textit{restricted permutations}, where permutations are forced to occur strictly within the levels of other factors. Both of these schemes can be performed on individual observations or on larger ``blocks'' of observations that act as ``exchangeable units'' \citep{anderson2003permutation}. In our methodology, we choose to use unrestricted permutations because they are faster, simpler to program, and maintain competitive statistical power \citep{camacho2023permutation}.

When using row-by-row unrestricted permutations, the ACF of the null distribution will be completely flat (Figure~\ref{fig:pollen_acf}A). However, this makes the null dataset devoid of any kind of dependence between consecutive observations. Consequently, if the data is autocorrelated, the natural dependence in the actual data will appear exceptional in comparison to the null, artificially inflating the Type I Error (i.e., making it more prone to falsely finding significance).

In correlated designs, the individual observations are no longer the correct exchangeable units; instead, larger groups or blocks of rows can become these exchangeable units. By grouping consecutive rows and permuting them as blocks, we inject comparable autocorrelation into the null distribution. The choice of block size is then derived from comparing the ACF of the residuals in our model with the ACF of the null distribution (Figure~\ref{fig:pollen_acf}B). This provides valid inference without the need to average rows and lose granularity.

Nonetheless, permuting multiple rows in unison may trade away some statistical power. When a data matrix has a total of $N$ rows and is permuted in blocks of size $n$, the number of possible permutations is given by $(N/n)!$. The larger the block size, the fewer permutations are available, which limits the minimum detectable p-value (Equation~\eqref{eq:pvalue}) and reduces statistical power if the number of available permutations becomes too small. Therefore, block size should be carefully tuned so the autocorrelation pattern in the residuals of the model closely aligns with the null ACF.

\section{Case studies}\label{sec:casestudies}

To demonstrate the applicability of the proposed methodology, we illustrate its use through two case studies based on real-world data. The first case study analyzes temperature records from various mountain lakes in Spain’s Sierra Nevada, a region recognized as a natural laboratory for global change research by LifeWatch ERIC, which was established by the European Commission \citep{SmartEcoMountains_LifeWatch_2025, ZamoraOliva2022}. The second focuses on the daily airborne pollen data set registered over 30 years in the city of Granada. Given that the airborne pollen concentrations are considered a form of biogenic pollution, recognized as one of the main causes of respiratory diseases \citep{Zeldin2006}, understanding the temporal evolution of pollen levels is of clinical interest.

All ASCA models were implemented in MATLAB using the \texttt{parglm} routine of the MEDA Toolbox \citep{MEDARep, camacho2015multivariate} and evaluated with 1,000 permutations. 
Additionally, to obtain more robust models, the \texttt{parglm} routine used for this paper computes the Moore-Penrose pseudoinverse of $\mathbf{D}^{\text{T}}\mathbf{D}$ in place of the standard inverse.

\subsection{Water temperature in Sierra Nevada lakes}\label{sec:lakes}

The first dataset contains measurements of water temperature in four high-mountain lakes in Sierra Nevada, Spain, located at altitudes between 2700 and 3100 meters: Aguas Verdes, La Caldera, Laguna Larga and Río Seco. Data collection was performed using seven sensors: one in Aguas Verdes, one in Río Seco, two in La Caldera, and three in Laguna Larga. Each sensor measured temperatures at three-hour intervals (eight times a day) from September 2009 to August 2021. Excluding leap days, each sensor generated 35,040 temperature measurements in total (12 years $\times$ 365 days per year $\times$ 8 readings per day). 

For this case study, our goal is to determine whether water temperature has risen in the past years, potentially due to climate change, and to examine the differences between lakes regarding these temperature changes.

The only non-temporal mode in this dataset is the `sensor'. Regarding temporal modes, two cyclostationary behaviors are identified: a daily (intra-day) cycle, where temperature fluctuates between day and night; and a yearly (intra-year) cycle, where temperature follows a seasonal pattern. This leads to the inclusion in the tensor of two cyclostationary temporal modes: `hour of the day' and `day of the year'. Since there is data for several years, a `year' mode was included to act as the evolution mode. This yielded a 4-mode tensor with dimensions 7 sensors $\times$ 8 hours $\times$ 365 days $\times$ 12 years.

When unfolding the tensor, the `sensor' mode is assigned to the rows because we are interested in testing for spatial differences in the data. The finest-grained temporal mode, the `hour of the day', exhibits strong autocorrelation from a measurement and the one taken three hours later. Therefore, it was assigned to the columns. The `day of the year', which also exhibits autocorrelation, is similarly placed in the columns. Thus, the complete intra-year cycle is contained within the columns. The evolution mode, `year', is placed in the rows, to assess the temporal evolution of temperatures, as marked by the objective. Consequently, the unfolding algorithm produces a 84 $\times$ 2920 matrix, where rows represent sensor-year combinations and columns correspond to the annual temperature readings.

\begin{figure}[p]
    \centering
    \small
    
    \begin{minipage}{0.45\linewidth}
        \textbf{A} \par \vspace{1pt}
        \centering
        \includegraphics[height=6cm, width=\linewidth, keepaspectratio]{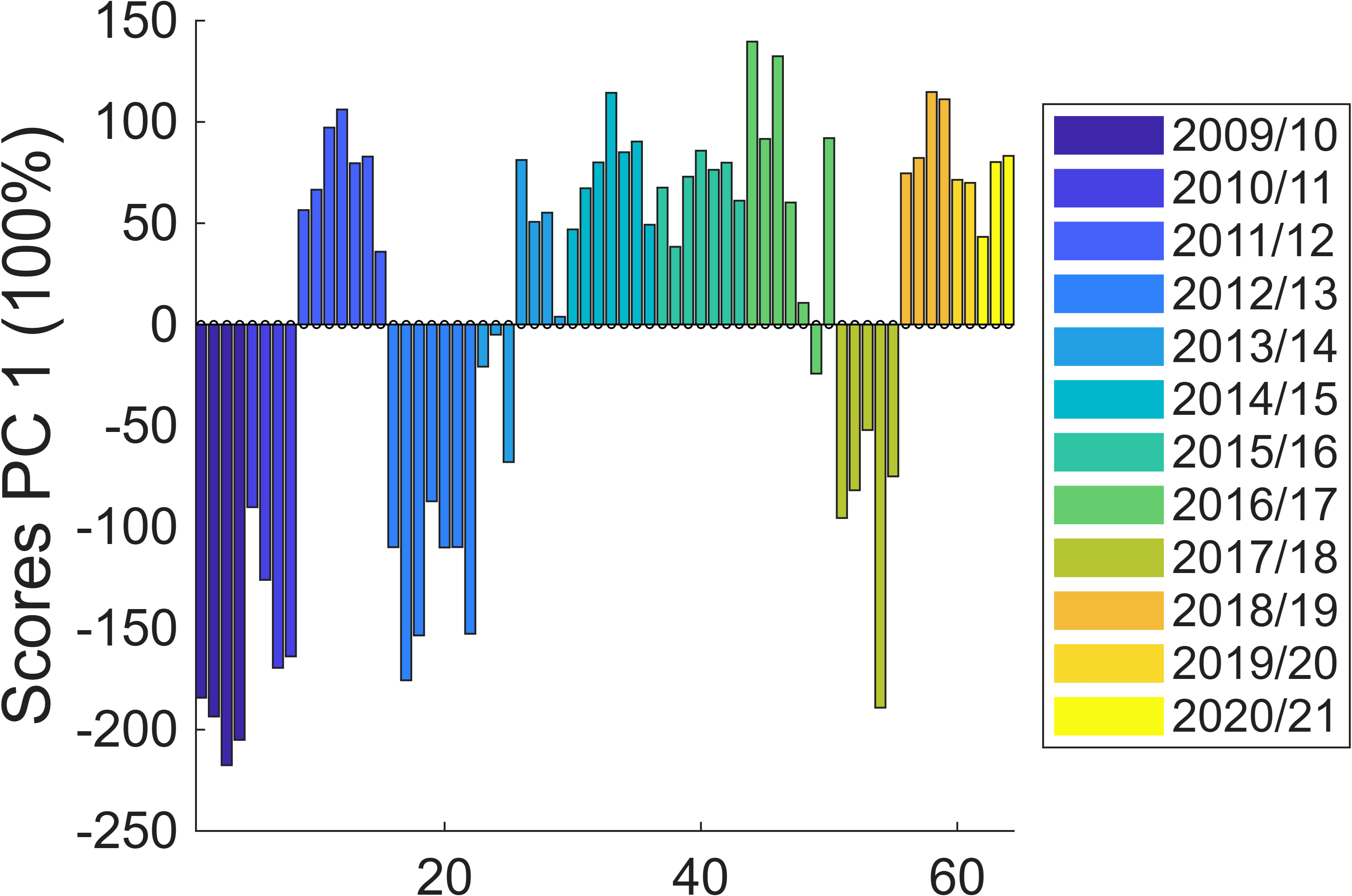}
    \end{minipage}
    \hspace{0.3cm}
    \begin{minipage}{0.45\linewidth}
        \textbf{B} \par \vspace{1pt}
        \centering
        \includegraphics[height=6cm, width=\linewidth, keepaspectratio]{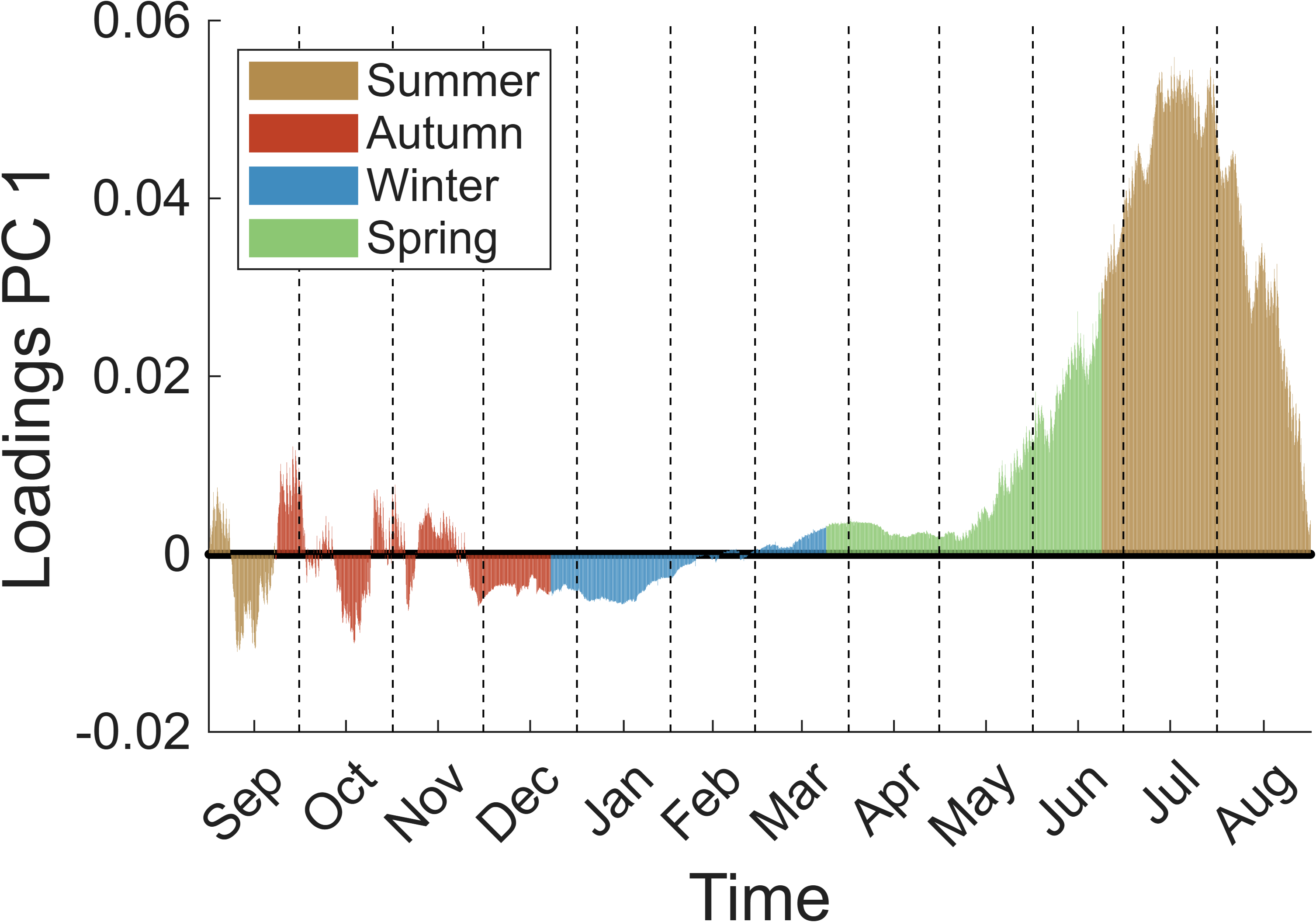}
    \end{minipage}


    \begin{minipage}{0.45\linewidth}
        \textbf{C} \par \vspace{1pt}
        \centering
        \includegraphics[height=6cm, width=\linewidth, keepaspectratio]{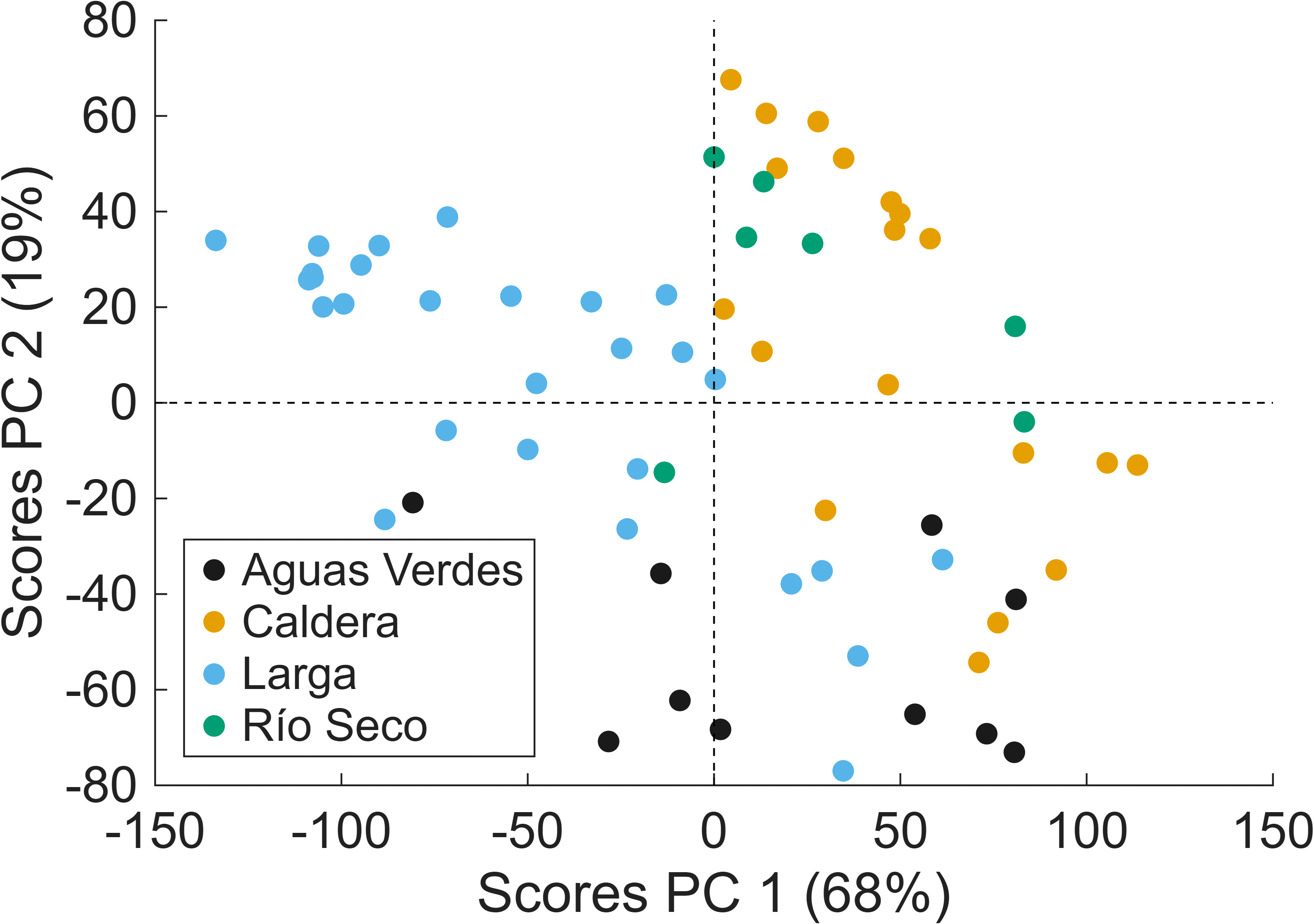}
    \end{minipage}
    \hspace{0.15cm}
    \begin{minipage}{0.45\linewidth}
        \textbf{D} \par \vspace{1pt}
        \centering
        \includegraphics[height=6cm, width=\linewidth, keepaspectratio]{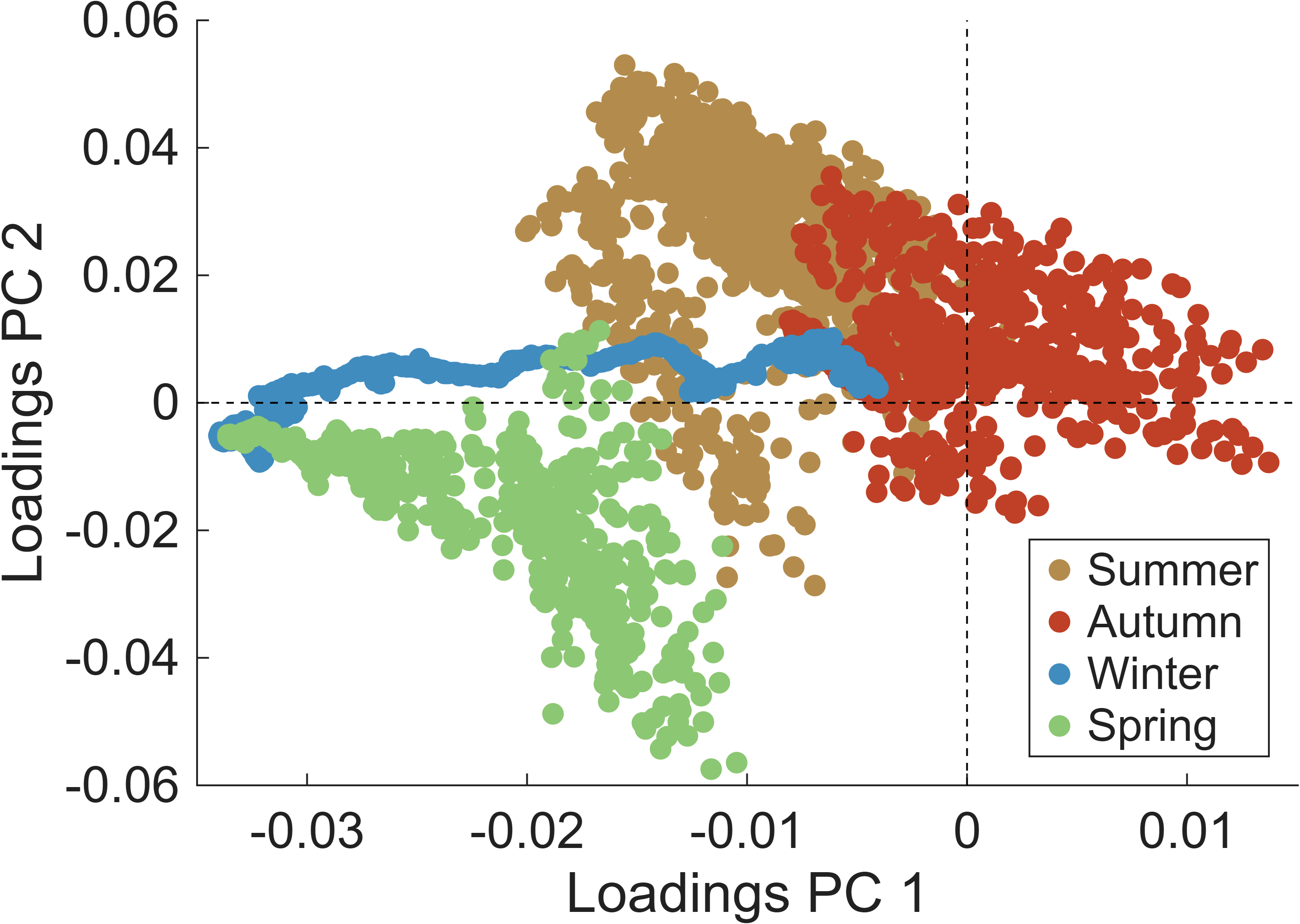}
    \end{minipage}


    \begin{minipage}{0.45\linewidth}
        \textbf{E} \par \vspace{1pt}
        \centering
        \includegraphics[height=6cm, width=\linewidth, keepaspectratio]{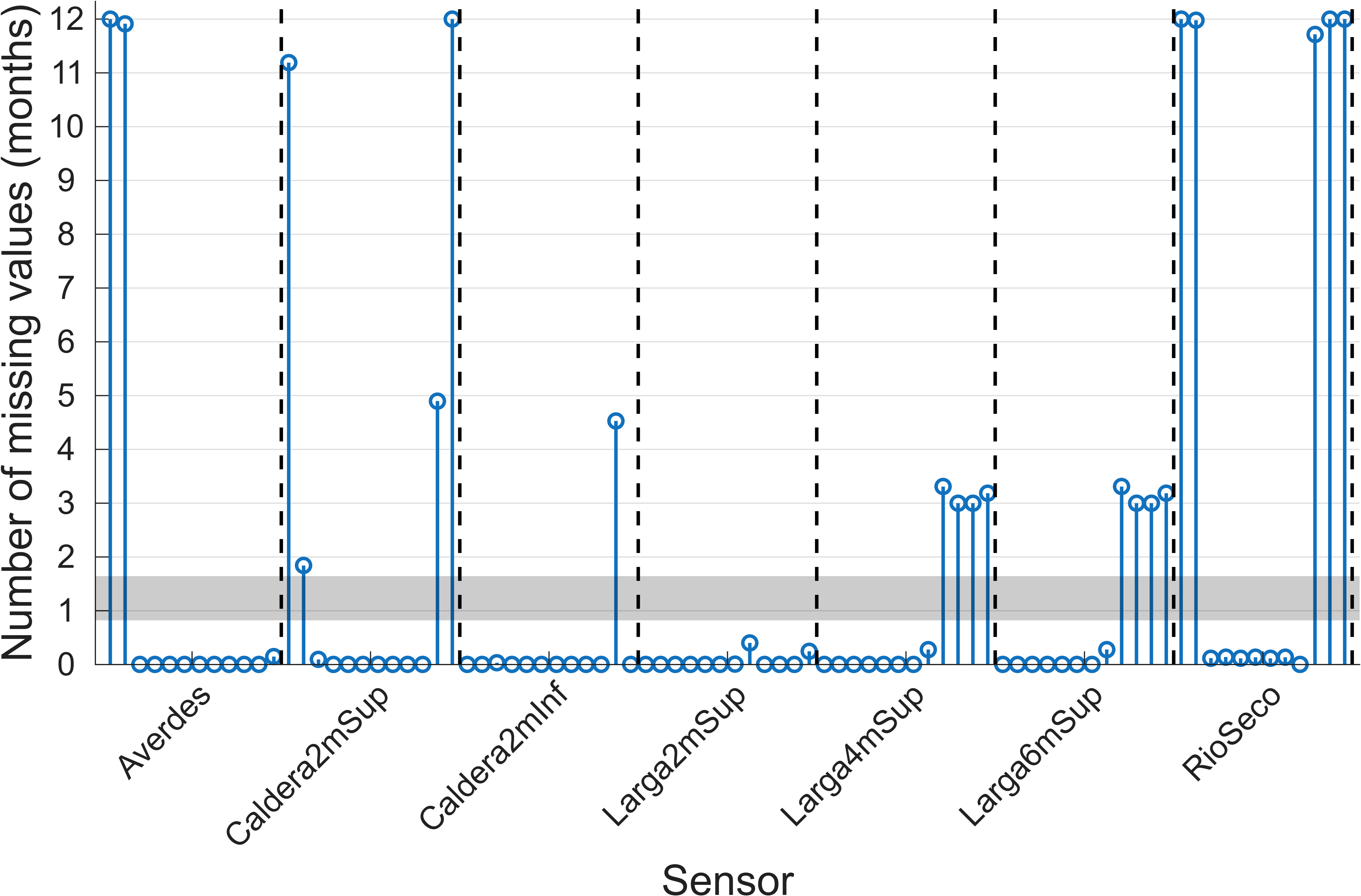}
    \end{minipage}
    \hspace{0.15cm}
    \begin{minipage}{0.45\linewidth}
        \textbf{F} \par \vspace{1pt}
        \centering
        \includegraphics[height=6cm, width=\linewidth, keepaspectratio]{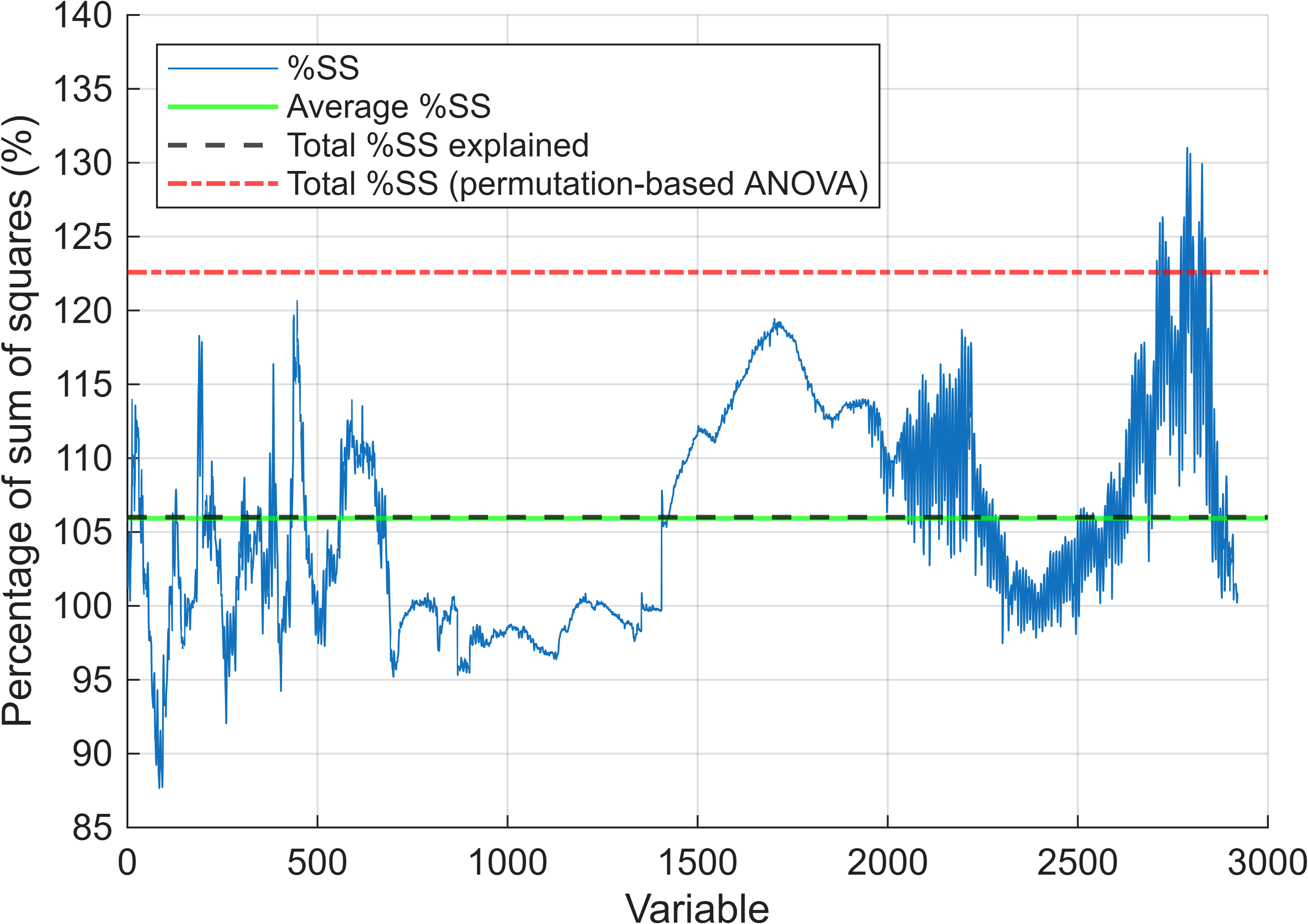}
    \end{minipage}


    \caption{\textbf{ASCA model of the Sierra Nevada lakes dataset.} \textbf{A} scores for factor `year'. \textbf{B} loadings for factor `year', colored by season. \textbf{C} scores for PCs 1 and 2 of factor `sensor', colored by lake. \textbf{D} loadings for PCs 1 and 2 of factor `sensor', colored by season. \textbf{E} number of missing values per sensor and year, expressed in months (a month is equivalent to 240 missing values). The gray band indicates the threshold above which rows were removed. \textbf{F} percentage of sum of squares of the ASCA factorization by variable (column of the data matrix), in blue. The mean of this sum-of-squares vector is represented in green. The total percentage of explained variance for ASCA and for the equivalent permutation-based ANOVA model are shown in black and red, respectively.}
    \label{fig:ASCA_lakes}
\end{figure}

Missing data varied considerably across lakes and years, with certain years missing entirely for some lakes (Figure~ \ref{fig:ASCA_lakes}E). Guided by visual inspection of this distribution, sensor-year combinations exceeding 300 missing values were treated as outliers and excluded, resulting in the removal of 20 observations. Following this exclusion, missing values accounted for 0.3\% of the dataset and were imputed using the column mean. An alternative approach involving data imputation within the permutation testing loop of ASCA \citep{Polushkina2025} was also tested, but it offered no additional insight compared to the original approach. Mean-centering was applied as a preprocessing step. ASCA was then performed on the $64 \times 2920$ matrix. Row-by-row unrestricted permutations were used, since the analysis of residuals (SM Figure~5) indicated that years are far apart enough that autocorrelation is not present in the residuals.

The two modes associated with the rows (i.e., `sensor' and `year') constitute the factors in the ASCA model. As one of the objectives was to evaluate the presence of a temperature trend over the years, the `year' factor was treated as ordinal. A preliminary ASCA analysis treating `year' as a nominal factor is available in Supplementary Materials (SM) Section~2.1.


Table~\ref{tab:lakes_anova}A presents the results of the significance tests for each factor. Both factors were statistically significant, with `year' being the dominant factor---approximately five times larger than the `sensor' factor in terms of mean squares. This confirms the presence of a temperature trend over the years. The year-sensor interaction was found to be non-significant, suggesting that the trend is consistent across all studied mountain lakes. The total sum of squares exceeds 100\% as a result of a moderately unbalanced design \citep{Camacho2026tutorial}. More details on this sum of squares are given later in this section.

The score and loading plots for the `year' factor are shown in Figure~\ref{fig:ASCA_lakes}A and \ref{fig:ASCA_lakes}B, respectively. These plots isolate the contribution of the year to the temperature variability from that of the sensor, effectively separating temporal from spatial effects. The scores for the latter years of the study are generally higher than those of the initial years, with two notable exceptions in 2011/2012 and 2017/2018. This behavior supports the existence of a significant rising trend in temperatures. The loadings associate high values of the first principal component with higher temperatures from May to August, particularly in July. Collectively, the scores and loadings indicate that summer temperatures have followed an upward trend over the years, barring the anomalous periods of 2011/2012 and 2017/2018, while temperatures throughout the rest of the year have not experienced significant changes.

Figures~\ref{fig:ASCA_lakes}C--D display the score and loading plots associated with the `sensor' factor for the first two principal components (PC1 and PC2). Coloring observations by lake highlights spatial differences in the intra-annual evolution of temperature, independent of interannual variability. PC1 accounts for $68\%$ of the variability associated with the `sensor' factor, and distinguishes autumn measurements (positive loadings) from those of late winter and early spring (negative loadings). Observations from Laguna Larga are positioned on the left side of the score plot, indicating that this lake generally exhibits higher temperatures during late winter and early spring relative to the other lakes. Conversely, La Caldera shows higher temperatures during autumn. PC2 ($19\%$ of explained variability of the factor) approximately separates late spring from late summer. Here, Aguas Verdes tends to experience higher temperatures in late spring, while Río Seco displays the opposite behavior. PC1 largely separates Laguna Larga from the other lakes. This differentiation may stem from geographical location: Laguna Larga, on the northern slope of Sierra Nevada, receives more sunlight in spring, while the other three lakes, on the southern slope, receive more sunlight in autumn, due to seasonal changes in the sun’s angle caused by the tilt of the Earth’s axis.

Addressing the objectives of this case study, the analysis with ASCA confirms a rise in water temperatures across the Sierra Nevada lakes from 2009 to 2021. This warming specifically occurs in the summer months, with no significant changes in other seasons. Regarding the difference between lakes, the analysis reveals that while each site exhibits its own unique behavior, possibly due to specific geography, the warming trend is consistent across all of them. While widespread temperature increases were observed across these high-mountain lakes, whether it is driven by anthropogenic climate change remains to be determined.

A parametric ANOVA model and a permutation-based ANOVA model were also applied to compare the performance of ASCA with other inferential approaches (Table~\ref{tab:lakes_anova}B--C). The permutation-based ANOVA is equivalent to the factorization and inference phases of ASCA. It was included as a non-parametric alternative to ANOVA, as it utilizes permutation testing, and therefore does not require the assumption of normality. Consistent with the ASCA setup, `sensor' and `year' (treated as an ordinal factor) were used as factors. To minimize autocorrelation and ensure comparability with the ASCA analysis, the input for these models was a univariate vector of 64 values, representing the yearly temperature average for each sensor-year combination. Parametric ANOVA computations were performed using the \texttt{anovan} function in MATLAB, while the permutation-based ANOVA utilized the \texttt{parglm} function of the MEDA Toolbox for MATLAB \citep{MEDARep, camacho2015multivariate}. An analysis of residuals is provided in SM Section~2.1 for all models.

\begin{table}[ht]
\small
\centering
\raggedright \hspace{1cm} \textbf{A} \\
\centering
\resizebox{0.75\textwidth}{!}{%
\begin{tabular}{lrrrrrr}
\hline
 & \textbf{SS ($\times 10^5$)} & \textbf{\%SS} & \textbf{df} & \textbf{MS ($\times 10^4$)} & \textbf{F} & \textbf{$p$-value} \\
\hline
\hline
Year (Ordinal)       & 1.81  & 14.69  & 1  & 18.10 & 10.97 & $<0.001$ \\
Sensor               & 2.28  & 18.49  & 6  & 3.80  & 2.30  & 0.016    \\
Year $\times$ Sensor & 0.72  & 5.86   & 6  & 1.21  & 0.73  & 0.882    \\
Residuals            & 8.26  & 66.96  & 50 & 1.65  & --    & --       \\
\hline
Total                & 12.3 & 106.00 & 63 & 1.96  & --    & --       \\
\hline
\hline
\end{tabular}
}
\vspace{0.3cm}

\raggedright \hspace{1cm} \textbf{B} \\
\centering
\resizebox{0.7\textwidth}{!}{%
\begin{tabular}{lrrrrrr}
\hline
 & \textbf{SS} & \textbf{\%SS} & \textbf{df} & \textbf{MS} & \textbf{F} & \textbf{$p$-value} \\
\hline
\hline
Year                 & 10.72 & 11.04 & 1  & 10.72 & 8.88 & 0.004 \\
Sensor               & 5.78  & 5.95  & 6  & 0.96  & 0.80 & 0.577 \\
Year $\times$ Sensor & 5.81  & 5.98  & 6  & 0.97  & 0.80 & 0.573 \\
Residuals            & 60.37 & 62.14 & 50 & 1.21  & --   & --    \\
\hline
Total                & 97.16 & 85.10 & 63 & 1.54  & --   & --    \\
\hline
\hline
\end{tabular}
}
\vspace{0.3cm}

\raggedright \hspace{1cm} \textbf{C} \\
\centering
\resizebox{0.7\textwidth}{!}{%
\begin{tabular}{lrrrrrr}
\hline
 & \textbf{SS} & \textbf{\%SS} & \textbf{df} & \textbf{MS} & \textbf{F} & \textbf{$p$-value} \\
\hline
\hline
Year                 & 15.12 & 15.56  & 1  & 15.12 & 12.19 & 0.006   \\
Sensor               & 36.80 & 37.87  & 6  & 6.13  & 4.94  & $<0.001$ \\
Year $\times$ Sensor & 5.16  & 5.31   & 6  & 0.86  & 0.69  & 0.736   \\
Residuals            & 62.03 & 63.85  & 50 & 1.24  & --    & --      \\
\hline
Total                & 97.16 & 122.59 & 63 & 1.54  & --    & --      \\
\hline
\hline
\end{tabular}

}
\caption{ASCA (\textbf{A}), parametric ANOVA (\textbf{B}) and permutation-based ANOVA (\textbf{C}) tables for the Sierra Nevada lakes dataset. The columns indicate, respectively: the sum of squares (SS), its percentage relative to the total (\%SS), degrees of freedom (df), mean squares (MS), the F-statistic (F), and the associated $p$-value. The deviation of the total \%SS from 100\% is caused by a loss of orthogonality due to a moderately unbalanced design.}
\label{tab:lakes_anova}
\end{table}

It should be noted that this comparison is not intended as an algorithmic benchmark, as applying ANOVA to yearly averages inherently discards within-year temporal information. Rather, we intentionally include this simplistic approach to represent another potential analysis strategy for this case study. This comparison highlights a key advantage of our proposed pipeline: the multivariate time series unfolding process. While the unfolded data generated by our pipeline could be analyzed using other multivariate techniques, we advocate for ASCA in this framework due to its capacity to produce interpretable visualizations of the loadings, which enable the joint interpretation of rows (years and lakes) and columns (intra-annual variation).

Mirroring the ASCA results, the `year' factor exhibited a larger MS than the `sensor' factor in both models. However, the significance levels differed: while the permutation-based model found both factors to be statistically significant, the parametric ANOVA only detected `year' as a significant factor. This reveals a loss of statistical power, as it fails to capture the spatial differences that were identified using ASCA. No significant interaction was detected in either case. It is also worth noting that these approaches rely on the aggregation of all temperature readings within a year. This process functions as a ``low-pass filter'' that discards intra-annual variability, which contains relevant information. By removing this variability, the model loses statistical power. Furthermore, averaging also hinders interpretability. For instance, even if the ANOVA models suggest a trend in temperatures, they obscure the fact that said trend only manifests during the summer---as observed with ASCA---rather than across the annual average. Ultimately, these results highlight the advantages of ASCA over traditional methods.

It is worth noting that the total percentage of the sum of squares explained is not $100\%$ for any of the models considered (neither the ASCA models nor both univariate ANOVA models). This discrepancy is a consequence of non-orthogonal factorization resulting from a moderately unbalanced design \citep{Camacho2026tutorial}, caused by missing data from several sensors across multiple years (Figure~\ref{fig:ASCA_lakes}E). In such unbalanced designs, factors lose their orthogonality, leading to an overlap in the variance associated with different factors. The degree of deviation from $100\%$ varies significantly across models, with both ANOVA models straying further than the ASCA model. A closer examination of the difference between the permutation-based ANOVA and ASCA, which employ the same factorization and inference methods, reveals a notable distinction in how both methods behave in this case study.
ASCA accounts for the sum of squares of every individual variable (each measurement in a year). In this dataset, this approach mitigated the impact of the non-orthogonal design on variance overlap, with the total percentage of sum of squares explained by ASCA being close to the average percentage of all variables (Figure~\ref{fig:ASCA_lakes}F). In the case of univariate ANOVA, by averaging all measurements in a year, the variance captured by the factors inherited the effects of the variables most affected by the unbalanced design, and inflated the total percentage of variance explained. In summary, beyond its advantages in interpretation and statistical power, ASCA outperformed ANOVA through its better handling of variance, resulting in a more precise separation of factors under the conditions evaluated in this case study.

Furthermore, the discrepancy between parametric and permutation-based ANOVA regarding this percentage stems from the different types of sum of squares utilized in the implementation of the \texttt{anovan} and \texttt{parglm} functions. The parametric ANOVA displayed in Table~\ref{tab:lakes_anova} uses a constrained type III sum of squares. Results for other types of sum of squares are provided in SM Table 4.

\subsection{Airborne pollen concentration in Granada}\label{sec:pollen}
The second dataset includes daily measurements of airborne pollen concentration for 44 different pollen types, collected over a period of 30 years (from 1993 to 2022). This dataset was provided by the Unit of Biological Air Quality of the University of Granada, Granada (Southern Spain), which works following the requirements of the European Aeroallergen Network \citep{Galan2014} and the Spanish Aerobiology Network \citep{Galan2007}.

The goal of this study case was to determine what changes in the pollen concentrations have taken place over the 30 years of measurements. We also sought to determine whether the seasonal behavior of the measured pollen types has shifted over the span of 30 years.

This dataset was interpreted as a tensor with one non-temporal mode (`pollen types') and two temporal modes: the fortnight (set of 14 days) of the year and the year. `Fortnight of the year' constitutes a cyclostationary temporal mode, whereas `year' is the evolution mode.

The fortnights are obtained by averaging sets of 14 consecutive days. We do this in order to account for the autocorrelation of adjacent daily measurements. In doing this, we also simplify the interpretation: if we were to avoid autocorrelation by moving the day mode to our columns, the interpretation of the loadings would be much trickier, since it would mix the effects of the day with the effects of the pollen types. It is worth noting that days with missing measurements were not included in the averaging. That way, if a fortnight contains a day with missing data, the corresponding average is computed using just the 13 measurements available. If all 14 days of measurements were missing for a given set, that fortnight is discarded from the study. This accounted for 26 of the 780 total fortnights ($3.33\%$).

One caveat to consider is that by splitting the 365 days of the year\footnote{The 29th of February was preemptively discarded from all leap years for better comparability between the different years.} into 26 fortnights, the last day of the year becomes isolated, as $14 \times 26 = 364$. To account for this, we include the last day of the year into the last fortnight of the year. This makes it so that this fortnight is composed of 15 days, instead of the usual 14. Because we calculate daily averages, data scaling remains unaffected by the addition of one day to the final fortnight, and the negligible difference in variance does not compromise comparability between periods. All in all, our final tensor is a 3-mode tensor with dimensions 30 years $\times$ 26 fortnights $\times$ 44 pollen types.

Once our tensor is defined, we move on to its unfolding. In order to answer the research questions we are interested in we need the temporal modes `fortnight of the year' and `year' in the rows of our data matrix so that we can analyze them as factors; while keeping the non-temporal mode `pollen type' as variables in the columns of the matrix. This results in a matrix with dimensions 754 $\times$ 44 after accounting for the discarded fortnights. We tried different unfoldings, but we found that this approach provided the richest insights.

To obtain a general overview of all pollen types and observe how both common and less abundant taxa evolve over time, the data was autoscaled before analysis. This preprocessing highlights the inner variability of each pollen type. Since different species produce widely different amounts of pollen, studying them with this preprocessing ensures each pollen type contributes equally to the model, and allows less abundant pollen types to be represented in the visualizations. For comparison, an ASCA analysis using mean-centering instead of autoscaling was also conducted and can be found in SM Section~2.2. We found that this additional model focused heavily on only two pollen types, \textit{Olea} and \textit{Cupressaceae}, and offered fewer meaningful insights than the autoscaled data.

Beyond preprocessing, we also need to choose the parameters to use for the ASCA model. For instance, we do not want to study fortnight as a nested factor of year. Doing so would mean that a fortnight of a specific year is not comparable to that same fortnight on a different year, which is not something we want to assume. Furthermore, nesting the two factors would disable us from studying their interaction. In this case study, we care about this interaction, as within it we will find the answer to whether or not the seasonal  behavior has veered over the years.  Therefore, we will not consider the two factors to be nested. Likewise, we will choose to study factor year as an ordinal one, since one of our objectives is to evaluate the presence of a trend across the years. An ASCA analysis treating factor year as nominal was also conducted and is available in SM Section~2.2.

We initially evaluated a standard random permutation scheme. However, model residuals exhibited autocorrelation across consecutive fortnights. Rather than further aggregating the data, which would lead to information loss, we adapted the permutation scheme to incorporate this temporal autocorrelation directly into the null distribution, thereby preserving the validity of the inference. Based on the residual correlogram (SM Figure~8), we implemented a block permutation scheme using 13-row blocks (corresponding to 13 consecutive fortnights, or 6 months). A comprehensive residual analysis is provided in SM Section~2.2.

\begin{table}[]
\centering
\resizebox{0.9\textwidth}{!}{%
\begin{tabular}{lrrrrrr}
\hline
 & \textbf{SS ($\times 10^3$)} & \textbf{\%SS} & \textbf{df} & \textbf{MS} & \textbf{F} & \textbf{$p$-value} \\
\hline
\hline
Year (Ordinal)          & 0.40  & 1.3   & 1   & 398.5 & 12.94 & $<0.001$ \\
Fortnight               & 7.64  & 24.2  & 25  & 305.6 & 9.93  & $<0.001$ \\
Fortnight $\times$ Year & 1.94  & 6.1   & 25  & 77.6  & 2.52  & $<0.001$ \\
Residuals               & 21.6 & 68.4  & 702 & 30.8  & --    & --       \\
\hline
Total                   & 31.6 & 100.0 & 753 & 42.0  & --    & --       \\
\hline
\hline
\end{tabular}
}
\caption{ASCA table of the pollen dataset: The columns indicate, respectively: the sum of squares (SS), its percentage relative to the total (\%SS), degrees of freedom (df), mean squares (MS), the F-statistic (F), and the associated $p$-value. }
\label{tab:ASCA_pollen}
\end{table}

Table \ref{tab:ASCA_pollen} displays the results for significance testing of each of the factors. The significance tests show clear significance for both factors and, to a lesser magnitude, their interaction.

\begin{figure}


    \begin{minipage}{0.49\linewidth}
        \textbf{A} \par \vspace{1pt}
        \centering
        \includegraphics[height=6.5cm, width=\linewidth, keepaspectratio]{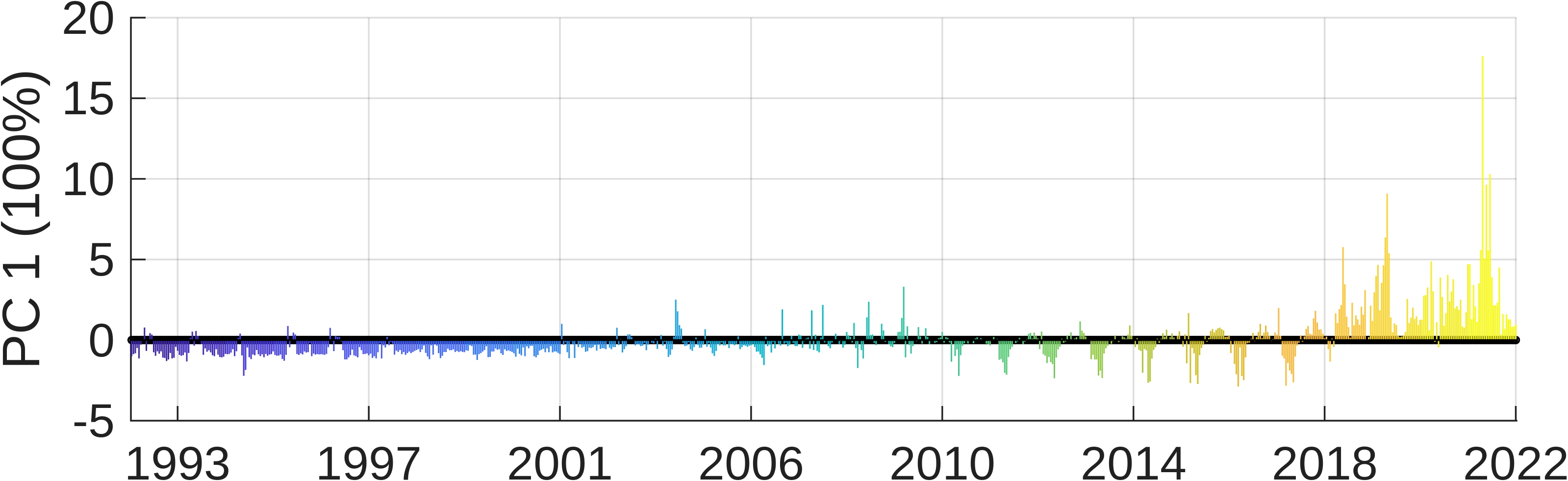}
        
    \end{minipage}
    \hfill
    \begin{minipage}{0.49\linewidth}
        \textbf{B} \par \vspace{1pt}
        \centering
        \includegraphics[height=6.5cm, width=\linewidth, keepaspectratio]{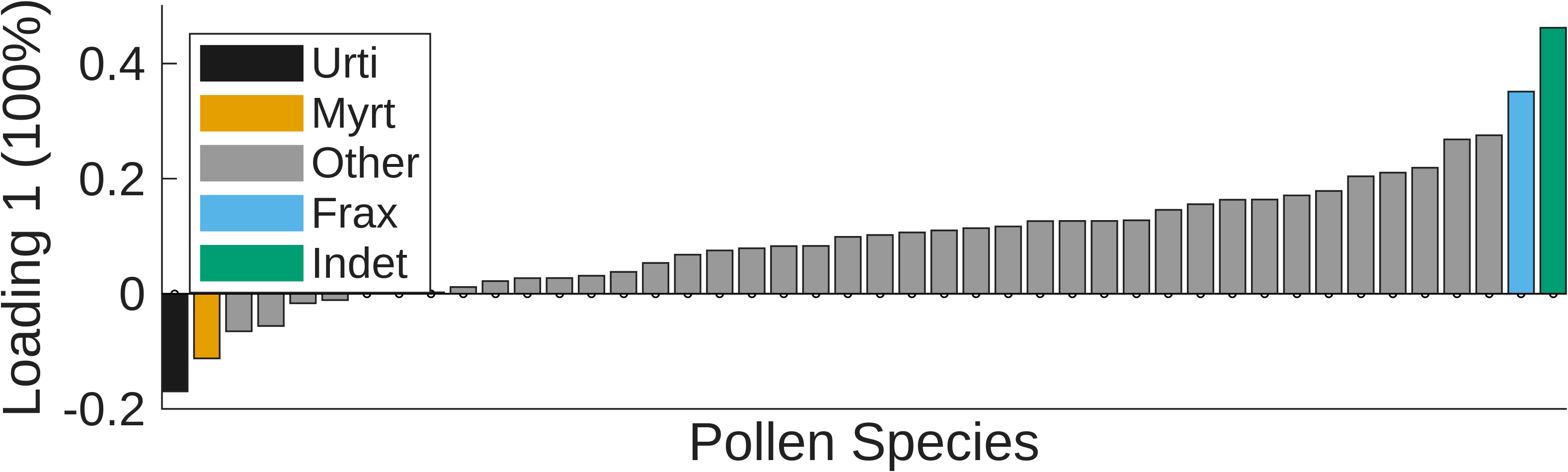}
    \end{minipage}

    \vspace{0.1cm}

    \begin{minipage}{0.69\linewidth}
        \textbf{C} \par \vspace{1pt}
        \centering
        \hspace{-1.8cm}
        \includegraphics[height=6.5cm, width=\linewidth, keepaspectratio]{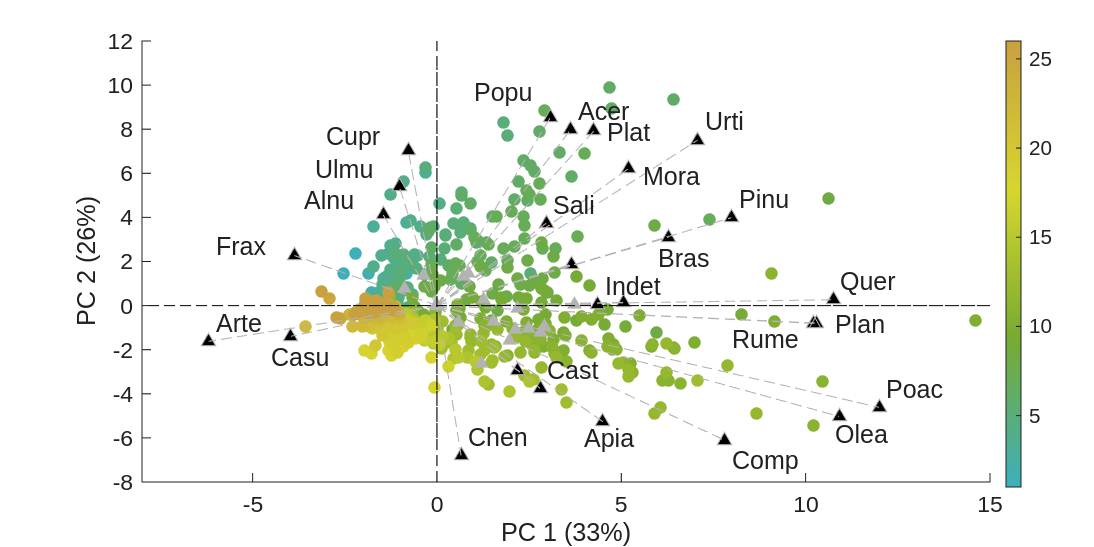}
    \end{minipage}
    \hspace{-0.8cm}
    \begin{minipage}{0.30\linewidth}
    \vspace{-0.7cm}
        \textbf{D} \par \vspace{1pt}
        \centering
        \includegraphics[width=\linewidth, keepaspectratio, clip]{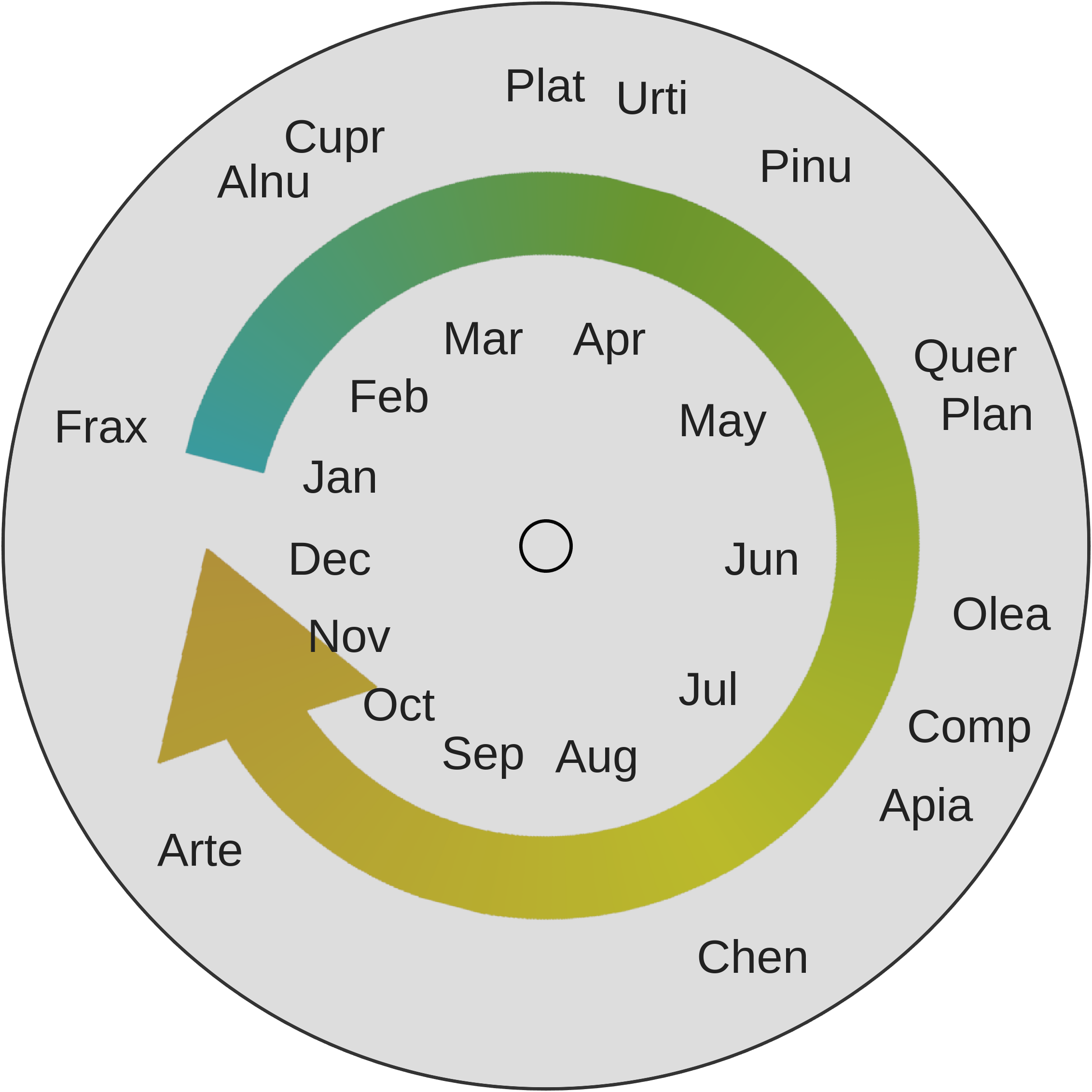}
    \end{minipage}

    \vspace{0.0cm}

    \begin{minipage}{0.49\linewidth}
        \textbf{E} \par \vspace{1pt}
        \centering
        \includegraphics[height=6.5cm, width=\linewidth, keepaspectratio]{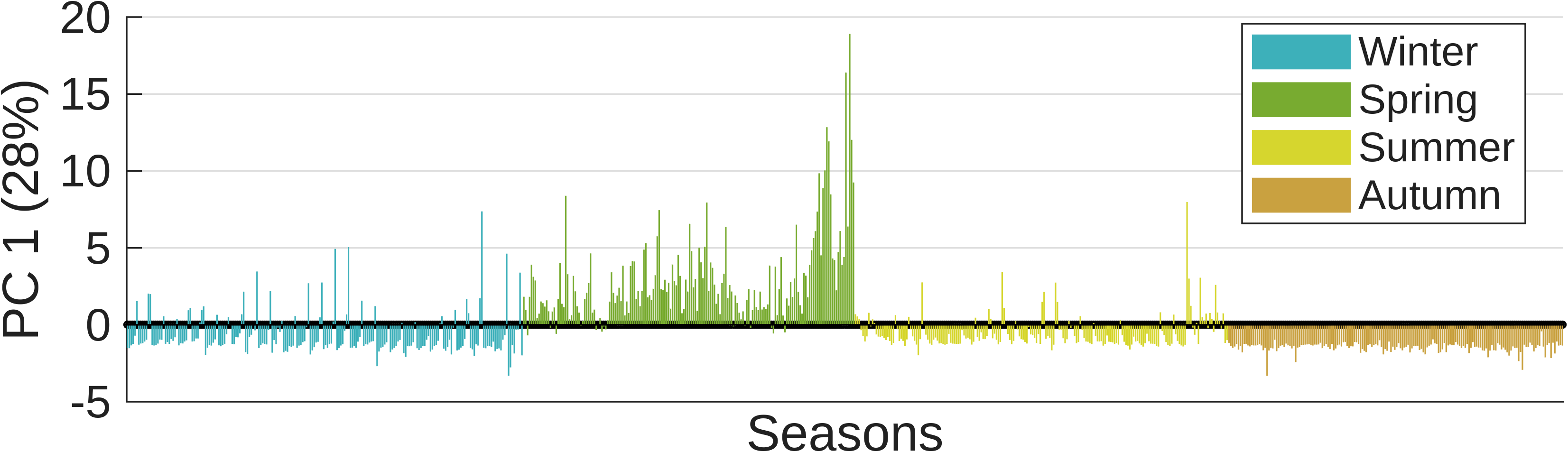}
        
    \end{minipage}
    \hfill
    \begin{minipage}{0.49\linewidth}
        \textbf{F} \par \vspace{1pt}
        \centering
        \includegraphics[height=6.5cm, width=\linewidth, keepaspectratio]{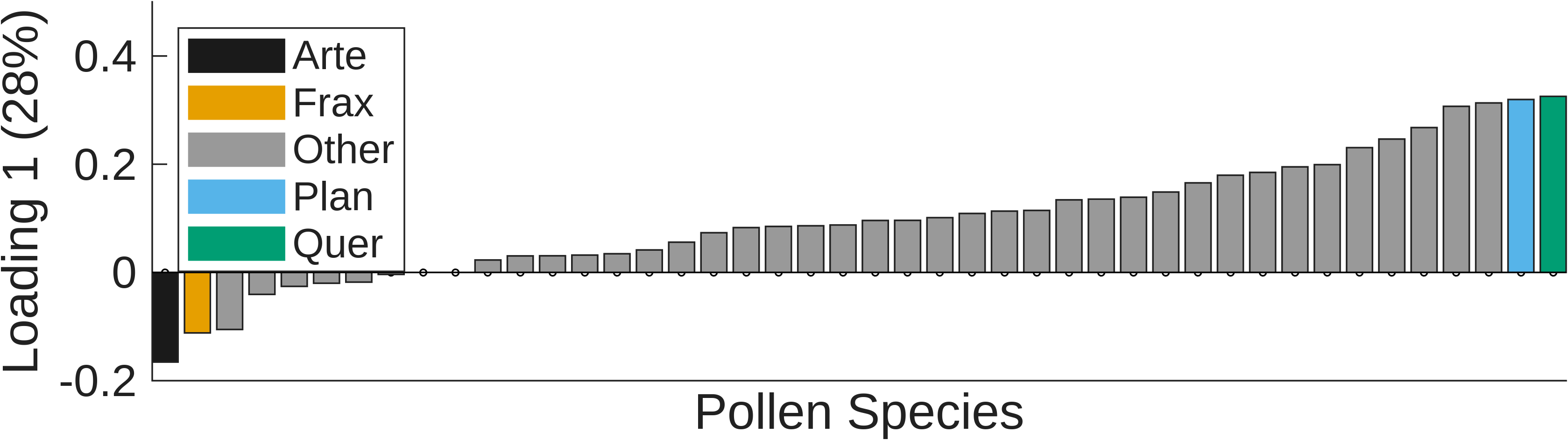}
    \end{minipage}

    \vspace{0.0cm}
    \centering
    \begin{minipage}{0.99\linewidth}
        \textbf{G} \par \vspace{1pt}
        \includegraphics[width=\linewidth]{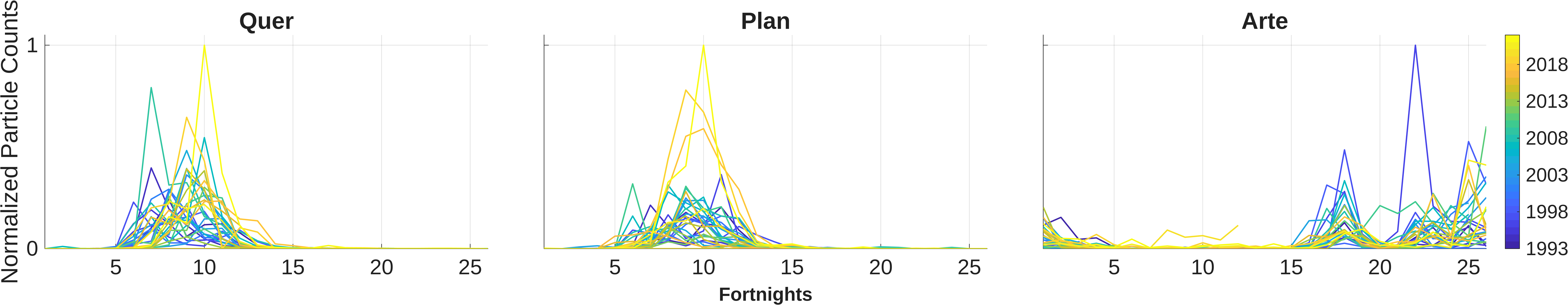}
    \end{minipage}

    \caption{\textbf{ASCA model of the pollen dataset. Main factors}. \textbf{A} scores of factor year colored by year. We can see a heavy increase in pollen concentrations from 2018 to 2022 compared to previous years. \textbf{B} loadings of factor year colored to highlight the pollen types with the largest growth and decay over the years. \textbf{C} biplot of factor fortnight with scores colored according to fortnight. The cyclic annual behavior of seasons is highlighted by the scores, while the loadings indicate which pollen types show the highest particle counts at each time of the year. \textbf{D} diagram of the yearly clock-like pattern present in the biplot of factor fortnight. \textbf{E} scores of the first component of the interaction between year and fortnight. The scores have been sorted according to season and year. We can see that all seasons remain constant over the years, with the exception of spring. Spring displays a growth in pollen concentrations over the years. \textbf{F} loadings of the first component of the interaction between year and fortnight, colored to showcase the pollen types with the largest and smallest increase during spring over the years. We can see that \textit{Quercus} and \textit{Plantago} pollen types have experienced the largest spring growth over the years, while \textit{Artemisia} and \textit{Fraxinus} pollen types show the smallest spring increase over the years. \textbf{G} normalized yearly counts of \textit{Quercus}, \textit{Plantago} and \textit{Artemisia} pollen types  across the 30 years. Both \textit{Quercus} and \textit{Plantago} display spring growth in the latter years, while \textit{Artemisia} pollen displays a decrease in counts in seasons other than spring. \newline
    \scriptsize The pollen types shown in the figures are named as acronyms: Arte, Artemisia; Casu, Casuarina; Frax, Fraxinus; Alnu, Alnus; Ulmu, Ulmus; Cupr, Cupressaceae; Popu, Populus; Acer, Acer; Plat, Platanus; Sali, Salix; Mora, Moraceae; Urti, Urticaceae; Pinu, Pinus; Bras, Brassicaceae; Indet, Indeterminate; Quer, Quercus; Plan, Plantago; Rume, Rumex; Poac, Poaceae; Cast, Castanea; Olea, Olea; Comp, Compositae; Apia, Apiaceae; Chen, Amaranthaceae.} 
    \label{fig:ASCA_pollen}
\end{figure}

Figures \ref{fig:ASCA_pollen}A and \ref{fig:ASCA_pollen}B show the scores and loadings of factor year in the ASCA model. These plots showcase a major increase in pollen concentrations during the last few years of measurements (from 2018 to 2022). Looking at the loadings of the model we can see that \textit{Fraxinus} and the group of ``Indeterminate'' pollen have increased the most in the last years of measurements, while \textit{Urticaceae} and \textit{Myrtaceae} pollen have decreased the most. Notably, the pollen type with the highest increase is the group of indeterminate type. 

After contrasting this finding with the original data, we discovered that it was due to an inconsistency in the data recording process. Looking at the original recordings, we detected that the cataloging that ran from 2021 to 2023---performed by less experienced staff---was partially incorrect, as it featured an overestimation of the ``Indeterminate pollen'' group. While this confirms that the result was due to a technical artifact, it underscores a major practical strength of ASCA: its high interpretability, which makes data quality issues like this one more apparent to the researcher and less likely to be mistaken with true biological effects.

Figure \ref{fig:ASCA_pollen}C shows the scores and loadings of the first two components of factor fortnight in a biplot. This plot showcases the cyclic nature of seasons and fortnights within a year. The passage of time can be spotted in the cyclic pattern of the scores: with time advancing clock-wise. The diagram in Figure \ref{fig:ASCA_pollen}D highlights this yearly pattern, specifying the pollen types that are the most present on each month of the year. For instance, we can see \textit{Pinus} pollen type being more dominant during early spring, while \textit{Olea} pollen type is more dominant during late spring. These visualizations serve as a baseline to the question of whether or not the seasonal behavior of the pollen has veered over the years, which we can answer by looking at the interaction of the two factors.

Figures \ref{fig:ASCA_pollen}E and \ref{fig:ASCA_pollen}F show the scores and loadings of the first component of the interaction of the `year' and 'fortnight' factors. Additional visualizations of this interaction can be found in SM Section~2.2. The scores in \ref{fig:ASCA_pollen}E show us that the increase along the years of measurements has not happened homogeneously across all four seasons. Rather, it is spring that has experienced a notable rise in the concentrations of pollen along the years. Figure \ref{fig:ASCA_pollen}F identifies the pollen types that have risen the most and the less during spring across the years. \textit{Quercus} and \textit{Plantago} pollen types have shown the greatest increases, while \textit{Artemisia} and \textit{Fraxinus} pollen types have decreased.

We can confirm these results by looking at the original data of the identified pollen types. Figure \ref{fig:ASCA_pollen}G shows the normalized pollen concentrations of \textit{Quercus}, \textit{Plantago} and \textit{Artemisia}, across all fortnights of all years. We can see that, indeed, the pollen types of \textit{Quercus} and \textit{Plantago} exhibit an increase in spring concentrations during the last years of measurements, while \textit{Artemisia} exhibits virtually no particle counts in spring, and has even seen a decrease in particle counts in the rest of the seasons across the years.

In response to the discovery of the problem during data collection, a review and curation process of the data has been initiated. Although a rigorous biological interpretation would need to wait until the curated records are available, this case study serves as an effective demonstration of ASCA’s value for data quality assessment.

\newgeometry{left=2cm, top=1.5cm, right=2cm}

\restoregeometry

\section{Conclusions}\label{sec:conclusion}

In this paper, we propose the use of ASCA, a multivariate extension of ANOVA, as a tool for the exploratory analysis of time series affected by cyclostationarity. ASCA combines statistical inference and visualization through score and loading plots that act similarly to a post-hoc test, facilitating the interpretation of group differences. While ASCA is traditionally applied to experimental data, our findings demonstrate its value for observational data. The proposed methodology models each potentially relevant cyclic time scale as a tensor mode. This multidimensional structure is then unfolded into a matrix that is used as input to ASCA, by assigning modes to the rows (as factors and their levels) or to the columns (as variables). Because a single tensor allows for multiple unfolding strategies, analysts can generate several models that complement each other by offering different perspectives on the same dataset.

The efficacy of the ASCA pipeline is demonstrated through insightful results from two real-world case studies. The analysis of water temperature trends in Sierra Nevada mountain lakes successfully isolated a significant warming trend during summer months that was consistent across different locations, which could not be found using traditional ANOVA. In that case study, ASCA also showed an improvement over ANOVA in the separation of variability across factors in an unbalanced design. Whether this behavior generalizes across broader conditions of design imbalance remains a subject for future work. Similarly, the study of airborne pollen concentrations in Granada identified a data quality problem, while proving its potential to identify changes in specific pollen types and shifts in seasonal behavior in further analyses.

Despite its strengths, this methodology presents some opportunities for improvement. Although the unfolding process accounts for autocorrelation to ensure the statistical integrity of the model, it limits the available unfolding possibilities. Moreover, the rapid increase in dimensionality that occurs when multiple modes are placed along the columns may elevate computational cost in highly multivariate datasets and hinder the interpretability of loading plots by displaying too many sources of variability simultaneously.

Overall, this work demonstrates that ASCA is a solid tool with great potential for the analysis of observational data, providing both statistical testing and intuitive, visual data exploration.

\section*{Acknowledgments}
This work was supported by grant no. PID2023-1523010B-IOO (MuSTARD), funded by the Agencia Estatal de Investigación in Spain, call no. MICIU/AEI/ 10.13039/501100011033, by the European Regional Development Fund; grant no. PID2024-161345NB-I00 from the MICIU/AEI/10.13039/501100011033 and FEDER, UE; grants BIOD22\_001 and BIOD22\_002, funded by Consejería de Universidad, Investigación e Innovación and Gobierno de España and Unión Europea – NextGenerationEU, and the University of Granada Plan Propio through Applied Research Projects C-EXP-167-UGR23 AEROFOUR and Laboratorio Singular UGR granted to the Air Biological Quality Unit, LS2024-3.

The authors would like to thank the anonymous reviewers for their constructive feedback, which greatly improved the quality of this manuscript.

\bibliographystyle{model5-names}
\biboptions{authoryear}
\begin{small}
\bibliography{references}
\end{small}

\end{document}


\begin{frontmatter}



\title{Supplementary Materials: Modeling cyclostationarity in time series using ASCA} 

\author[1]{Daniel Vallejo-España\corref{cor1}}
\ead{dvalesp@ugr.es}
\author[1]{Jesús García Sánchez}
\ead{gsus@ugr.es}
\author[2]{Manuel Villar-Argaiz}
\ead{mvillar@ugr.es}
\author[3]{Concepción De Linares}
\ead{delinare@ugr.es}
\author[1]{José Camacho}
\ead{josecamacho@ugr.es}
\affiliation[1]{organization={Department of Signal Theory, Telematics and Communications, University of Granada},
            addressline={Calle Periodista Daniel Saucedo Aranda},
            city={Granada},
            postcode={18014},
            state={Andalucía},
            country={Spain}}
\affiliation[2]{organization={Department of Ecology, University of Granada},
            addressline={Avenida de Fuentenueva},
            city={Granada},
            postcode={18071},
            state={Andalucía},
            country={Spain}}
\affiliation[3]{organization={Department of Botany, University of Granada},
            addressline={Avenida de Fuentenueva},
            city={Granada},
            postcode={18071},
            state={Andalucía},
            country={Spain}}
\cortext[cor1]{Corresponding author.}
\end{frontmatter}

\section{ASCA analysis of the simulated example dataset}

In this section, we present and discuss the results of the application of our methodology to a simulated dataset used as an example throughout Section~\ref{sec:methodology} of the main text. This dataset consists of hourly electricity consumption data from primary schools in various cities of the United States over the course of a typical meteorological year. Two different unfoldings were applied, corresponding to the two-factor and three-factor matrices displayed in Figure~\ref{fig:tensor_unfolding}B--C of the main text. As a result, two distinct analyses with ASCA were performed.

\subsection{Unfolding with two factors: `city' and `week'}

Starting from the a tensor with modes `city', 'hour of the day', `day of the week' and `week', a matrix was constructed by assigning the `city' and `week' modes to the rows as factors, and the `hour of the day' and `day of the week' to the columns. After mean-centering, ASCA was applied to this matrix. We applied a 10-row block permutation scheme (10 consecutive weeks) to account for autocorrelation between one week and the next. This block length was empirically chosen after inspecting a correlogram of the residuals. An analysis of residuals is presented later in this section.

Both factors were deemed significant, with `city' being approximately four times larger than `week'---the temporal evolution factor---in terms of mean squares (Table~\ref{tab:energy_2f}). This suggests that spatial variability in electricity demand is considerably more pronounced than seasonal or temporal fluctuations in the time scale considered. 

\begin{table}[ht]
\centering
\resizebox{0.7\textwidth}{!}{%
\begin{tabular}{lrrrrrr}
\hline
 & \textbf{SS ($\times 10^7$)} & \textbf{\%SS} & \textbf{df} & \textbf{MS ($\times 10^6$)} & \textbf{F} & \textbf{$p$-value} \\
\hline
\hline
City      & 2.30  & 26.97  & 7   & 3.28 & 55.45 & $<0.001$ \\
Week      & 4.11  & 48.23  & 51  & 0.81 & 13.61 & $<0.001$ \\
Residuals & 2.11  & 24.81  & 357 & 0.06 & --    & --       \\
\hline
Total     & 8.52  & 100.00 & 415 & 0.21 & --    & --       \\
\hline
\hline
\end{tabular}
}
\caption{ASCA table for the energy consumption dataset, considering factors `city' and `week'. The columns indicate, respectively: the sum of squares (SS), its percentage relative to the total (\%SS), degrees of freedom (df), mean squares (MS), the F-statistic (F), and the associated $p$-value.}
\label{tab:energy_2f}
\end{table}

\begin{figure}[p]  
    \centering
    
    \begin{minipage}{0.47\linewidth}
        \textbf{A} \par \vspace{1pt}
        \centering
        \includegraphics[height=6.5cm, width=\linewidth, keepaspectratio]{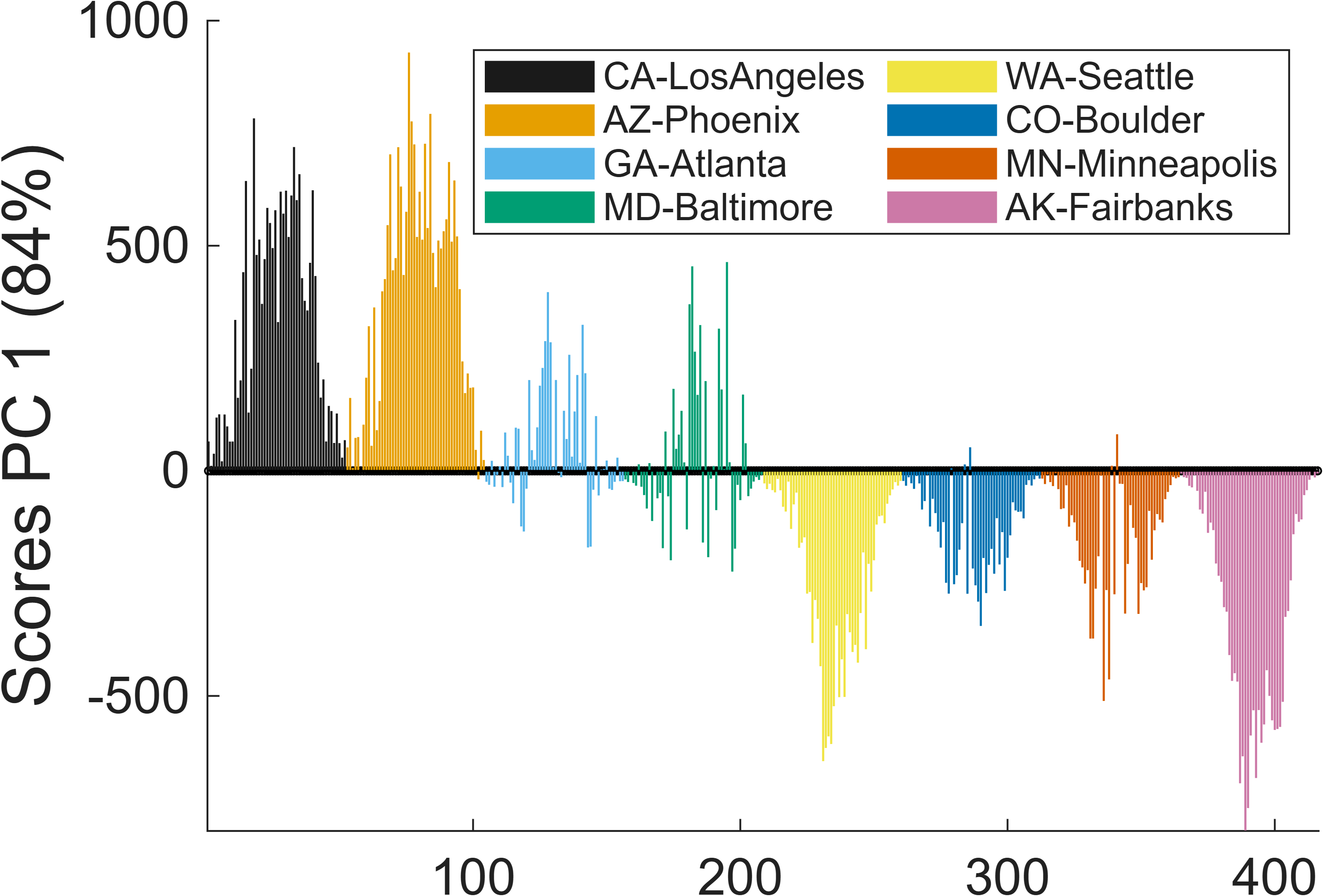}
    \end{minipage}
    \hfill
    \begin{minipage}{0.47\linewidth}
        \textbf{B} \par \vspace{1pt}
        \includegraphics[width=0.95\linewidth]{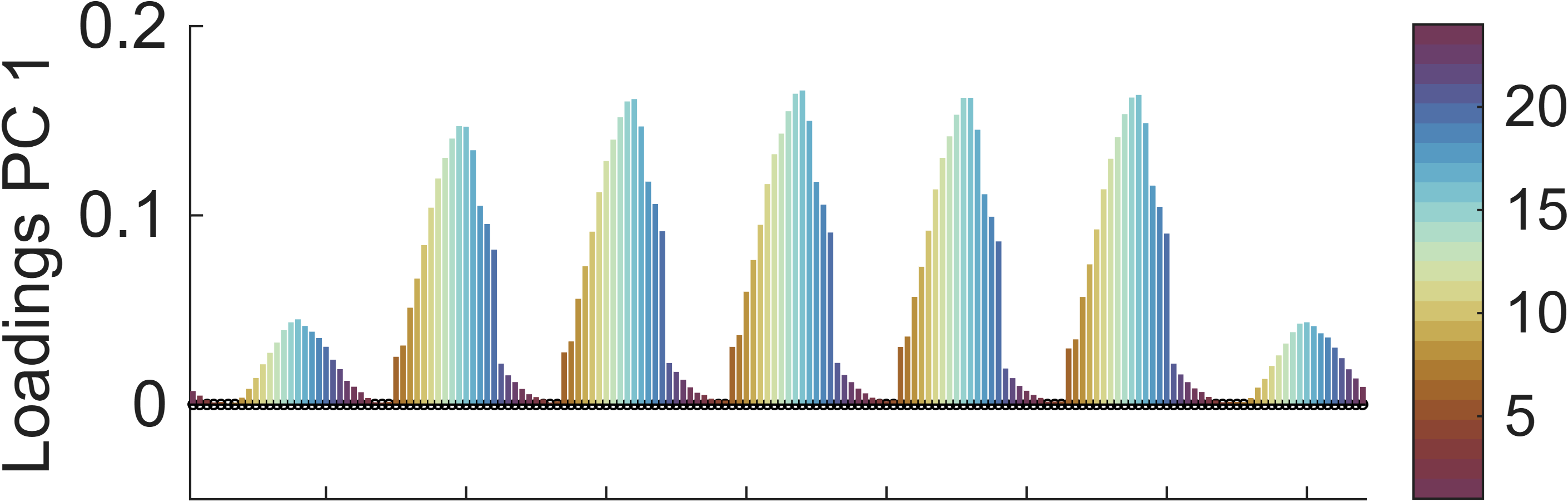}
        \vspace{-0.05cm}
        \raggedright \textbf{C} \par \vspace{1pt}
        \centering
        \includegraphics[width=\linewidth]{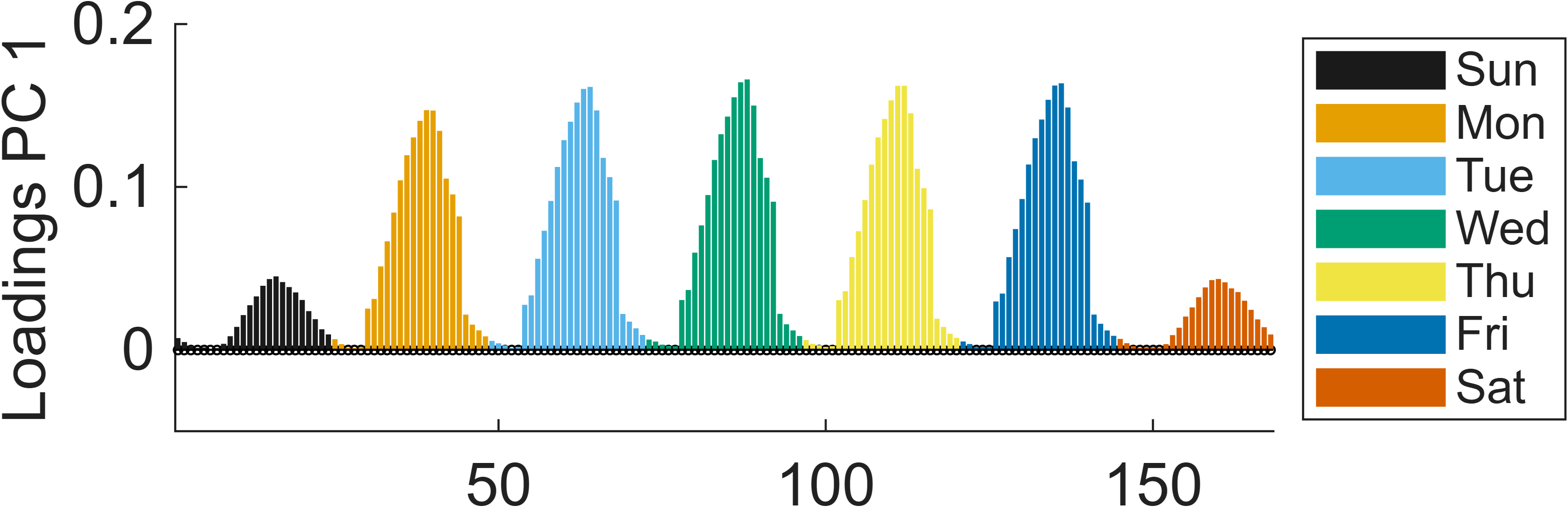}
    \end{minipage}

    \begin{minipage}{0.47\linewidth}
        \textbf{D} \par \vspace{1pt}
        \centering
        \includegraphics[height=6.5cm, width=\linewidth, keepaspectratio]{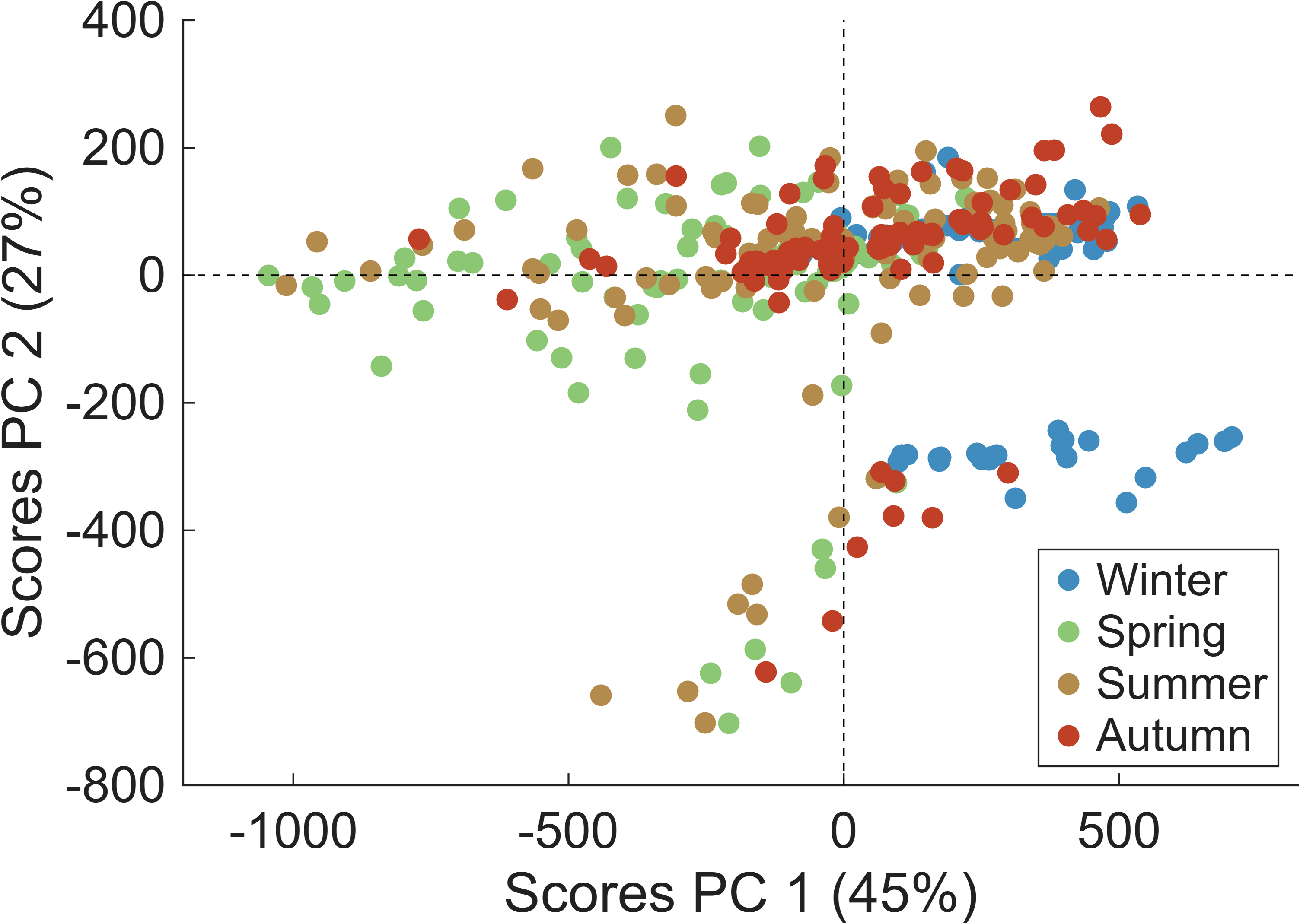}
    \end{minipage}
    \hspace{0.2cm}
    \begin{minipage}{0.47\linewidth}
        \textbf{E} \par \vspace{1pt}
        \centering
        \includegraphics[height=6.5cm, width=\linewidth, keepaspectratio]{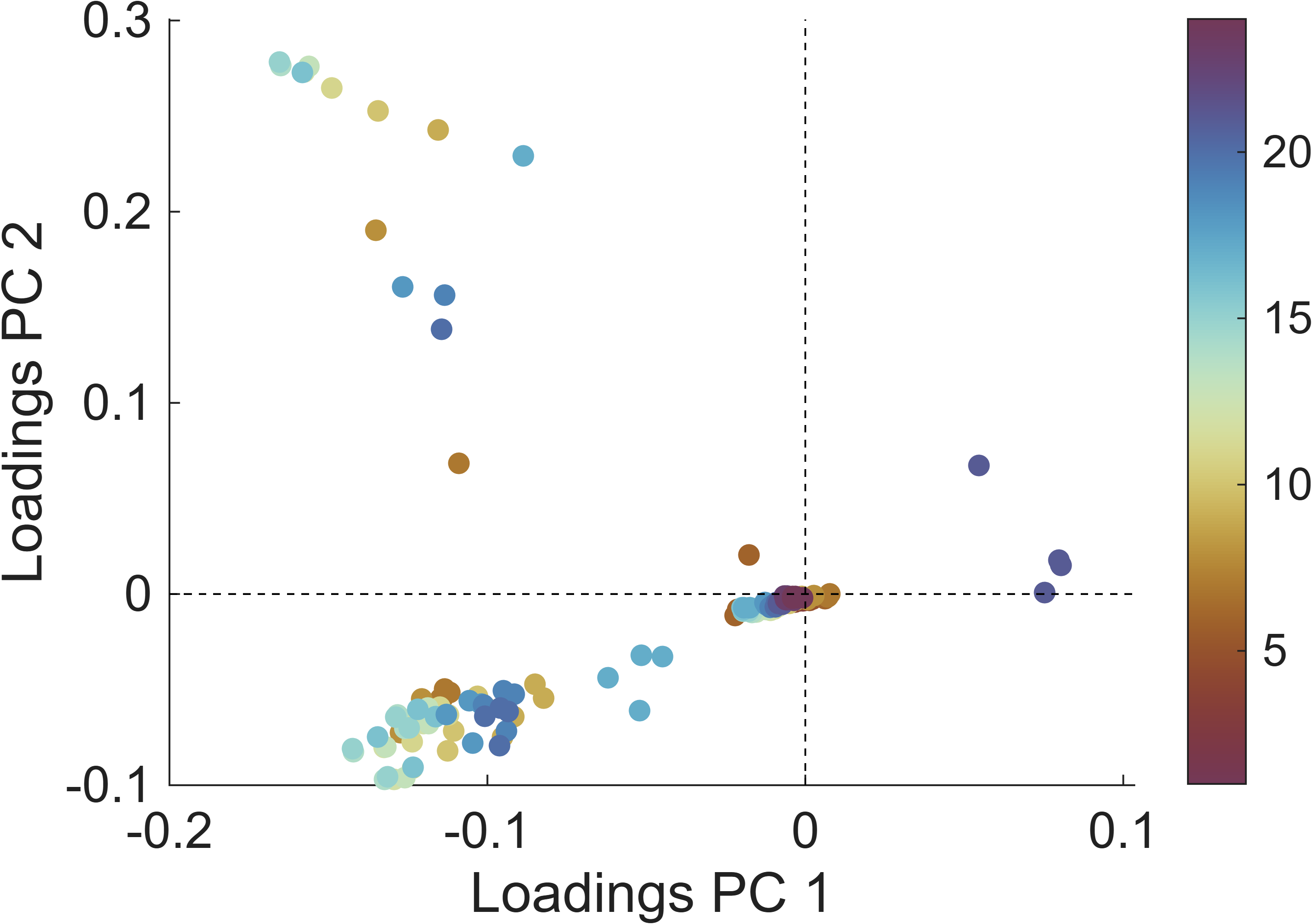}
    \end{minipage}

    \begin{minipage}{0.47\linewidth}
        \textbf{F} \par \vspace{1pt}
        \centering
        \includegraphics[height=6.5cm, width=\linewidth, keepaspectratio]{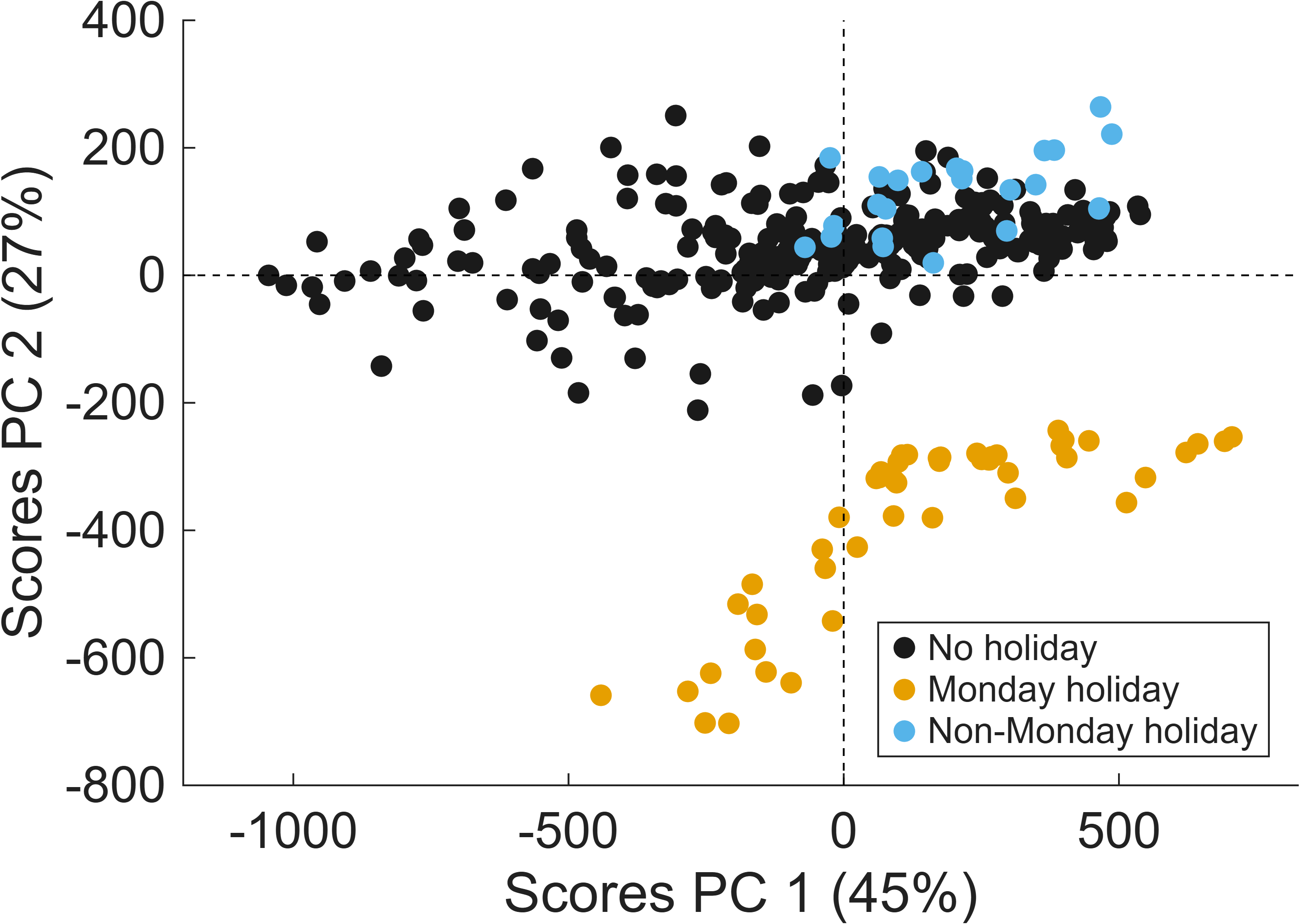}
    \end{minipage}
    \hspace{0.2cm}
    \begin{minipage}{0.47\linewidth}
        \textbf{G} \par \vspace{1pt}
        \centering
        \includegraphics[height=6.5cm, width=\linewidth, keepaspectratio]{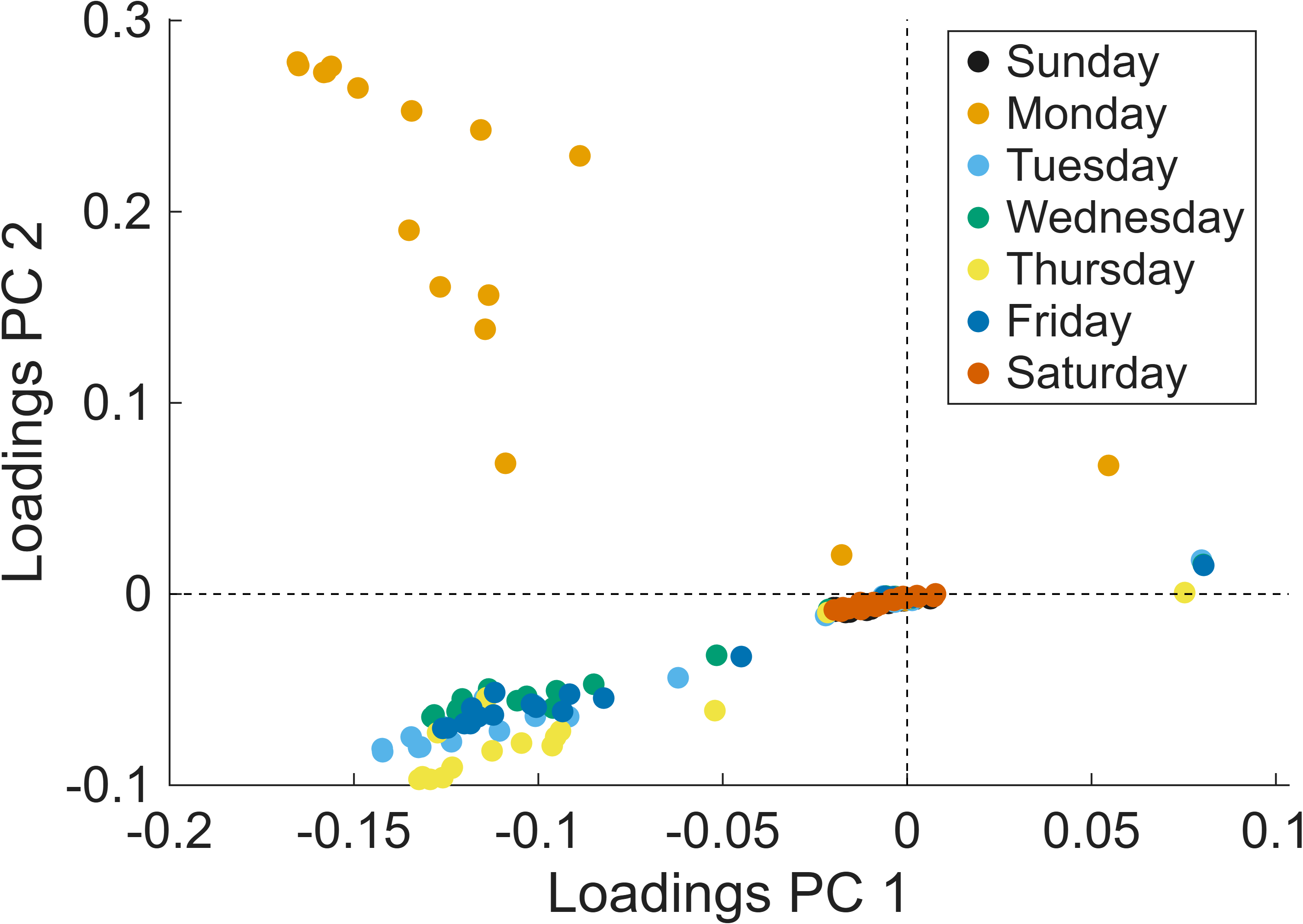}
    \end{minipage}

    \caption{\textbf{ASCA model of the energy consumption dataset, considering factors `city' and `week'.} \textbf{A} scores for PC 1 of factor `city'. \textbf{B} loadings for PC 1 of factor `city', colored by hour of the day. \textbf{C} loadings for PC 1 of factor `city', colored by day of the week. \textbf{D} scores for PCs 1 and 2 of factor `week', colored by season. \textbf{E} loadings for PCs 1 and 2 of factor `week', colored by hour of the day. \textbf{F} scores for PCs 1 and 2 of factor `week'. Colors indicate whether an observation corresponds to a Monday public holiday, a different public holiday or a non-holiday. \textbf{G} loadings for PCs 1 and 2 of factor `week', colored by day of the week.}
    \label{fig:energy_2f}
\end{figure}

The score plot for the `city' factor (Figure~\ref{fig:energy_2f}A) shows that the first principal component (PC1, accounting for 84\% of the factor's variance) separates schools in warmer climates (Phoenix, Los Angeles), with positive scores, from schools in cooler regions (Fairbanks, Seattle), with negative scores. Figures~\ref{fig:energy_2f}B and \ref{fig:energy_2f}C show the corresponding loading plot colored in two different ways: by hour of the day, and by day of the week, respectively. The loadings indicate that warmer cities experience intensified consumption during the late morning and afternoon of weekdays, as those times display higher loadings. Conversely, consumption patterns during weekends show remarkable similarity across all cities, as the weekend loadings are smaller in magnitude, indicating less contribution to variance between the cities. Since all buildings considered are primary schools, this behavior is expected, given that schools are mostly in use from Monday through Friday.

Regarding factor `week', two main patterns can be seen in the scores and loadings of the first two PCs. To illustrate this, the scores for PCs 1 and 2 have been colored in two different ways (Figures~\ref{fig:energy_2f}D and \ref{fig:energy_2f}F). Like in factor `city', the loadings have also been colored by hour of the day (Figure~\ref{fig:energy_2f}E) and by day of the week (Figure~\ref{fig:energy_2f}G). First, the scores from PC1 (45\% of factor variability) differentiate the time of the year, as seen in Figure~\ref{fig:energy_2f}D. Spring is on the left, which corresponds in the loadings to higher consumption in the morning and afternoon (Figure~\ref{fig:energy_2f}E), particularly on weekdays (Figure~\ref{fig:energy_2f}G). Winter, however, is on the right, exhibiting lower consumption in said hours, and a higher peak between 20:00 and 21:00 on weekdays. These results, paired with the fact that factor `week' is significant, indicate that the weekly cycle experiences changes throughout the year, particularly on weekdays, while consumption on weekends remains mostly constant. Furthermore, PC2 (27\% of variability) identifies the impact of public holidays. The main pattern captured in PC2 involves the contrast between Monday holidays and regular school Mondays. PC2 separates weeks containing a Monday holiday from standard school weeks (Figure~\ref{fig:energy_2f}F). A lower consumption is observed on Monday holidays between 07:00 and 20:00, due to they not being a school day. Because more public holidays in the United States occur on Mondays than on other days, Monday holidays account for more variability than mid-week holidays, which could explain this behavior being more visible on Mondays.

\begin{figure}[p]
    \centering
    
    \begin{minipage}{0.48\linewidth}
        \textbf{A} \par \vspace{1pt}
        \centering
        \includegraphics[height=6.5cm, width=\linewidth, keepaspectratio]{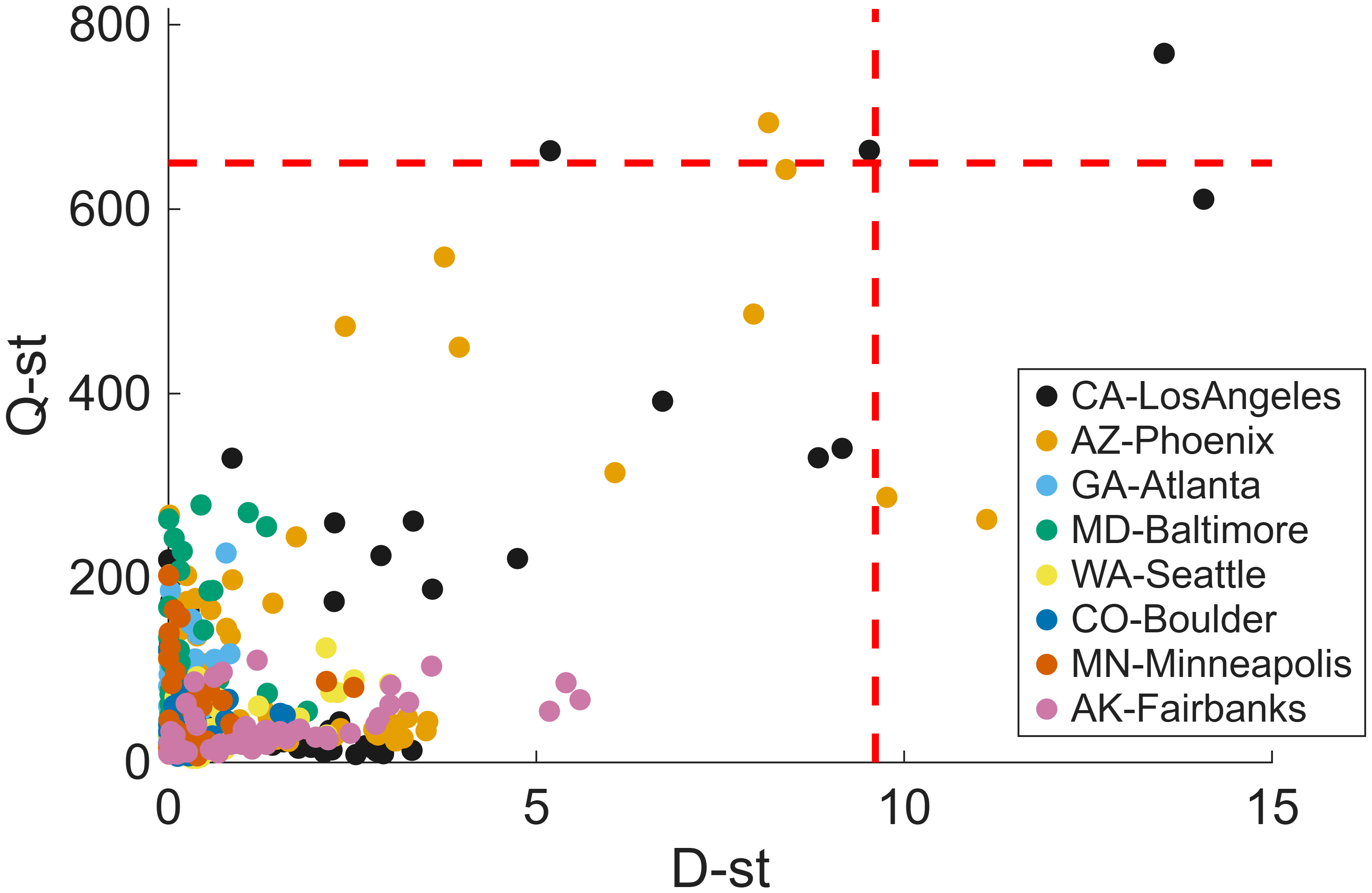}
    \end{minipage}
    \hfill
    \begin{minipage}{0.48\linewidth}
        \textbf{B} \par \vspace{1pt}
        \centering
        \includegraphics[height=6.5cm, width=\linewidth, keepaspectratio]{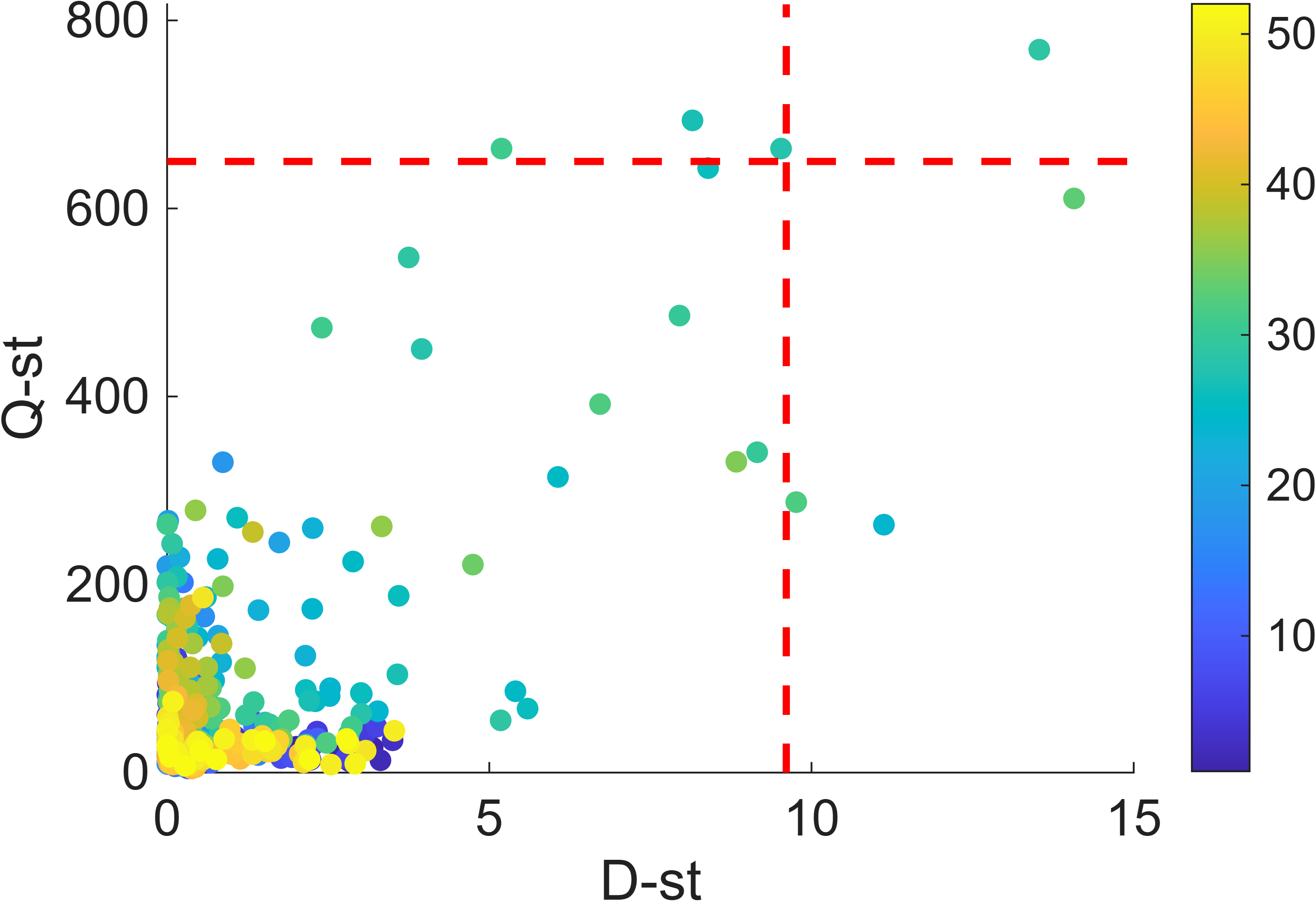}
    \end{minipage}
    
    \begin{minipage}{0.48\linewidth}
        \textbf{C} \par \vspace{1pt}
        \centering
        \includegraphics[height=6.5cm, width=\linewidth, keepaspectratio]{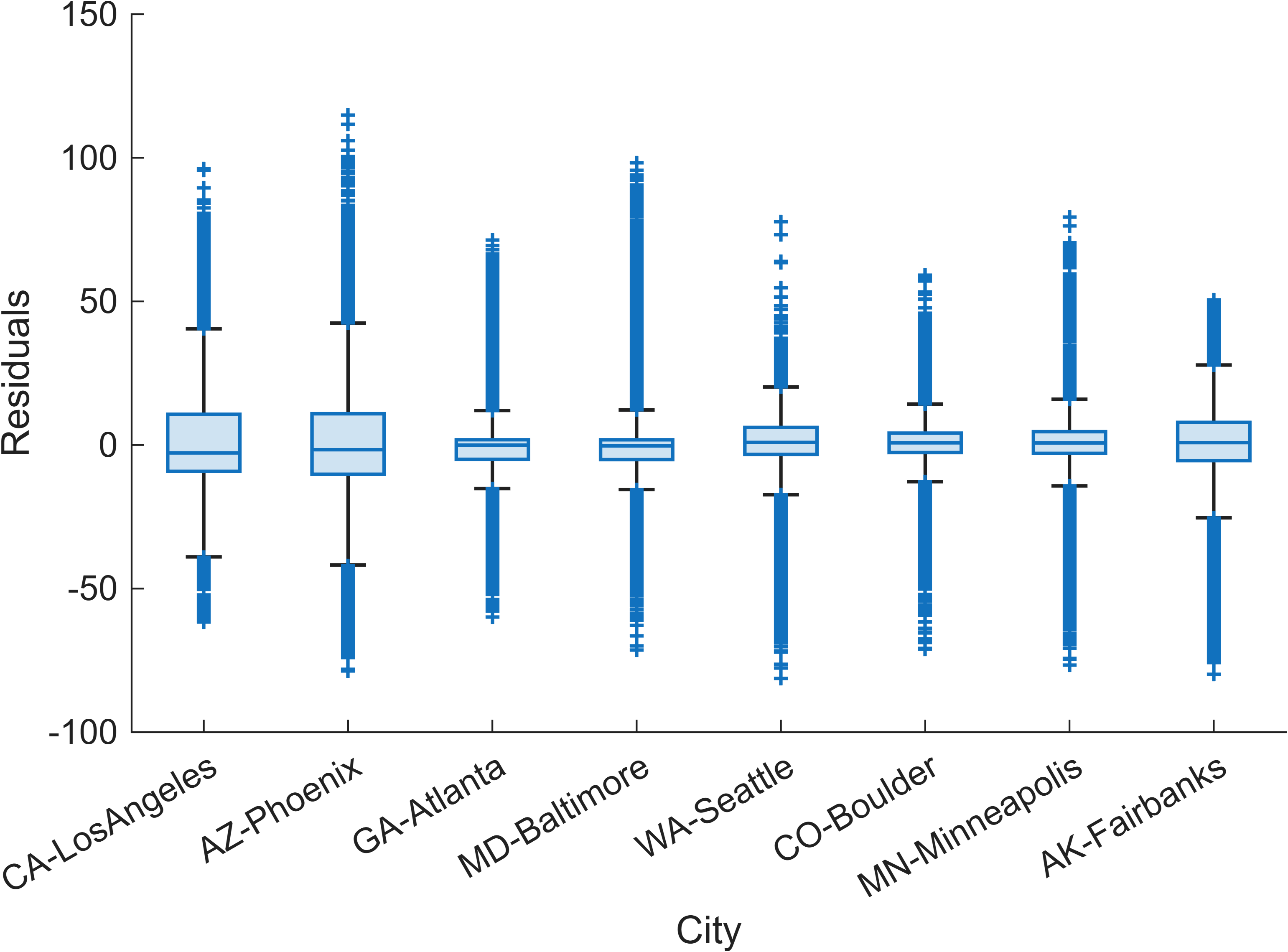}
    \end{minipage}
    \hfill
    \begin{minipage}{0.48\linewidth}
        \textbf{D} \par \vspace{1pt}
        \centering
        \includegraphics[height=6.5cm, width=\linewidth, keepaspectratio]{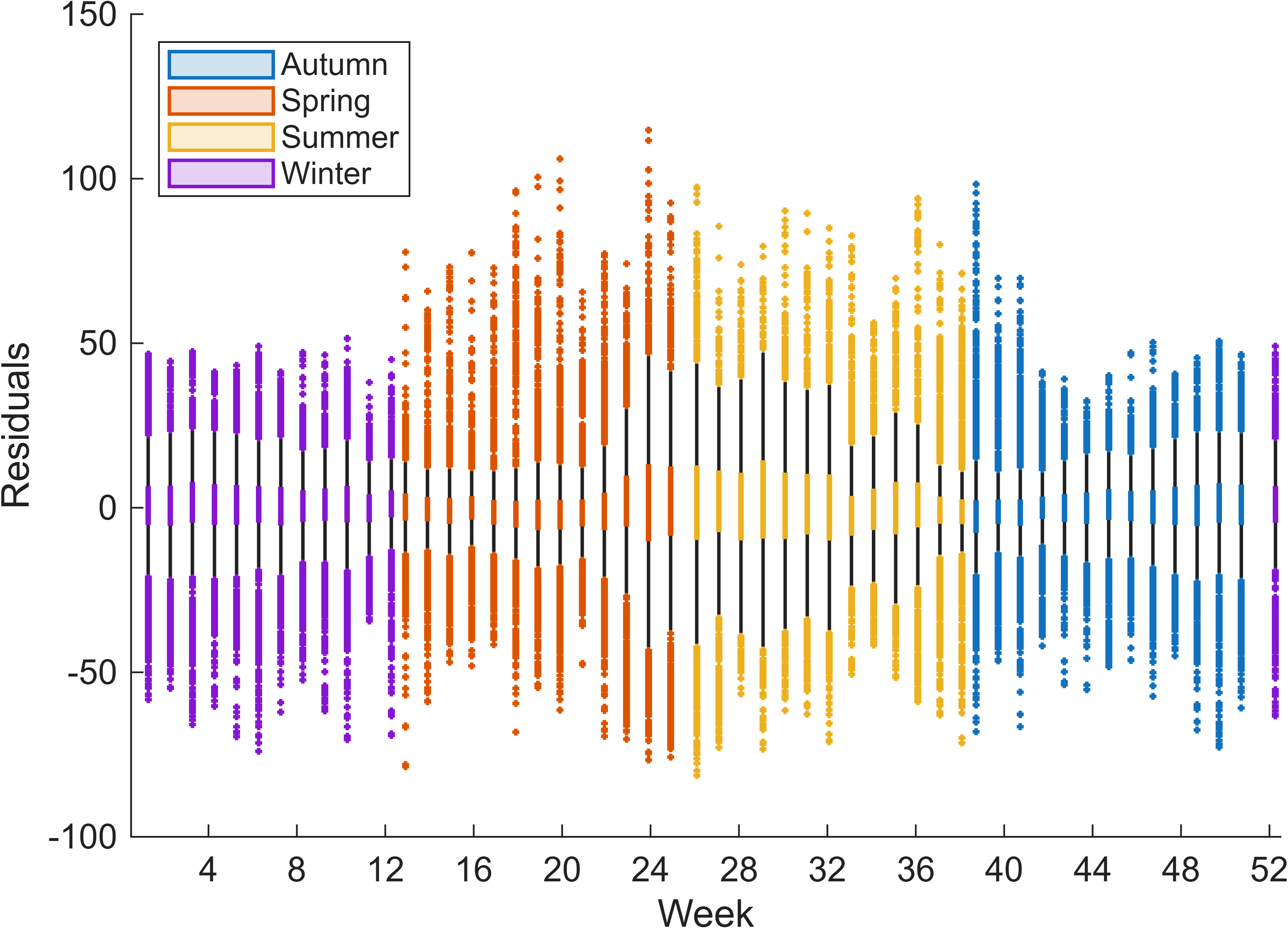}
    \end{minipage}

    \begin{minipage}{\linewidth}
        \textbf{E} \par \vspace{1pt}
        \centering
        \includegraphics[height=6.5cm, width=\linewidth, keepaspectratio]{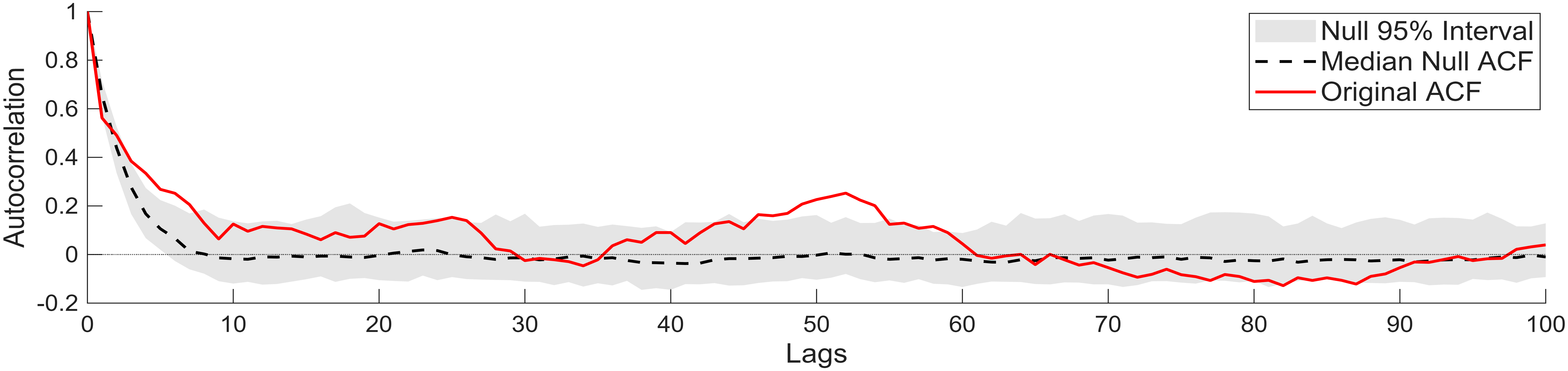}
    \end{minipage}

    \caption{\textbf{Residuals of the ASCA model of the energy consumption dataset, considering factors `city' and `week'.} \textbf{A} Multivariate Statistical Process Control (MSPC) chart (Q-statistic vs. D-statistic) of the model, colored by city. The control limits (red dotted lines) are percentile 99 for both D and Q statistics. \textbf{B} MSPC chart of the model, colored by week of the year. \textbf{C} boxplot of residuals by city. \textbf{D} boxplot of residuals by week, colored by season. \textbf{E} sample autocorrelation function (ACF) of the Q-statistic. The red line represents the proposed model, while the black dotted line shows the median across 100 block permutations (10-row blocks). The shaded region denotes the 95\% empirical interval across all permutations.}
    \label{fig:energy_2f_residuals}
\end{figure}

An analysis of the model's residuals was performed to detect heteroscedasticity and dependence. The residuals were first explored using Multivariate Statistical Process Control (MSPC; \citealp{Ferrer2007MSPC}). Some observations from Los Angeles and Phoenix, from certain weeks of the middle of the year, appear as outliers in the MSPC charts (Figure~\ref{fig:energy_2f_residuals}A--B). As shown in the boxplots of Figure~\ref{fig:energy_2f_residuals}C--D, these cities and weeks also exhibit slightly higher variance than the rest. However, this minor heteroscedasticity does not significantly impact the model's overall results. To assess autocorrelation, we constructed a correlogram (Figure~\ref{fig:energy_2f_residuals}E) of the Q-statistic---the sum of square residuals across all variables. The autocorrelation coefficients roughly align with the median of 100 random block permutations approximating the null distribution.

\subsection{Unfolding with three factors: `city', `day of the week' and `week'}

To explore a different perspective, an alternative unfolding of the same four-mode tensor was performed. In this configuration, the `city', `day of the week', and `week' modes were assigned to the rows as factors, with respectively 8, 7 and 52 levels. The `hour of the day' remained in the columns. This setup resulted in an ASCA model with three main factors. Preprocessing consists on mean-centering, and no interaction terms were included. While the previous model relied on loading plots to visualize differences across days of the week, this version allows for statistical testing of said differences. To account for expected day-to-day autocorrelation within the null distribution, we performed block permutations using 5-row blocks (corresponding to 5 consecutive days).

\begin{figure}[p]
    \centering
    \begin{minipage}{0.48\linewidth}
        \textbf{A} \par \vspace{1pt}
        \centering
        \includegraphics[height=6.5cm, width=\linewidth, keepaspectratio]{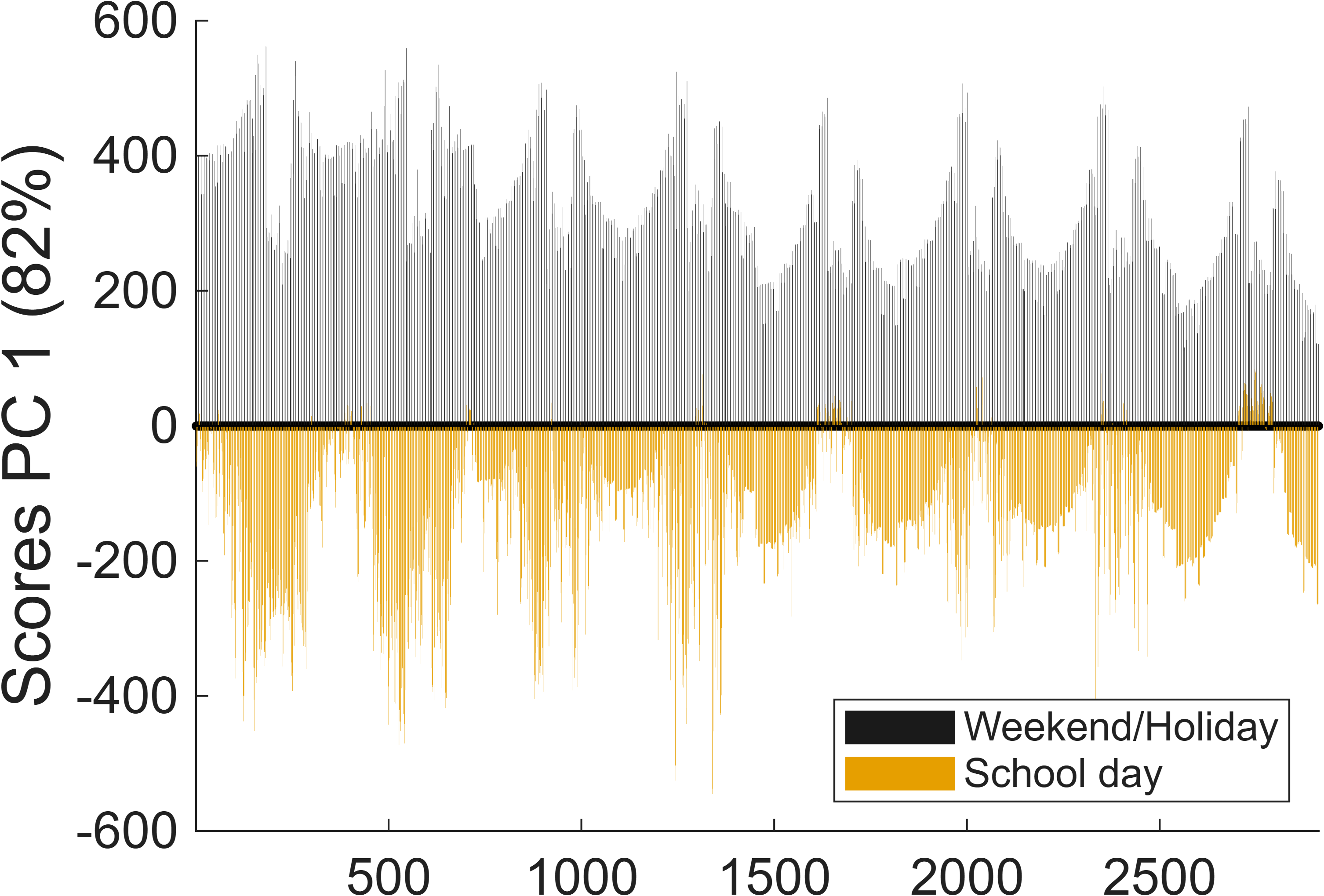}
    \end{minipage}
    \hfill
    \begin{minipage}{0.48\linewidth}
        \textbf{B} \par \vspace{1pt}
        \centering
        \includegraphics[height=6.5cm, width=\linewidth, keepaspectratio]{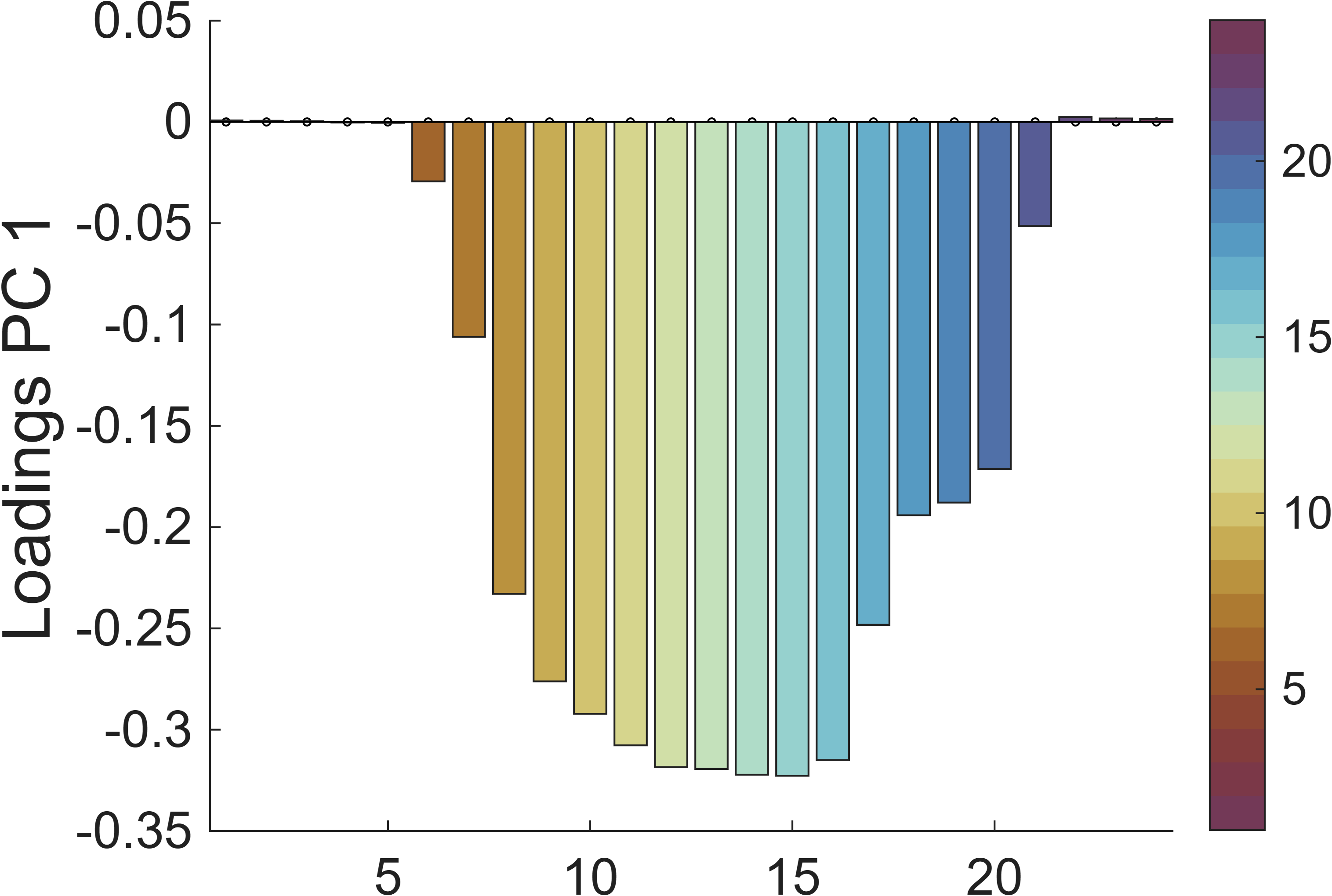}
    \end{minipage}
    
    \begin{minipage}{0.48\linewidth}
        \textbf{C} \par \vspace{1pt}
        \centering
        \includegraphics[height=6.5cm, width=\linewidth, keepaspectratio]{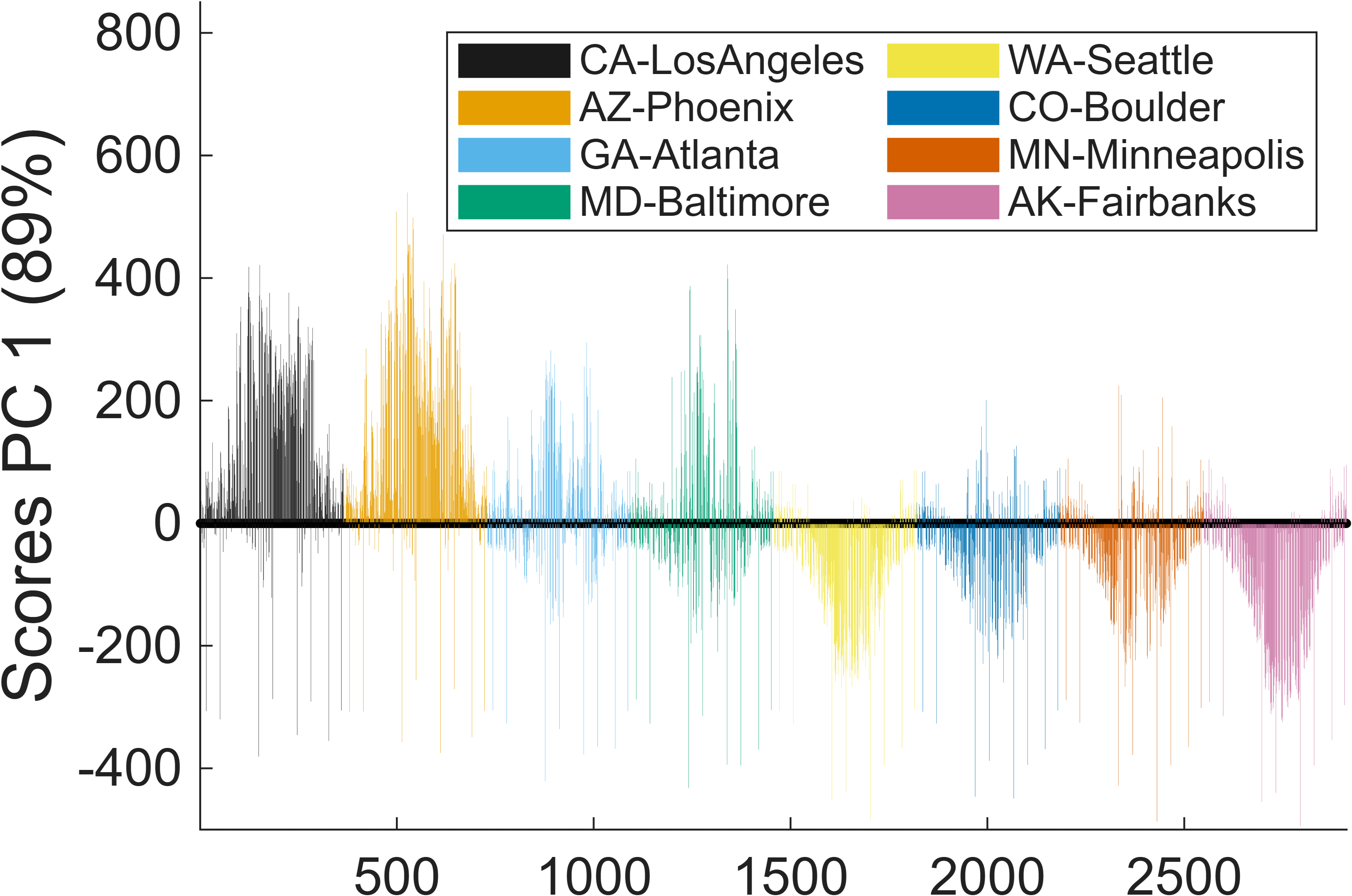}
    \end{minipage}
    \hfill
    \begin{minipage}{0.48\linewidth}
        \textbf{D} \par \vspace{1pt}
        \centering
        \includegraphics[height=6.5cm, width=\linewidth, keepaspectratio]{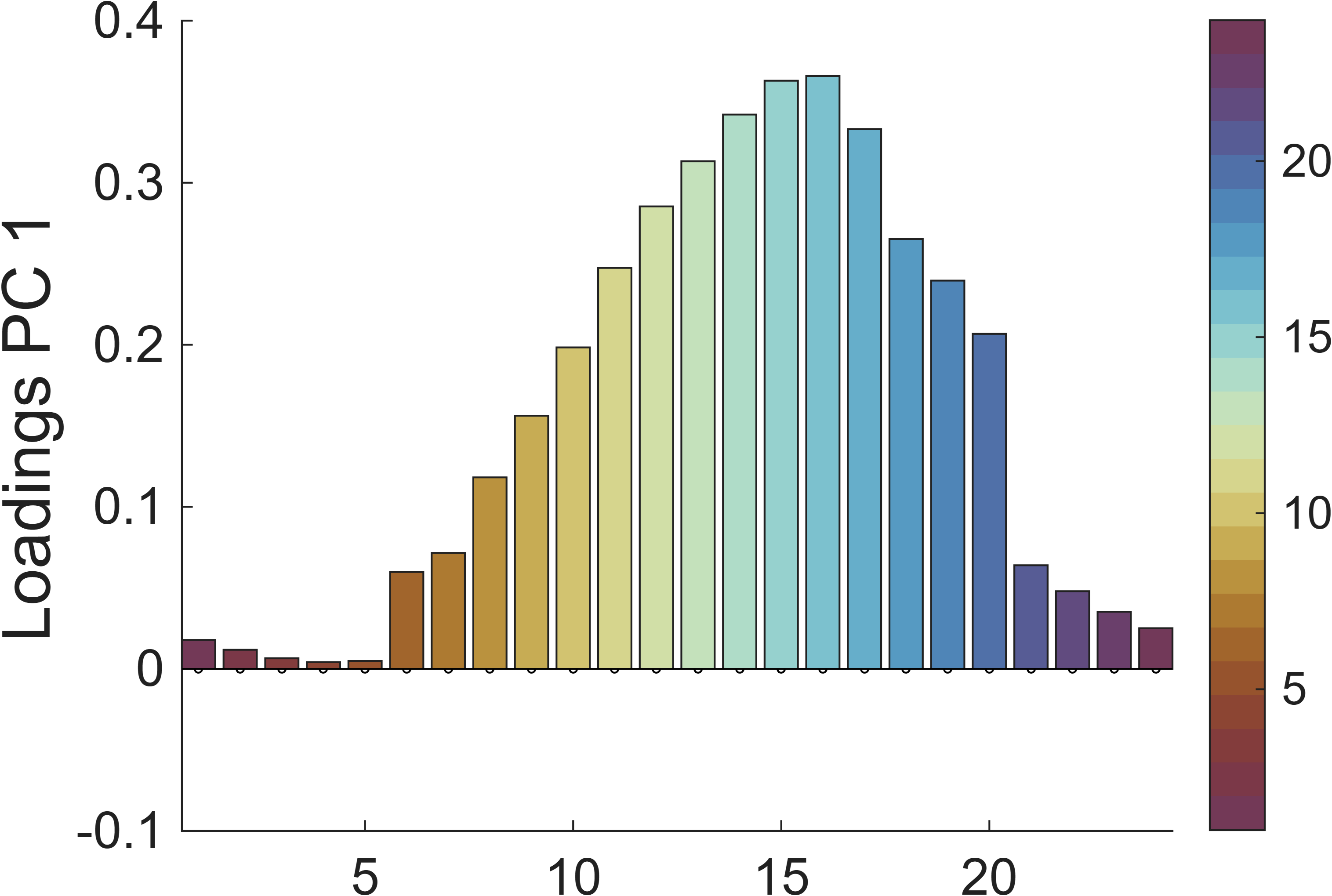}
    \end{minipage}

    \begin{minipage}{0.48\linewidth}
        \textbf{E} \par \vspace{1pt}
        \centering
        \includegraphics[height=6.5cm, width=\linewidth, keepaspectratio]{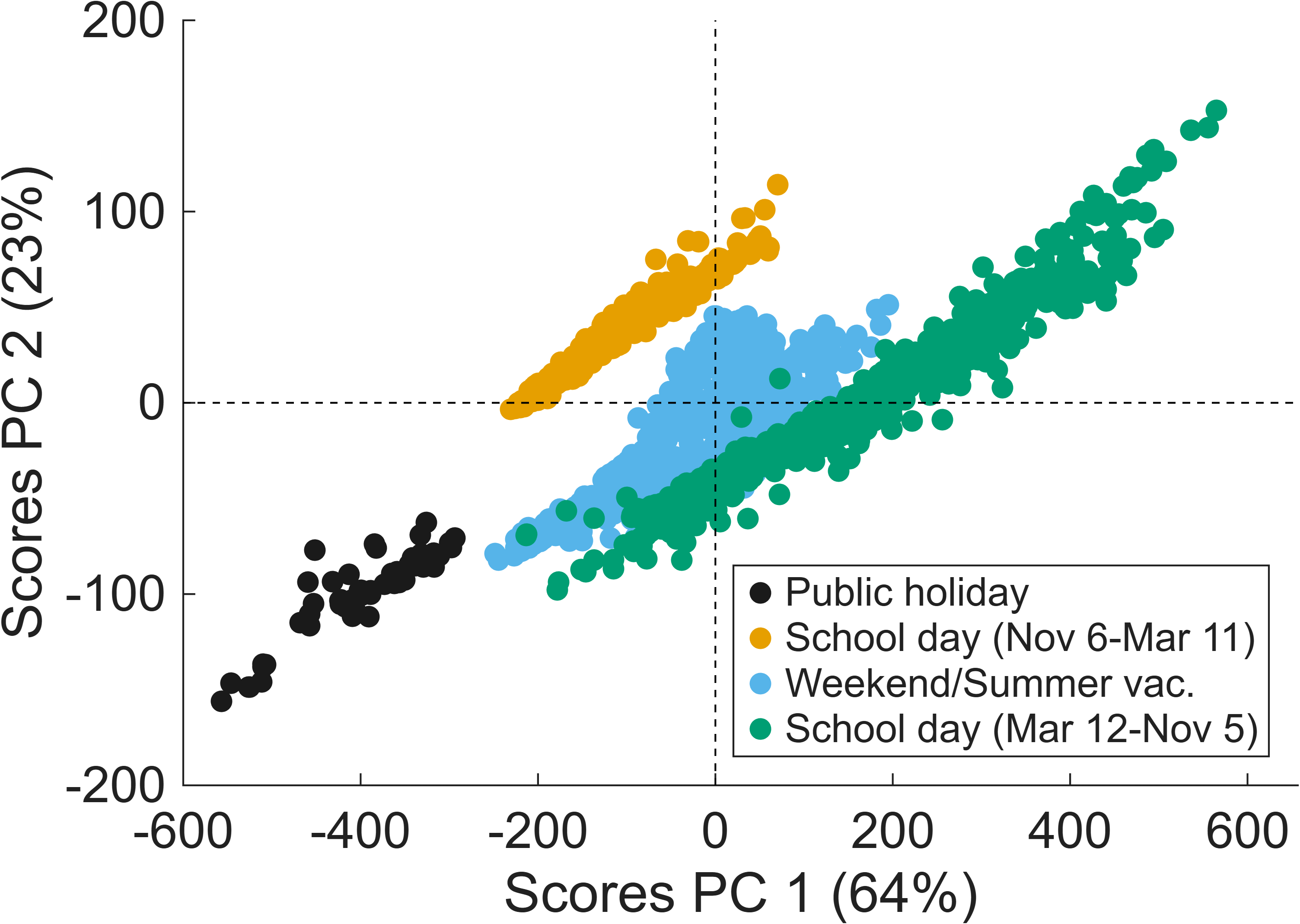}
    \end{minipage}
    \hfill
    \begin{minipage}{0.48\linewidth}
        \textbf{F} \par \vspace{1pt}
        \centering
        \includegraphics[height=6.5cm, width=\linewidth, keepaspectratio]{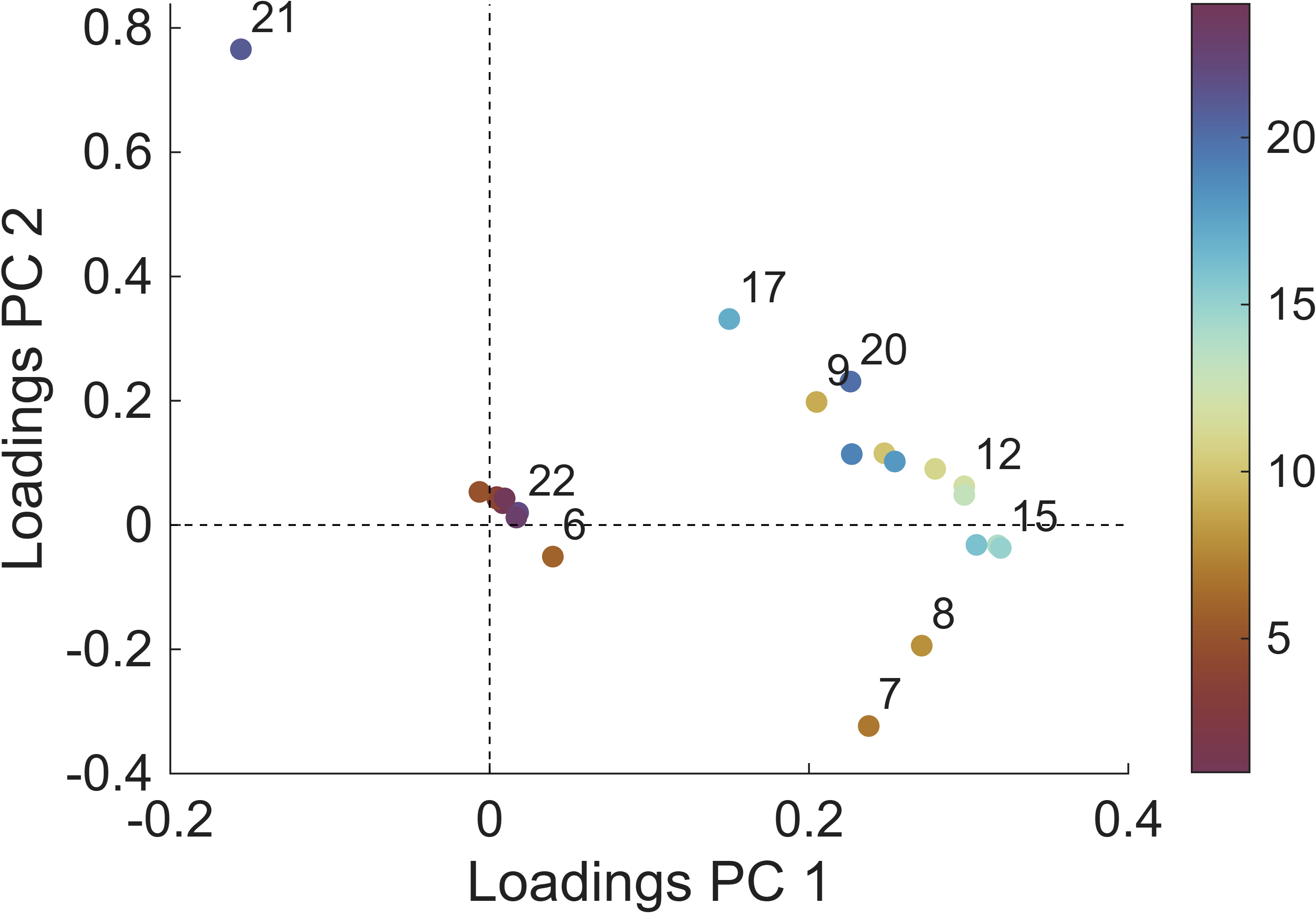}
    \end{minipage}

    \caption{\textbf{ASCA model of the energy consumption dataset, considering factors `city', `week' and `day of the week'.} \textbf{A} scores for the first component of factor `day of the week'. Colors indicate whether or not an observation corresponds to a working day. \textbf{B} loadings for the first component of factor `day of the week', colored by hour of the day. \textbf{C} scores for PC 1 of factor `city'. \textbf{D} loadings for PC 1 of factor `city', colored by hour of the day. \textbf{E} scores for PCs 1 and 2 of factor `week'. Observations are colored according to four different types of days in relation to work. \textbf{F} loadings for PCs 1 and 2 of factor `week', colored by hour of the day. }
    \label{fig:energy_3f}
\end{figure}

The results indicate that all factors are statistically significant (Table~\ref{tab:energy_3f}); however, `day of the week' is by far the most dominant factor, followed by `city' and `week'. The importance of the day of the week could also be observed in the previous model's loadings, but it could not be quantified as it was not one of the factors. The prominence of the `day of the week' factor stems from the contrast in energy demand between school days and weekends or holidays. This contrast is much more pronounced than spatial differences and temporal differences at a larger time scale (weeks). This is clearly captured in the score plot of the first principal component (PC1, 82\% of factor variance) in Figure~\ref{fig:energy_3f}A. The corresponding loadings (Figure~\ref{fig:energy_3f}B) demonstrate that school days exhibit substantially higher consumption between 06:00 and 21:00, especially from 09:00 to 16:00. Conversely, nighttime consumption remains largely indistinguishable from weekend and holiday levels, as the building is not in use.

\begin{table}[htbp]
\centering
\resizebox{0.75\textwidth}{!}{%
\begin{tabular}{lrrrrrr}
\hline
 & \textbf{SS ($\times 10^7$)} & \textbf{\%SS} & \textbf{df} & \textbf{MS ($\times 10^6$)} & \textbf{F} & \textbf{$p$-value} \\
\hline
\hline
City            & 1.93  & 9.40   & 7    & 2.80  & 178.74 & $<0.001$ \\
Week            & 2.20  & 10.71  & 51   & 0.40  & 27.95  & $<0.001$ \\
Day of the week & 12.00 & 58.50  & 6    & 20.00 & 1297.6 & $<0.001$ \\
Residuals       & 4.39  & 21.39  & 2847 & 0.02  & --     & --       \\
\hline
Total           & 20.52 & 100.00 & 2911 & 0.07  & --     & --       \\
\hline
\hline
\end{tabular}
}
\caption{ASCA table for the energy consumption dataset, considering factors `city', `week' and `day of the week'. The columns indicate, respectively: the sum of squares (SS), its percentage relative to the total (\%SS), degrees of freedom (df), mean squares (MS), the F-statistic (F), and the associated $p$-value.}
\label{tab:energy_3f}
\end{table}

Factor `city' (Figure~\ref{fig:energy_3f}C--D) exhibits behavior consistent with the previous model: schools in warmer cities maintain higher consumption on average during the morning and afternoon compared with those in cooler cities. However, because this model separates the variability of `day of the week' and `city' into distinct factors, we cannot directly observe city differences across days of the week. To capture those differences, an interaction term (`city' $\times$ `day of the week') would need to be incorporated into the model.

The first component of the week factor (PC1, 64\% of factor variance), shown in Figure~\ref{fig:energy_3f}E–F, separates groups of weeks based on the academic calendar. It distinguishes standard school weeks (characterized by high daytime demand) from weeks containing holidays, which exhibit significantly lower consumption. The second component (23\% of variability) reveals a seasonal shift in standard school weeks. Two groups of weeks can be observed: one from November 6th to March 11th, with high demand at 20:00--21:00, and one from March 12th to November 5th, with less consumption in that same time window. This abrupt shift in patterns likely stems from the specific heating and cooling cycles defined in the simulation. This short time window of higher demand in winter was also observed in the previous model. Finally, a fourth group, formed by weekends and the summer vacation period considered in the simulation (July and August), shows minimal deviation from the global average hourly consumption. This group’s midday consumption remains higher than on public holidays, likely due to extracurricular activities. 

The distinction between Monday holidays and other holidays is absent from the main effects of `week' and `day of the week.' Consequently, an interaction term between these factors would be required. This highlights that assigning a mode to the columns and analyzing the loading plots (as in the previous model) is functionally equivalent to visualizing the interaction between that mode assigned to the rows (as in the current model) and the other factors.

\begin{figure}[p]
    \centering
    
    \begin{minipage}{0.32\linewidth}
        \textbf{A} \par \vspace{1pt}
        \centering
        \includegraphics[height=6.5cm, width=\linewidth, keepaspectratio]{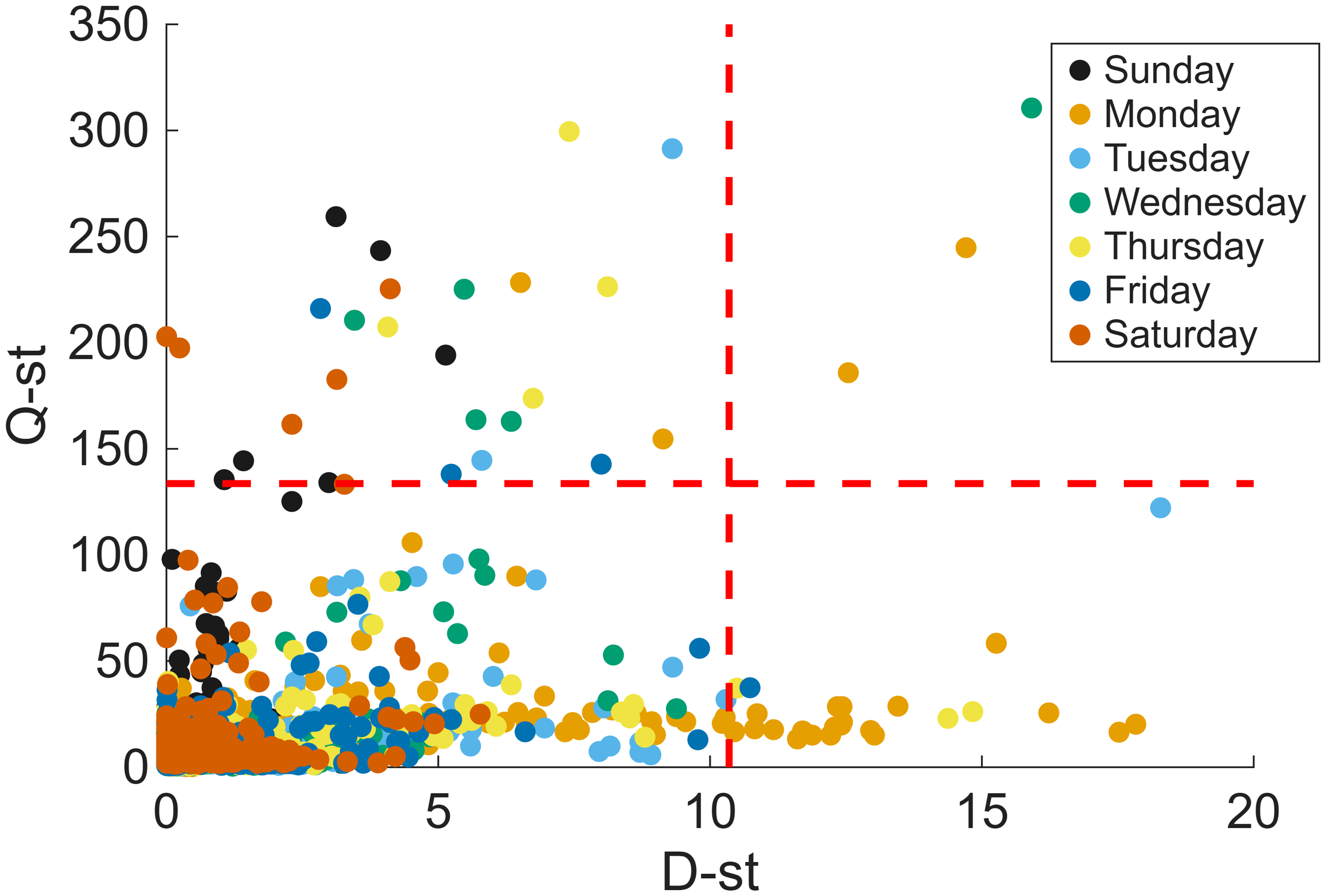}
    \end{minipage}
    \hfill
    \begin{minipage}{0.32\linewidth}
        \textbf{B} \par \vspace{1pt}
        \centering
        \includegraphics[height=6.5cm, width=\linewidth, keepaspectratio]{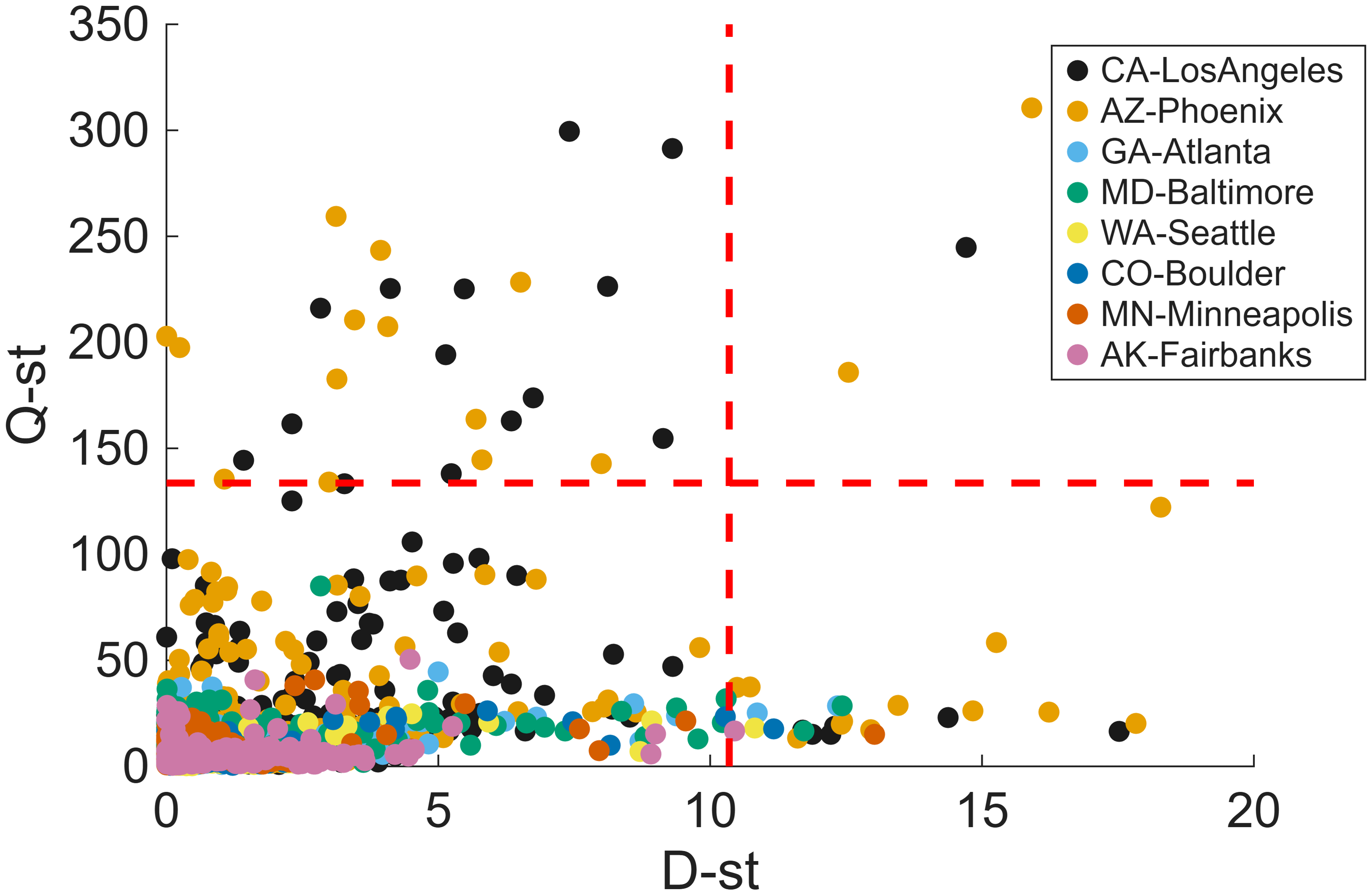}
    \end{minipage}
    \hfill
    \begin{minipage}{0.32\linewidth}
        \textbf{C} \par \vspace{1pt}
        \centering
        \includegraphics[height=6.5cm, width=\linewidth, keepaspectratio]{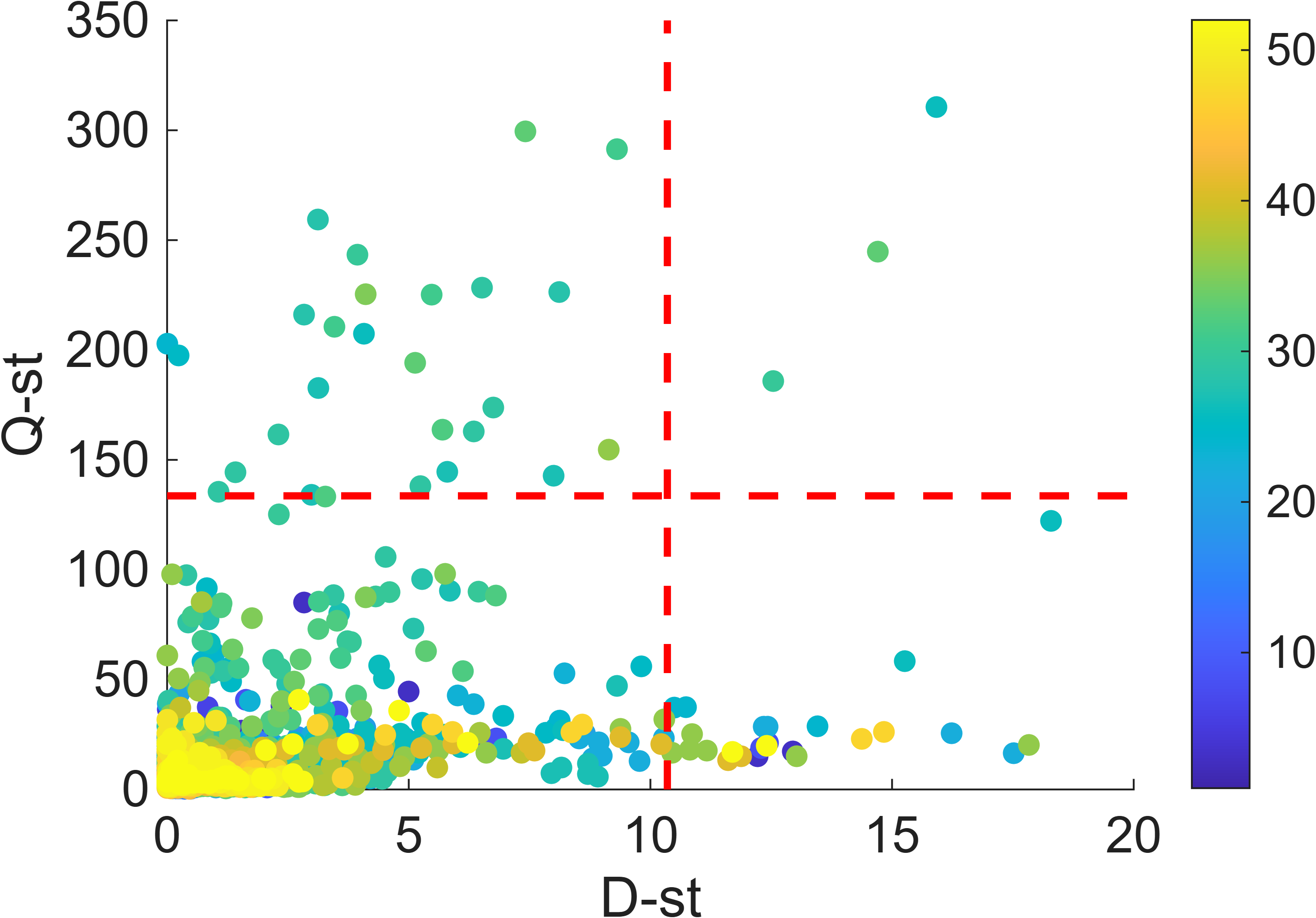}
    \end{minipage}
    
    \begin{minipage}{0.32\linewidth}
        \textbf{D} \par \vspace{1pt}
        \centering
        \includegraphics[height=6.5cm, width=\linewidth, keepaspectratio]{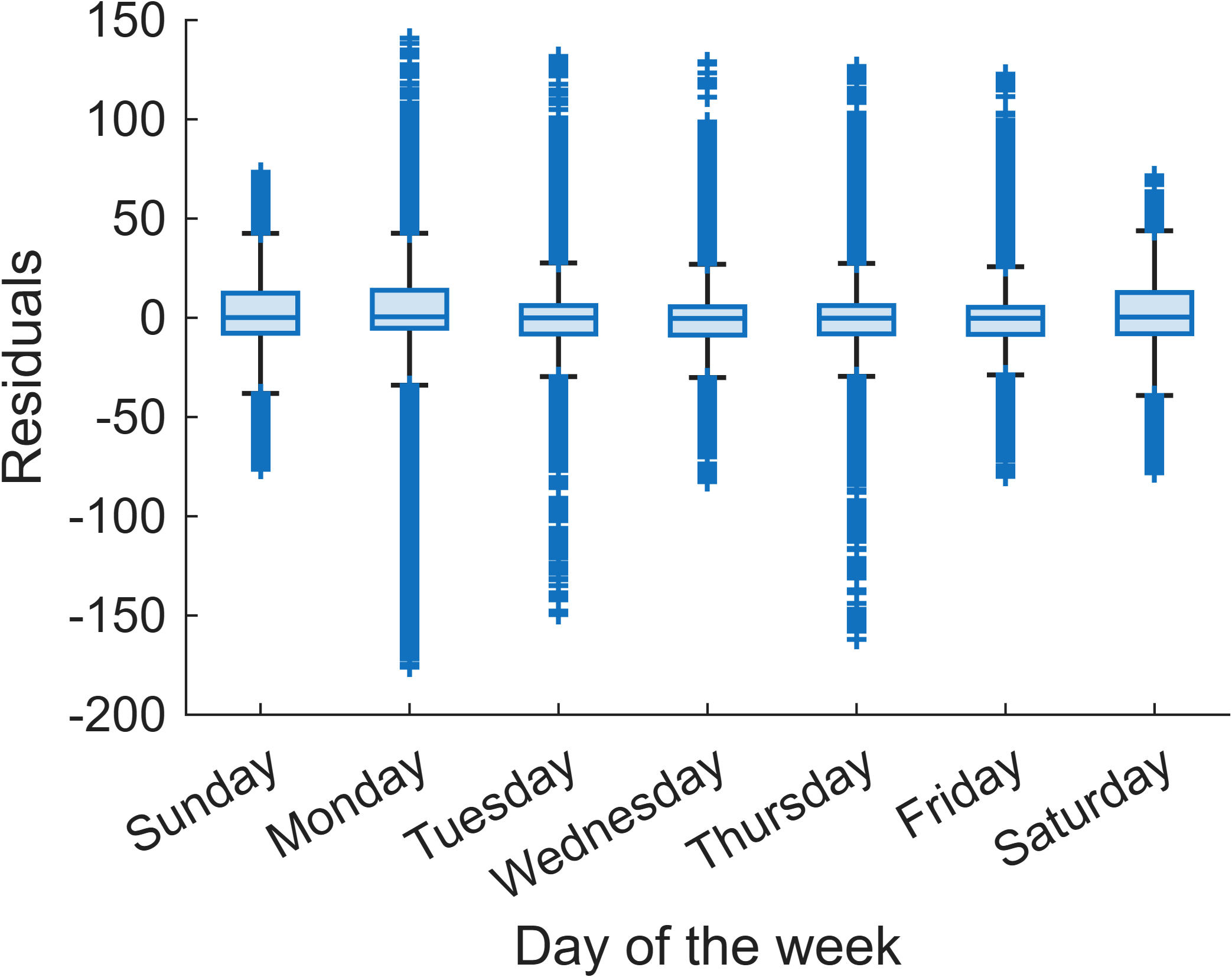}
    \end{minipage}
    \hfill
    \begin{minipage}{0.32\linewidth}
        \textbf{E} \par \vspace{1pt}
        \centering
        \includegraphics[height=6.5cm, width=\linewidth, keepaspectratio]{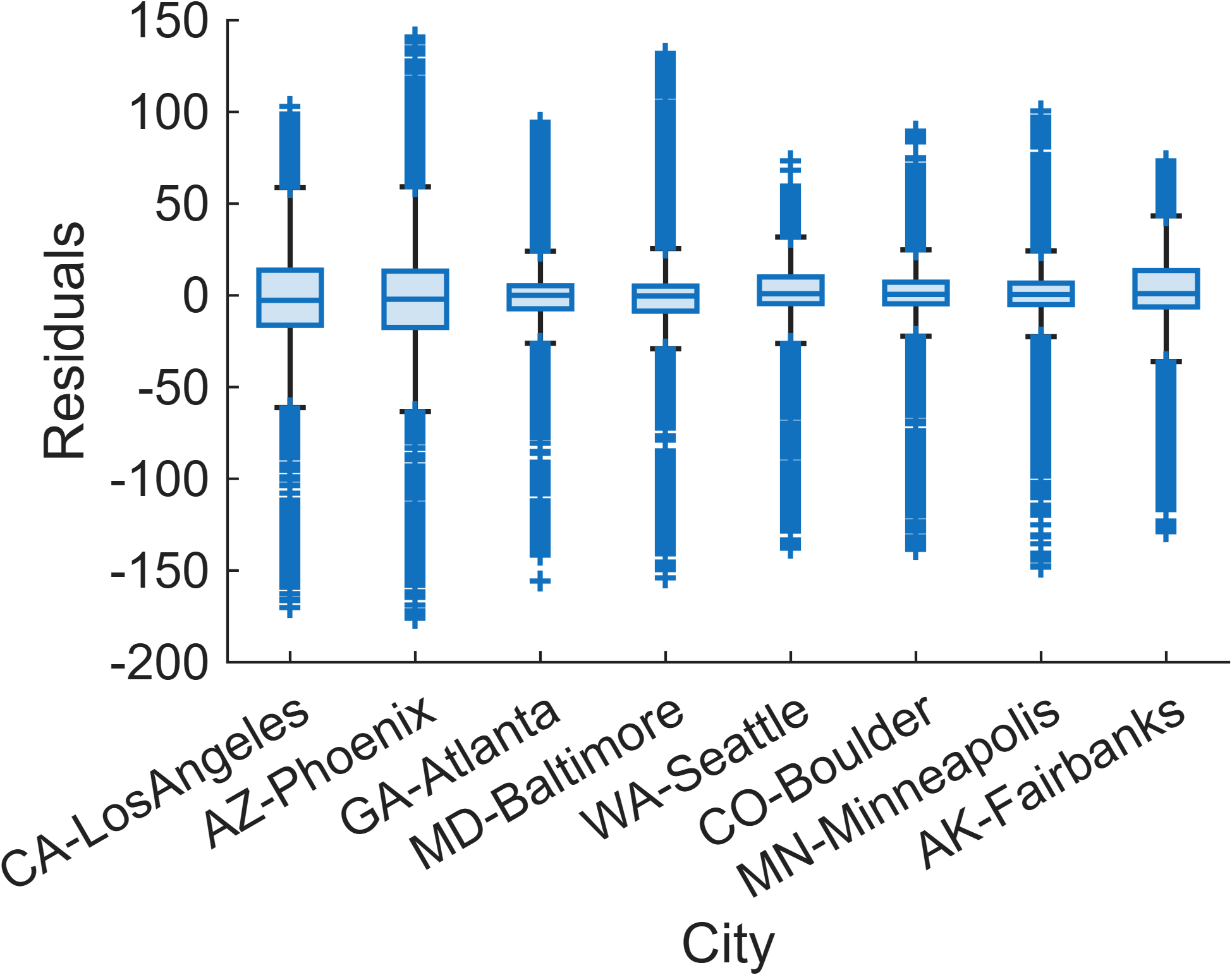}
    \end{minipage}
    \hfill
    \begin{minipage}{0.32\linewidth}
        \textbf{F} \par \vspace{1pt}
        \centering
        \includegraphics[height=6.5cm, width=\linewidth, keepaspectratio]{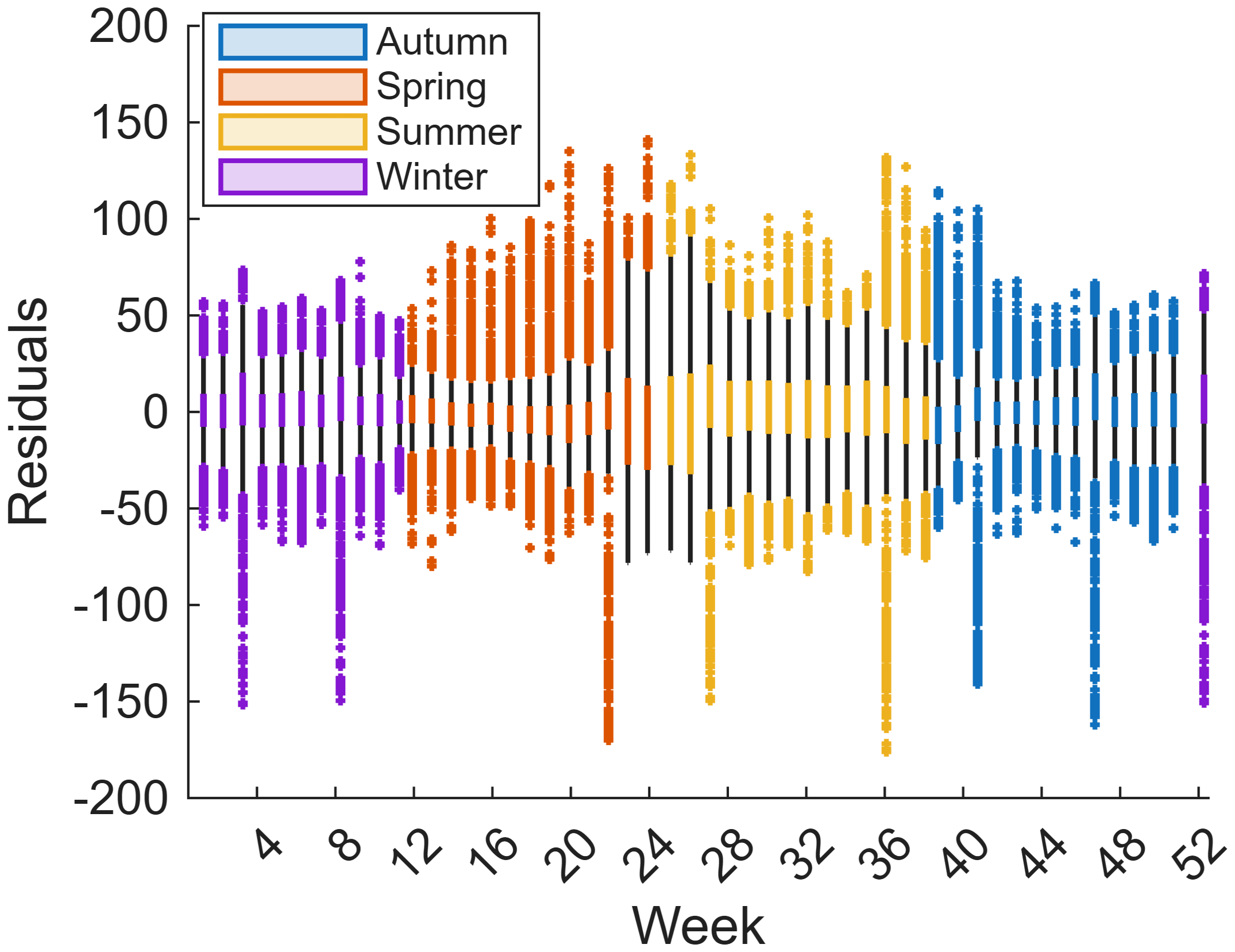}
    \end{minipage}

    \begin{minipage}{\linewidth}
        \textbf{G} \par \vspace{1pt}
        \centering
        \includegraphics[height=6.5cm, width=\linewidth, keepaspectratio]{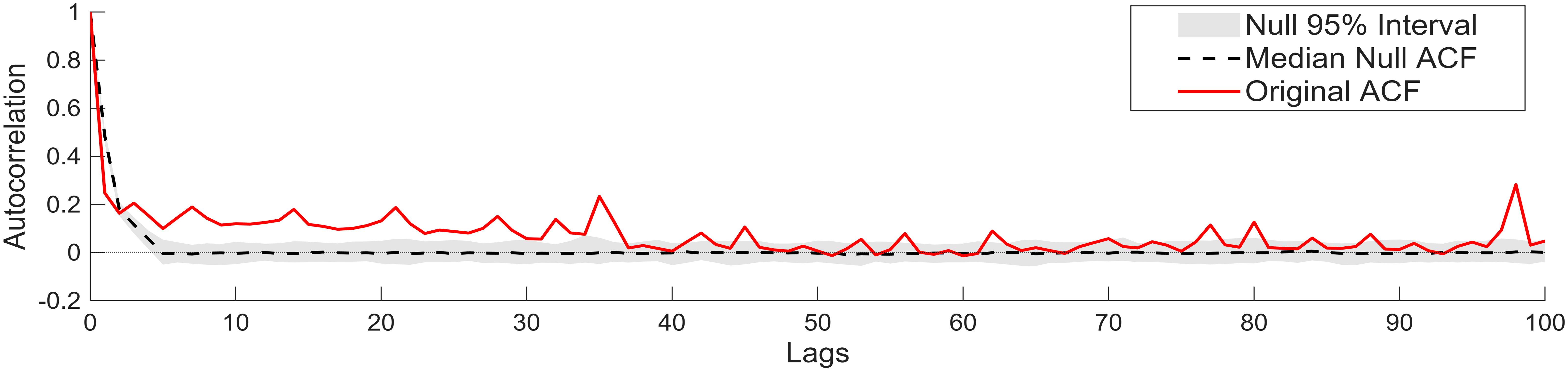}
    \end{minipage}

    \caption{\textbf{Residuals of the ASCA model of the energy consumption dataset, considering factors `city', `week' and `day of the week'.} \textbf{A} Multivariate Statistical Process Control (MSPC) chart (Q-statistic vs. D-statistic) of the model, colored by day of the week. The control limits (red dotted lines) are percentile 99 for both D and Q statistics. \textbf{B} MSPC chart of the model, colored by city. \textbf{C} MSPC chart of the model, colored by week of the year. \textbf{E} boxplot of residuals by day of the week. \textbf{E} boxplot of residuals by city. \textbf{F} boxplot of residuals by week, colored by season. \textbf{G} sample autocorrelation function (ACF) of the Q-statistic. The red line represents the proposed model, while the black dotted line shows the median across 100 block permutations (5-row blocks). The shaded region denotes the 95\% empirical interval across all permutations.}
    \label{fig:energy_3f_residuals}
\end{figure}

Consistent with the previous model, an analysis of residuals was conducted to monitor for heteroscedasticity and dependence. Using MSPC charts, we again identified mid-year outliers originating from Los Angeles and Phoenix (Figure~\ref{fig:energy_3f_residuals}A–C), although these outliers do not align with specific days of the week. The boxplots in Figure~\ref{fig:energy_3f_residuals}E–F confirm that these specific locations and time frames exhibit a variance slightly above the rest of the cities and weeks of the year. Nonetheless, this marginal heteroscedasticity does not significantly affect the conclusions extracted from the model.

A correlogram of the Q-statistic (Figure~\ref{fig:energy_3f_residuals}G) reveals strong autocorrelation across the first two lags. In contrast, autocorrelation coefficients between 0.1 and 0.2 are present from lags 3 through 36 which, despite their small magnitude, deviate from the null distribution. Permutation in larger blocks would make the exchangeable units contain entire weeks, and inference at the finer 'day of the week' scale would be compromised. Ultimately, this unfolding might not be appropriate, and the previous two-factor approach would be a more reliable choice for the analysis of this dataset. Nevertheless, we believe this analysis effectively showcases how a different unfolding affects the interpretation of the data.

\section{Case studies}

This section provides supplementary models and visualizations supporting the analysis of the case studies presented in Section~\ref{sec:casestudies} of the main text.

\subsection{Water temperature in Sierra Nevada lakes}

\begin{figure}[p]
    \centering
    
    \begin{minipage}{0.48\linewidth}
        \textbf{A} \par \vspace{1pt}
        \centering
        \includegraphics[height=6.5cm, width=\linewidth, keepaspectratio]{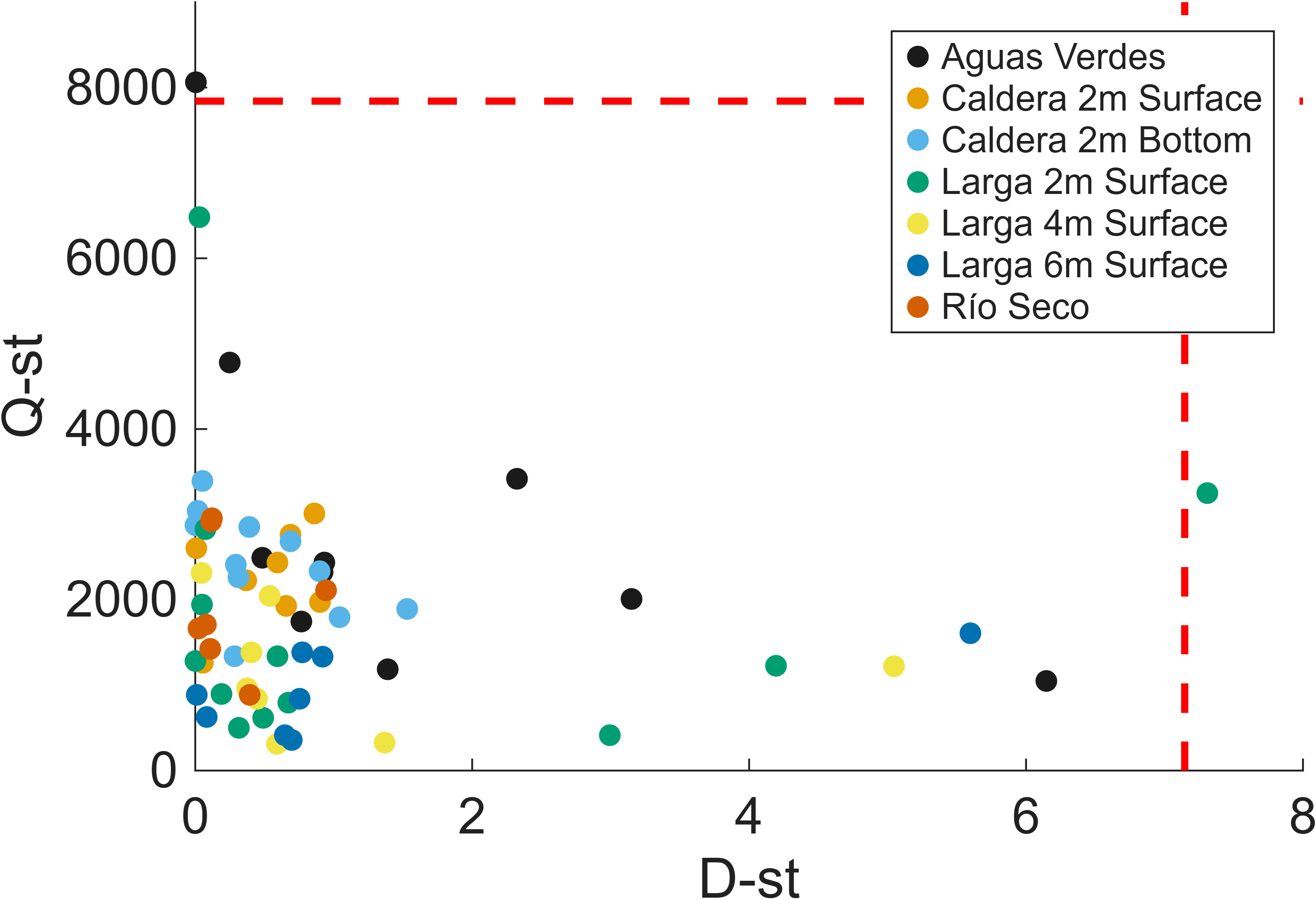}
    \end{minipage}
    \hfill
    \begin{minipage}{0.48\linewidth}
        \textbf{B} \par \vspace{1pt}
        \centering
        \includegraphics[height=6.5cm, width=\linewidth, keepaspectratio]{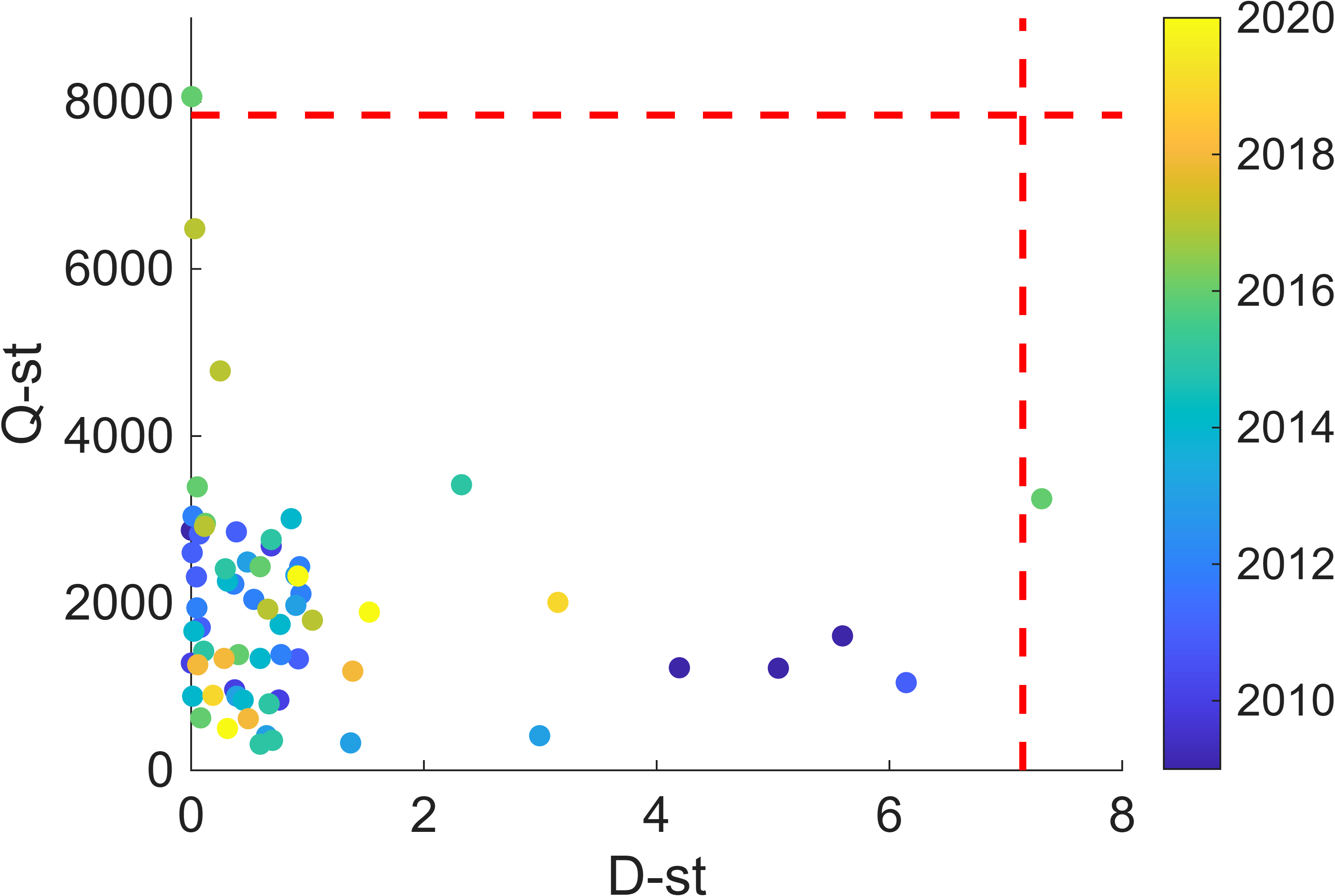}
    \end{minipage}
    
    \begin{minipage}{0.48\linewidth}
        \textbf{C} \par \vspace{1pt}
        \centering
        \includegraphics[height=6.5cm, width=\linewidth, keepaspectratio]{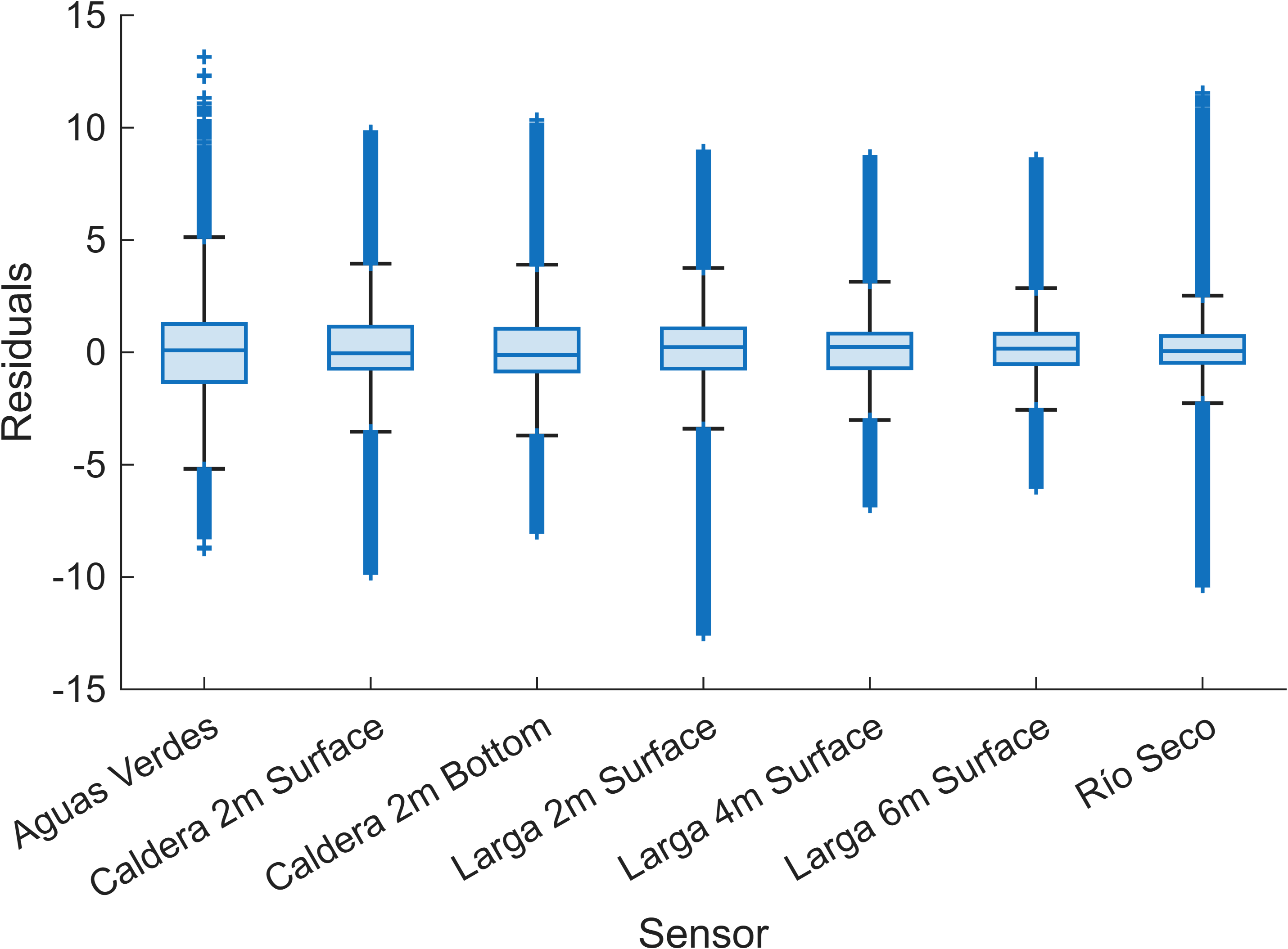}
    \end{minipage}
    \hfill
    \begin{minipage}{0.48\linewidth}
        \textbf{D} \par \vspace{1pt}
        \centering
        \includegraphics[height=6.5cm, width=\linewidth, keepaspectratio]{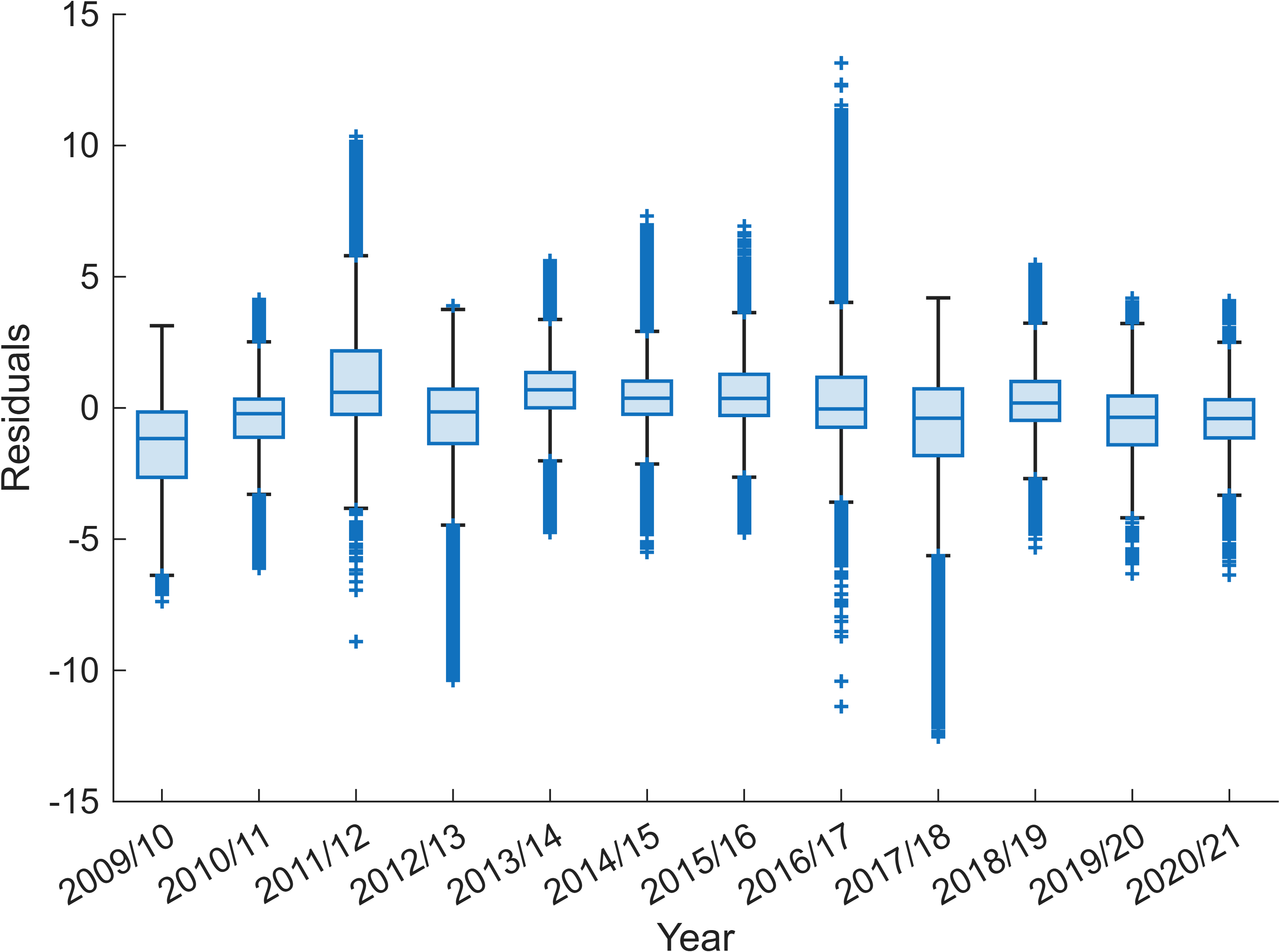}
    \end{minipage}

    \begin{minipage}{\linewidth}
        \textbf{E} \par \vspace{1pt}
        \centering
        \includegraphics[height=6.5cm, width=\linewidth, keepaspectratio]{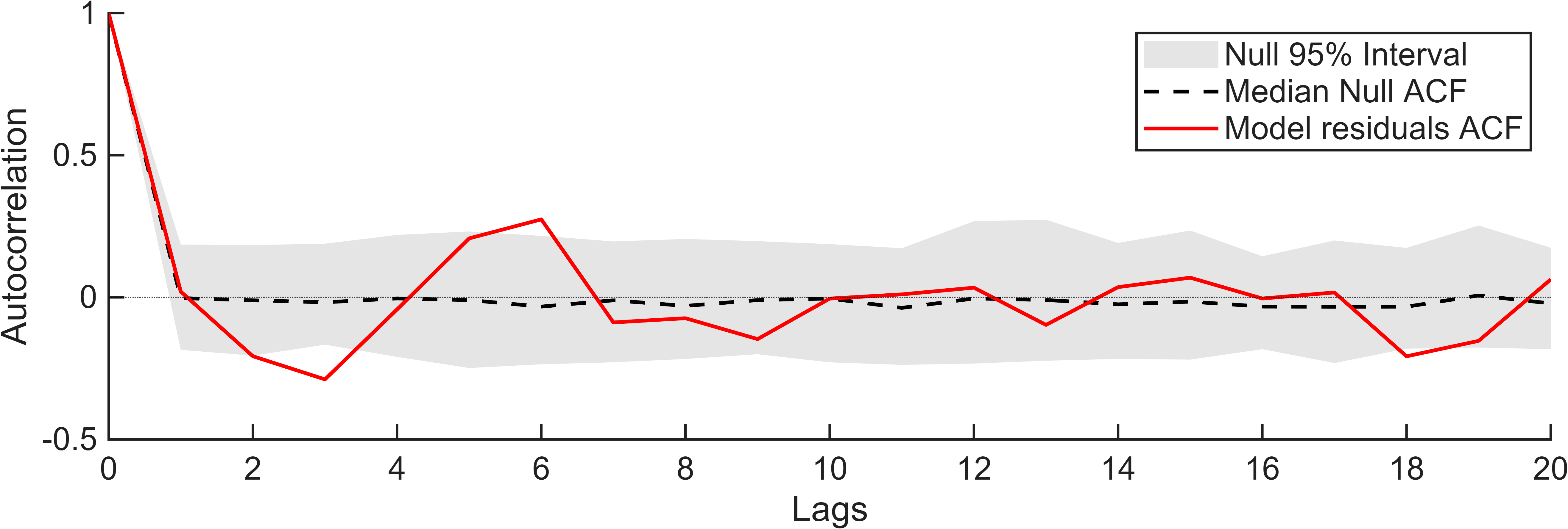}
    \end{minipage}

    \caption{\textbf{Residuals of the ASCA model of the Sierra Nevada lakes dataset.} \textbf{A} Multivariate Statistical Process Control (MSPC) chart (Q-statistic vs. D-statistic) of the model, colored by sensor. The control limits (red dotted lines) are percentile 99 for both D and Q statistics. \textbf{B} MSPC chart of the model, colored by year. \textbf{C} boxplot of residuals by sensor. \textbf{D} boxplot of residuals by year. \textbf{E} sample autocorrelation function (ACF) of the Q-statistic.}
    \label{fig:lakes_residuals}
\end{figure}

To ensure the reliability of the ASCA model inference, the residuals were evaluated for heteroscedasticity and dependence. First, MSPC plots were utilized to visualize the distribution of the residuals. As shown in Figure~\ref{fig:lakes_residuals}A--B, the distribution appears structureless, with no discernible patterns. Boxplots were used for each factor to check for heteroscedasticity (Figure~\ref{fig:lakes_residuals}C--D). The variance remains largely consistent across levels for both factors, with minor differences between years. Finally, the lack of autocorrelation in the residuals was confirmed via a correlogram of the Q-statistic, shown in Figure~\ref{fig:lakes_residuals}E.

A preliminary model was developed prior to the primary analysis, treating the factor `year' as nominal rather than ordinal. This approach allows us to assess whether the expectation of a trend is valid. The interaction between factors was omitted due to limited degrees of freedom. As detailed in Table~\ref{tab:lakes_nominal}, both `year' and `sensor' factors are statistically significant, with the former exhibiting approximately double the influence of the latter in terms of mean squares.

\begin{figure}[ht]
    \centering
    
    \begin{minipage}{0.48\linewidth}
        \textbf{A} \par \vspace{1pt}
        \centering
        \includegraphics[height=6.5cm, width=\linewidth, keepaspectratio]{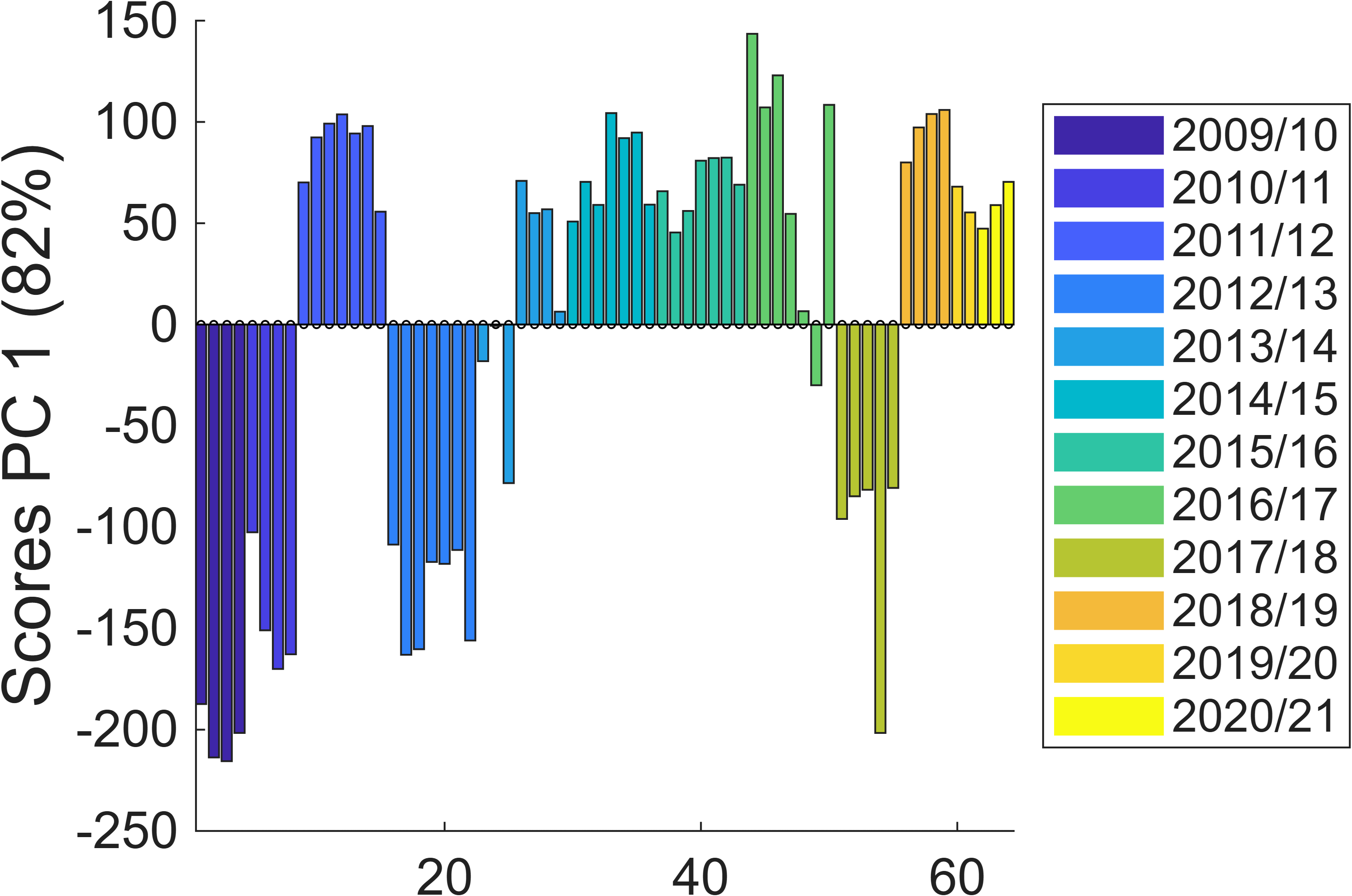}
    \end{minipage}
    \hfill
    \begin{minipage}{0.48\linewidth}
        \textbf{B} \par \vspace{1pt}
        \centering
        \includegraphics[height=6.5cm, width=\linewidth, keepaspectratio]{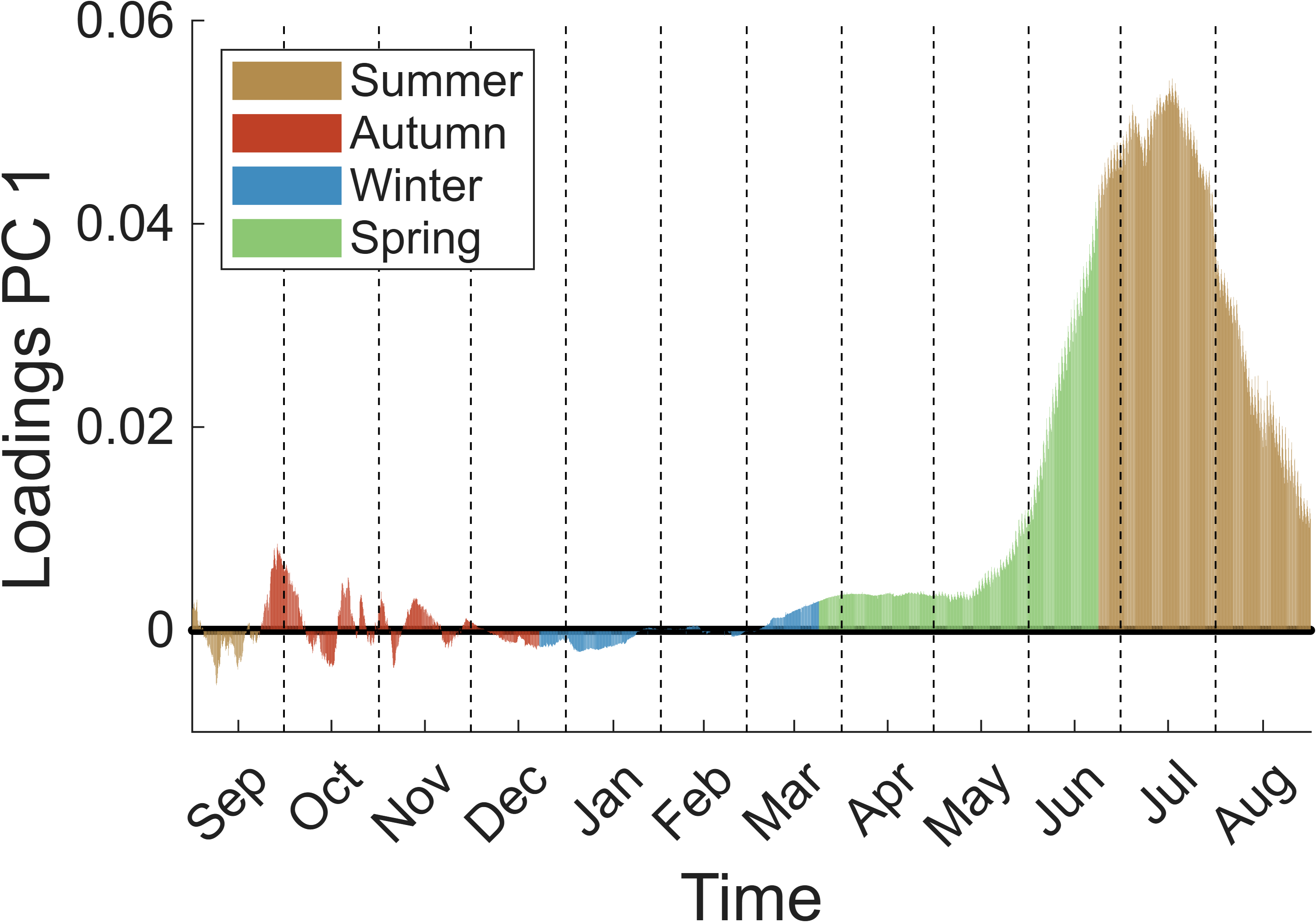}
    \end{minipage}
    
    \begin{minipage}{0.48\linewidth}
        \textbf{C} \par \vspace{1pt}
        \centering
        \includegraphics[height=6.5cm, width=\linewidth, keepaspectratio]{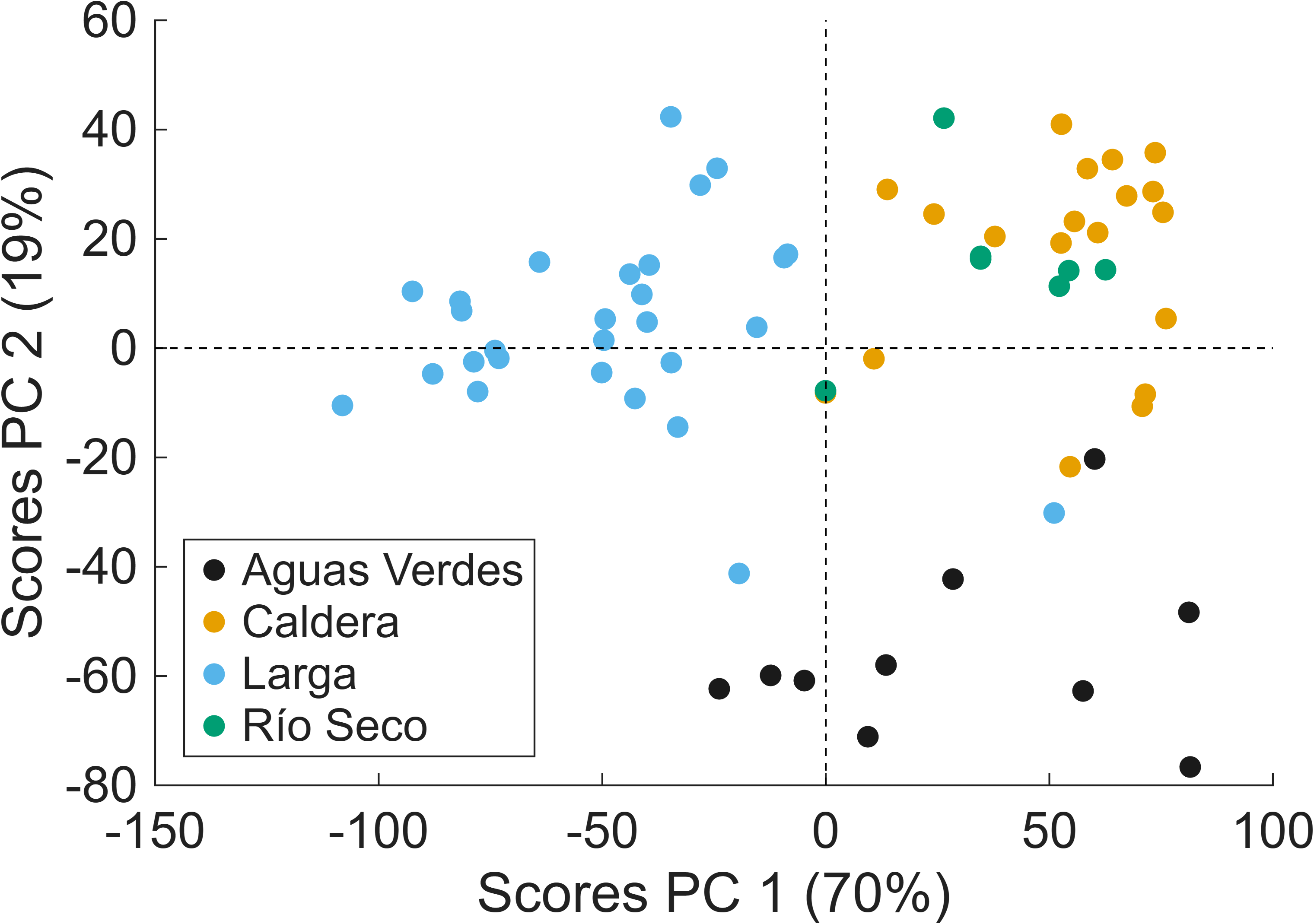}
    \end{minipage}
    \hfill
    \begin{minipage}{0.48\linewidth}
        \textbf{D} \par \vspace{1pt}
        \centering
        \includegraphics[height=6.5cm, width=\linewidth, keepaspectratio]{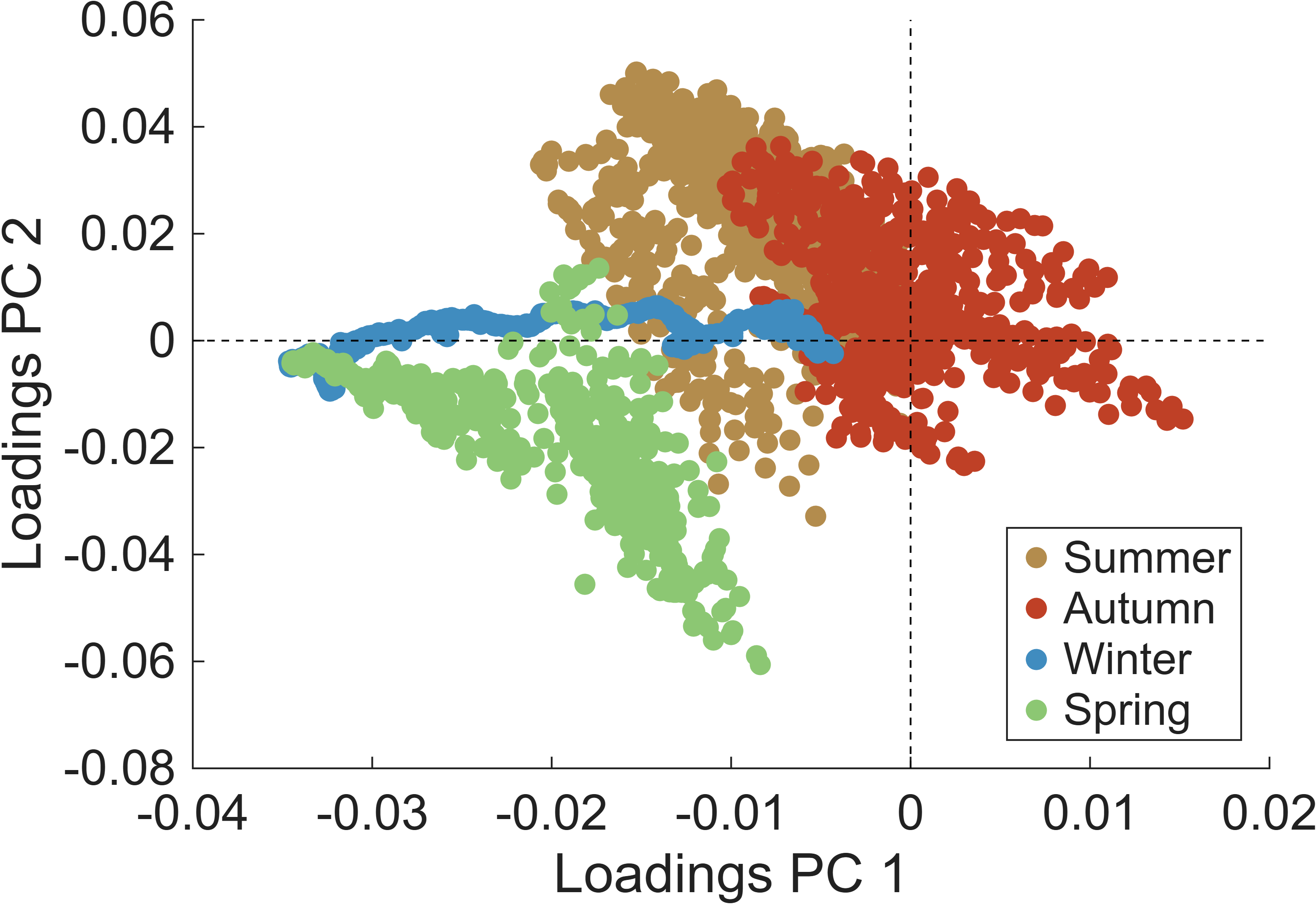}
    \end{minipage}

    \caption{\textbf{ASCA model of the Sierra Nevada lakes dataset, considering the `year' factor as a nominal factor.} \textbf{A} scores for the first component of factor `year'. \textbf{B} loadings for the first component of factor `year', colored by season. \textbf{C} scores for PCs 1 and 2 of factor `sensor', colored by lake. \textbf{D} loadings for PCs 1 and 2 of factor `sensor', colored by season.}
    \label{fig:lakes_nominal}
\end{figure}

Figure~\ref{fig:lakes_nominal}A--B illustrates the score and loading plots for the first component of factor `year' (explaining $82\%$ of the variance) derived from the ASCA model. High positive scores in this component correlate with elevated temperatures during the summer months, specifically from May to August. With the exception of the 2011/2012 period, earlier years display negative scores, while recent years (excluding 2017/2018) show positive scores. This distribution implies a general increase in summer temperatures over the study period, barring two anomalous years, and justifies testing for a trend. The model discussed in Subsection~\ref{sec:lakes} of the main text formally tests the significance of this trend. Finally, for completeness, Figure~\ref{fig:lakes_nominal}C--D presents the score and loading plots for the first two principal components of factor `sensor', which exhibit a pattern closely resembling that of the model with the ordinal `year' factor, although not identical due to the moderate imbalance.

\begin{table}[h]
\centering
\resizebox{0.75\textwidth}{!}{%
\begin{tabular}{lrrrrrr}
\hline
 & \textbf{SS ($\times 10^5$)} & \textbf{\%SS} & \textbf{df} & \textbf{MS ($\times 10^3$)} & \textbf{F} & \textbf{$p$-value} \\
\hline
\hline
Year      & 8.05  & 65.21  & 11 & 73.16 & 12.88 & $<0.001$ \\
Sensor    & 2.07  & 16.75  & 6  & 34.46 & 6.07  & $<0.001$ \\
Residuals & 2.61  & 21.17  & 46 & 5.68  & --    & --       \\
\hline
Total     & 12.3 & 103.13 & 63 & 19.59 & --    & --       \\
\hline
\hline
\end{tabular}
}
\caption{ASCA table for the Sierra Nevada lakes dataset, treating factor `year' as nominal. The columns indicate, respectively: the sum of squares (SS), its percentage relative to the total (\%SS), degrees of freedom (df), mean squares (MS), the F-statistic (F), and the associated $p$-value. The deviation of the total \%SS from 100\% is caused by a loss of orthogonality due to a moderately unbalanced design.}
\label{tab:lakes_nominal}
\end{table}

\begin{figure}[ht]
    \centering
    
    \begin{minipage}{0.48\linewidth}
        \textbf{A} \par \vspace{1pt}
        \centering
        \includegraphics[height=7cm, width=\linewidth, keepaspectratio]{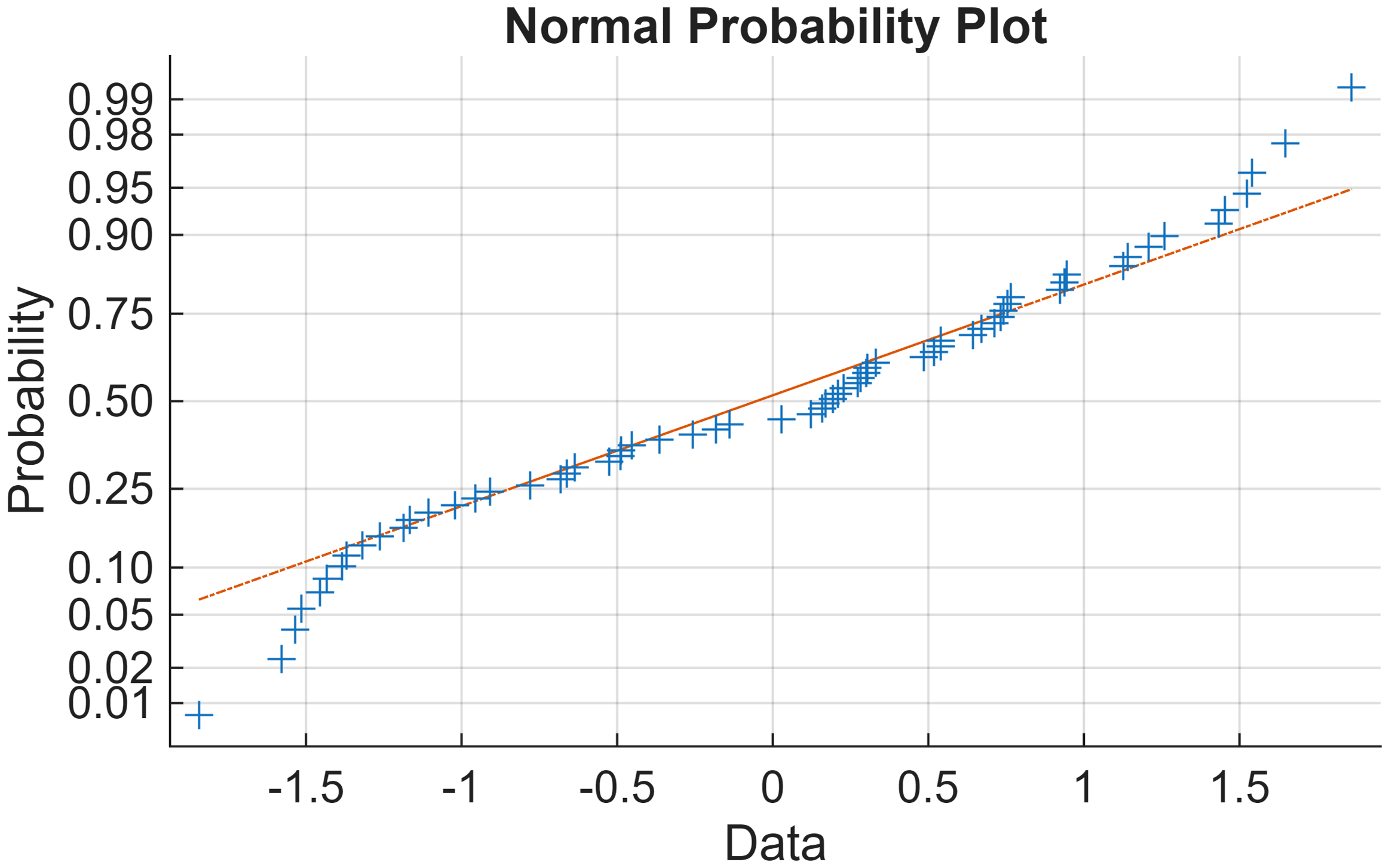}
    \end{minipage}
    \hfill
    \begin{minipage}{0.48\linewidth}
        \textbf{B} \par \vspace{1pt}
        \centering
        \includegraphics[width=\linewidth]{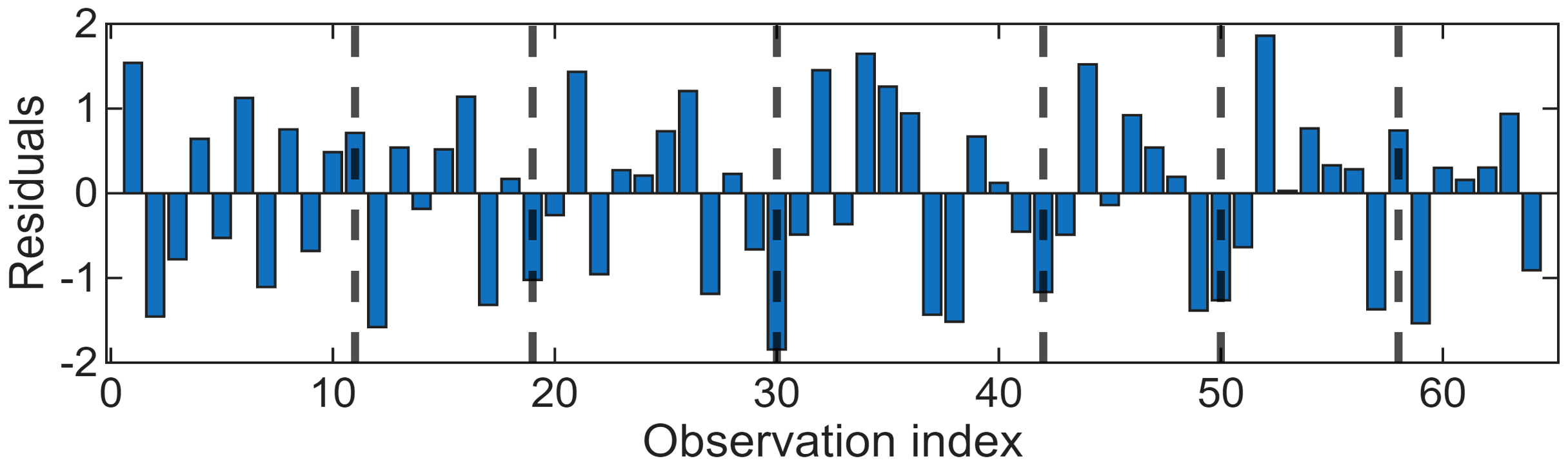}
        \raggedright \textbf{C} \par \vspace{1pt}
        \includegraphics[width=\linewidth]{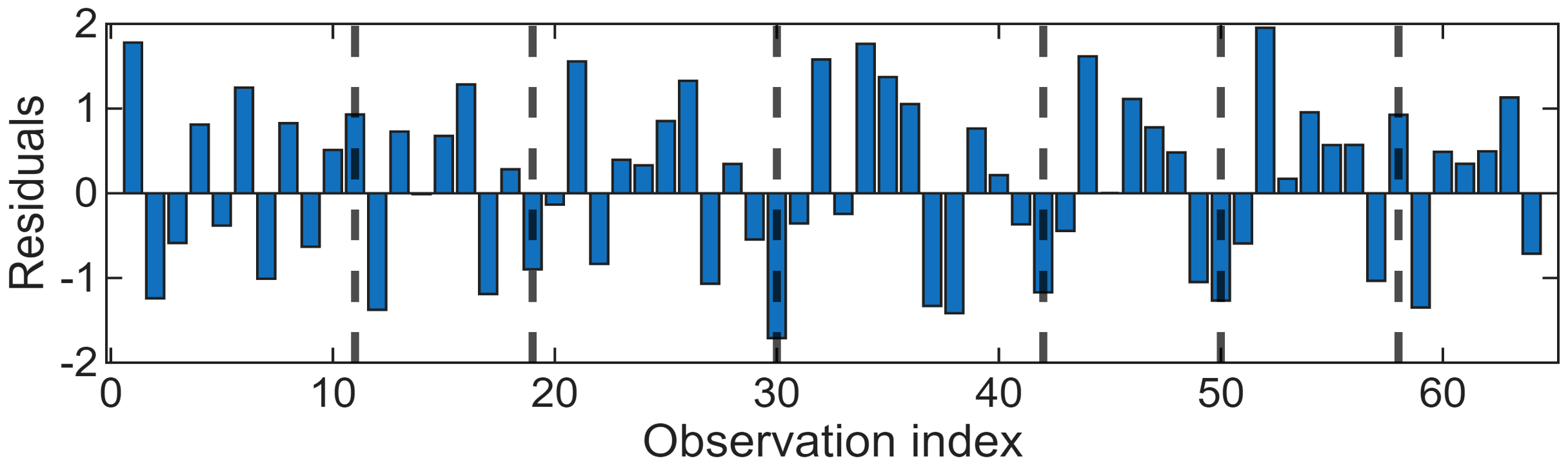}
    \end{minipage}

    \caption{\textbf{Analysis of residuals of the ANOVA models for the Sierra Nevada lakes dataset.} \textbf{A} Normal Q-Q plot of the residuals of the parametric ANOVA model. \textbf{B} Residuals of the parametric ANOVA model. The dotted lines separate observations from different sensors. \textbf{C} Residuals of the permutation-based ANOVA model.}
    \label{fig:residuals}
\end{figure}

The assumptions of ANOVA were verified to ensure the reliability of the statistical inference of the models in the main text, as illustrated in Figure~\ref{fig:residuals}. The slight deviation from normality in the residuals (Figure~\ref{fig:residuals}A) is tolerated by the permutation-based ANOVA model, which utilizes permutation testing to circumvent the normality requirement. Figure~\ref{fig:residuals}B--C shows that autocorrelation associated with the `year' factor is undetectable in the residuals of both models.

In addition to the ANOVA models in the main text, two extra models were applied, employing different forms of sum of squares calculation for unbalanced designs---like the one in this case study. Table~\ref{tab:lakes_anova_sm} summarizes the results of two models, one with type I sum of squares and another with a type II sum of squares in which the hierarchy of terms accounts for ordinal and nominal factors (option \texttt{`h'} of the \texttt{`sstype'} parameter of the \texttt{anovan} function in MATLAB R2025b).

\begin{table}[ht]
\small
\centering

\raggedright \hspace{1cm} \textbf{A} \\
\centering
\resizebox{0.7\textwidth}{!}{%
\begin{tabular}{lrrrrrr}
\hline
 & \textbf{SS} & \textbf{\%SS} & \textbf{df} & \textbf{MS} & \textbf{F} & \textbf{$p$-value} \\
\hline
\hline
Year (Ordinal)              & 3.67  & 3.78  & 1  & 3.67  & 3.04 & 0.087 \\
Sensor               & 27.31 & 28.11 & 6  & 4.55  & 3.77 & 0.004 \\
Year $\times$ Sensor & 5.81  & 5.98  & 6  & 0.97  & 0.80 & 0.573 \\
Residuals            & 60.37 & 62.14 & 50 & 1.21  & --   & --    \\
\hline
Total                & 97.16 & 100.0 & 63 & 1.54  & --   & --    \\
\hline
\hline
\end{tabular}
}
\vspace{0.3cm}

\raggedright \hspace{1cm} \textbf{B} \\
\centering
\resizebox{0.7\textwidth}{!}{%
\begin{tabular}{lrrrrrr}
\hline
 & \textbf{SS} & \textbf{\%SS} & \textbf{df} & \textbf{MS} & \textbf{F} & \textbf{$p$-value} \\
\hline
\hline
Year (Ordinal)            & 10.08 & 10.37 & 1  & 10.08 & 8.35 & 0.006 \\
Sensor               & 27.31 & 28.11 & 6  & 4.55  & 3.77 & 0.004 \\
Year $\times$ Sensor & 5.81  & 5.98  & 6  & 0.97  & 0.80 & 0.573 \\
Residuals            & 60.37 & 62.14 & 50 & 1.21  & --   & --    \\
\hline
Total                & 97.16 & 106.59& 63 & 1.54  & --   & --    \\
\hline
\hline
\end{tabular}

}
\caption{Parametric ANOVA tables for the Sierra Nevada lakes dataset, using different types of sum of squares. The columns indicate, respectively: the sum of squares (SS), its percentage relative to the total (\%SS), degrees of freedom (df), mean squares (MS), the F-statistic (F), and the associated $p$-value. \textbf{A} Type I Sum of Squares. \textbf{B} Hierarchical model based on Type II Sum of Squares (option \texttt{`h'} of the \texttt{anovan} function in MATLAB). The deviation of the total \%SS from 100\% is caused by a loss of orthogonality due to a moderately unbalanced design.}
\label{tab:lakes_anova_sm}
\end{table}

\subsection{Airborne pollen concentration in Granada}

\subsubsection{Analysis of residuals}

\begin{figure}[p]
    \centering
    
    \begin{minipage}{0.48\linewidth}
        \textbf{A} \par \vspace{1pt}
        \centering
        \includegraphics[height=6.5cm, width=\linewidth, keepaspectratio]{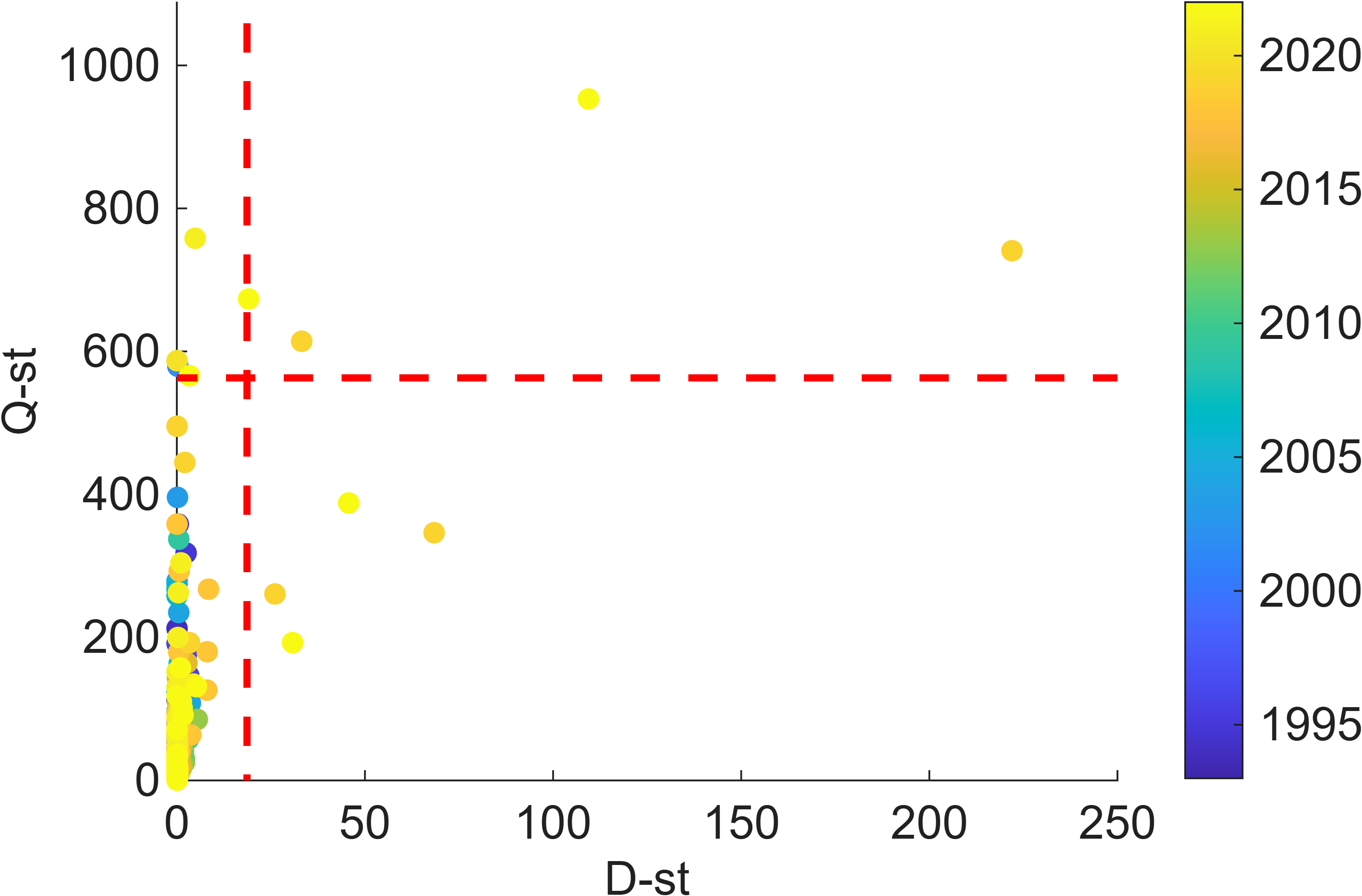}
    \end{minipage}
    \hfill
    \begin{minipage}{0.48\linewidth}
        \textbf{B} \par \vspace{1pt}
        \centering
        \includegraphics[height=6.5cm, width=\linewidth, keepaspectratio]{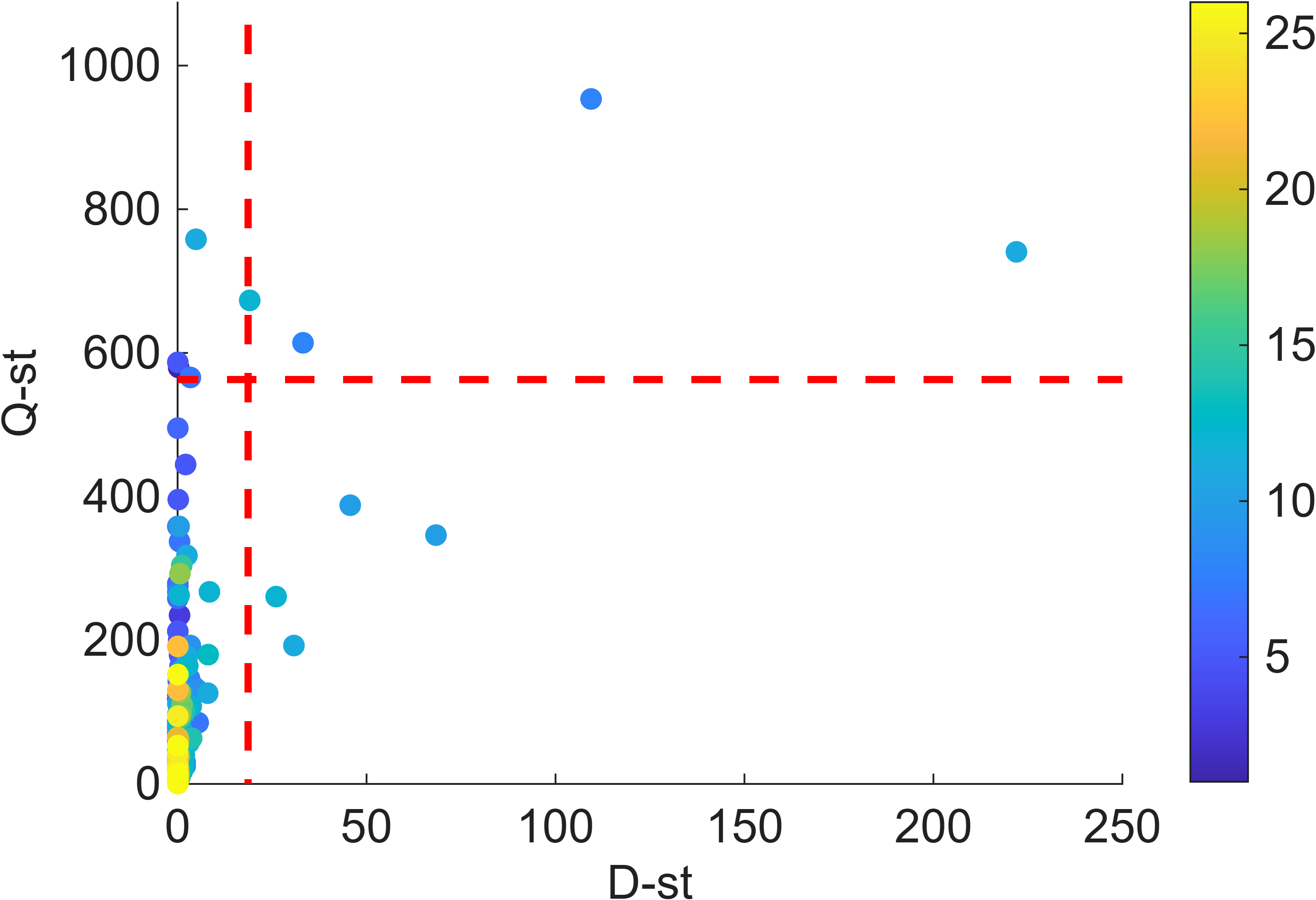}
    \end{minipage}
    
    \begin{minipage}{0.48\linewidth}
        \textbf{C} \par \vspace{1pt}
        \centering
        \includegraphics[height=6.5cm, width=\linewidth, keepaspectratio]{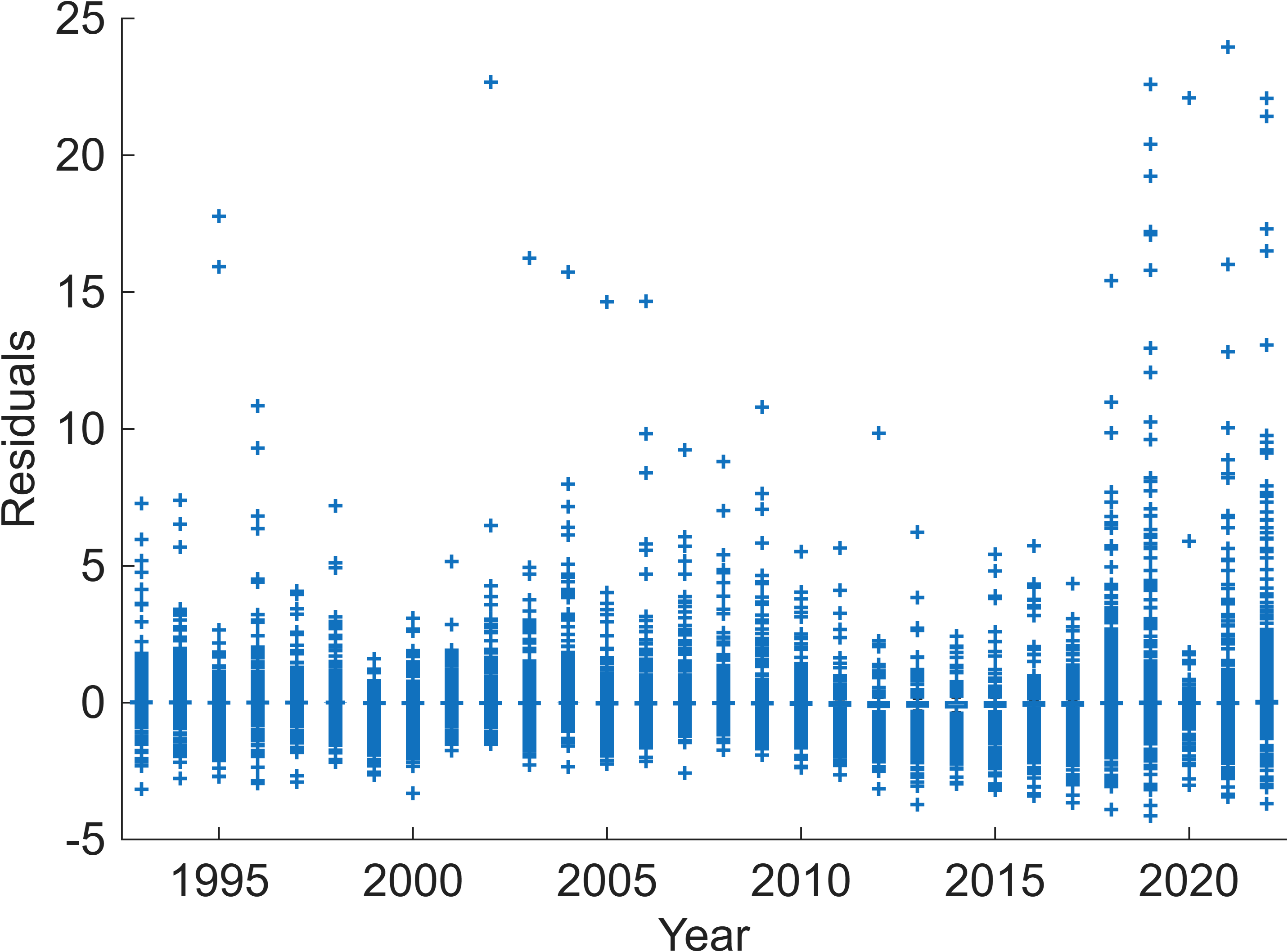}
    \end{minipage}
    \hfill
    \begin{minipage}{0.48\linewidth}
        \textbf{D} \par \vspace{1pt}
        \centering
        \includegraphics[height=6.5cm, width=\linewidth, keepaspectratio]{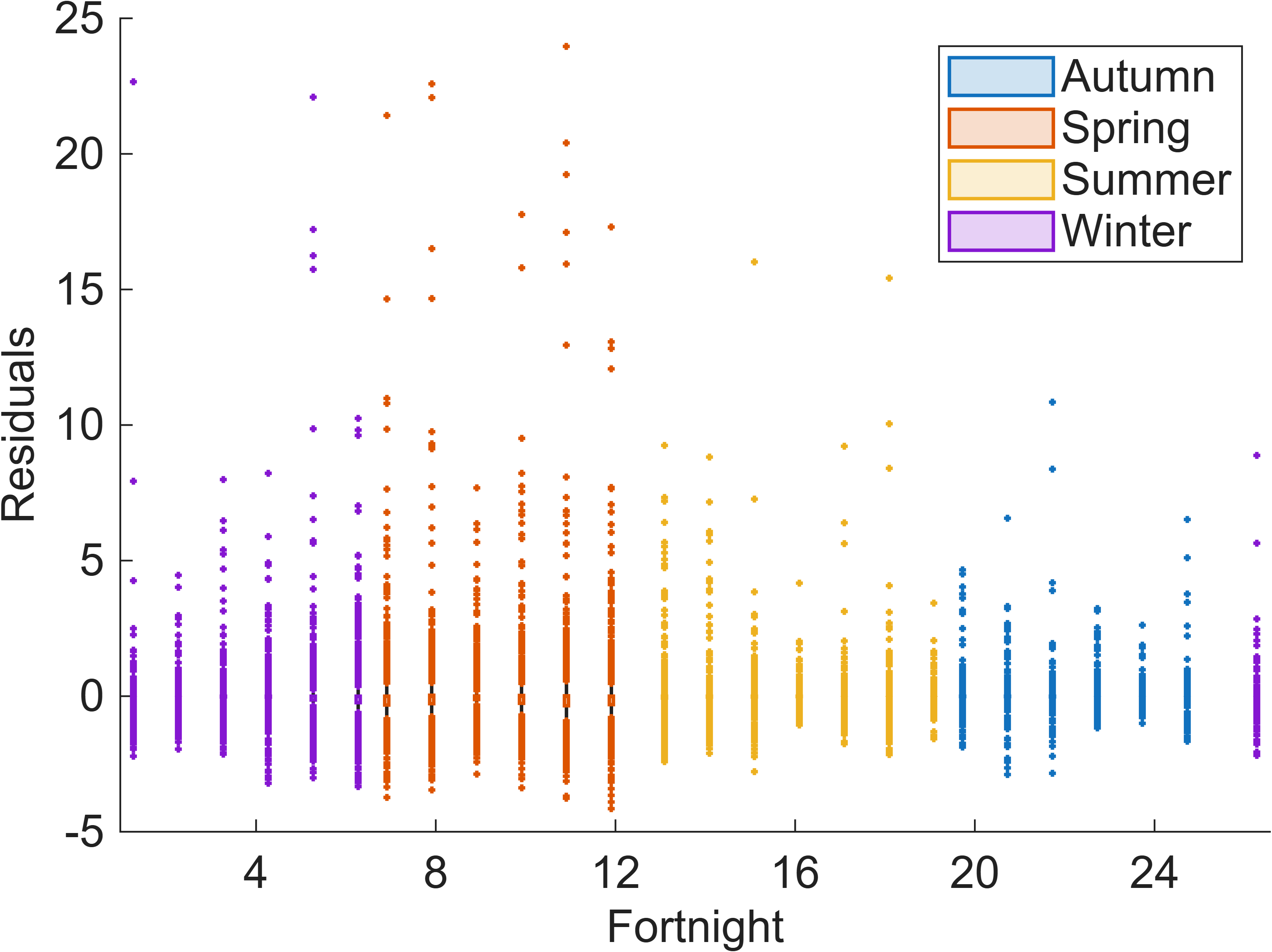}
    \end{minipage}

    \begin{minipage}{\linewidth}
        \textbf{E} \par \vspace{1pt}
        \centering
        \includegraphics[height=6.5cm, width=\linewidth, keepaspectratio]{pollen_ASCA_2_ACF_k2.png}
    \end{minipage}

    \caption{\textbf{Residuals of the ASCA model of the airborne pollen dataset.} \textbf{A} Multivariate Statistical Process Control (MSPC) chart (Q-statistic vs. D-statistic) of the model, colored by year. The control limits (red dotted lines) are percentile 99 for both D and Q statistics. \textbf{B} MSPC chart of the model, colored by fortnight. \textbf{C} boxplot of residuals by year. \textbf{D} boxplot of residuals by fortnight. \textbf{E} sample autocorrelation function (ACF) of the Q-statistic. The red line represents the proposed model, while the black dotted line shows the median across 100 block permutations (6-row blocks). The shaded region denotes the 95\% empirical interval across all permutations.}
    \label{fig:pollen_residuals}
\end{figure}

Residuals of the ASCA model were evaluated to assess homoscedasticity and independence. First, MSPC plots were used to examine the residual distribution. As shown in Figure~\ref{fig:pollen_residuals}A--B, a few outliers corresponding to spring in recent years are present. Next, factor-wise boxplots were inspected to check for heteroscedasticity (Figure~\ref{fig:pollen_residuals}C--D). Variance remained largely consistent across years, despite slightly higher variability during late winter and spring fortnights relative to the rest of the year. Finally, a correlogram of the residual Q-statistic revealed autocorrelation exceeding $0.2$ across the first four lags. To account for this temporal autocorrelation within the null distribution, we implemented a block permutation scheme using 13-row blocks (equivalent to 13 consecutive fortnights). The correlogram in Figure~\ref{fig:pollen_residuals}E displays the correlogram for the main model alongside the median of 100 random block permutations, indicating only minor deviations from the null distribution.

\subsubsection{Alternative models}

Several different ASCA models were developed for the airborne pollen dataset. Here, we will present some of the preliminary models of the one discussed in the main text, as well as some extra plots of the interaction factor of the main text's model.  In total, we will showcase three different models. These models utilize the same block permutation scheme as the main model.

The first of these models uses mean centering preprocessing instead of the autoscaling used in the main text's model. Also unlike the main text's model, it considers factor year to be nominal. Table \ref{tab:sup_pollen_tables}A displays the results of significance testing for the ASCA model. In this case, the year factor is not deemed significant, implying that the most common pollen types (\textit{Olea} and \textit{Cupressaceae}) that dominate the model experience minimal change across years. By contrast, the main text's model, which applies autoscaling, reveals significant temporal trends for less abundant pollen types. Factor fortnight remains statistically significant.

\begin{figure}


    \begin{minipage}{0.49\linewidth}
        \textbf{A} \par \vspace{1pt}
        \centering
        \includegraphics[height=6.5cm, width=\linewidth, keepaspectratio]{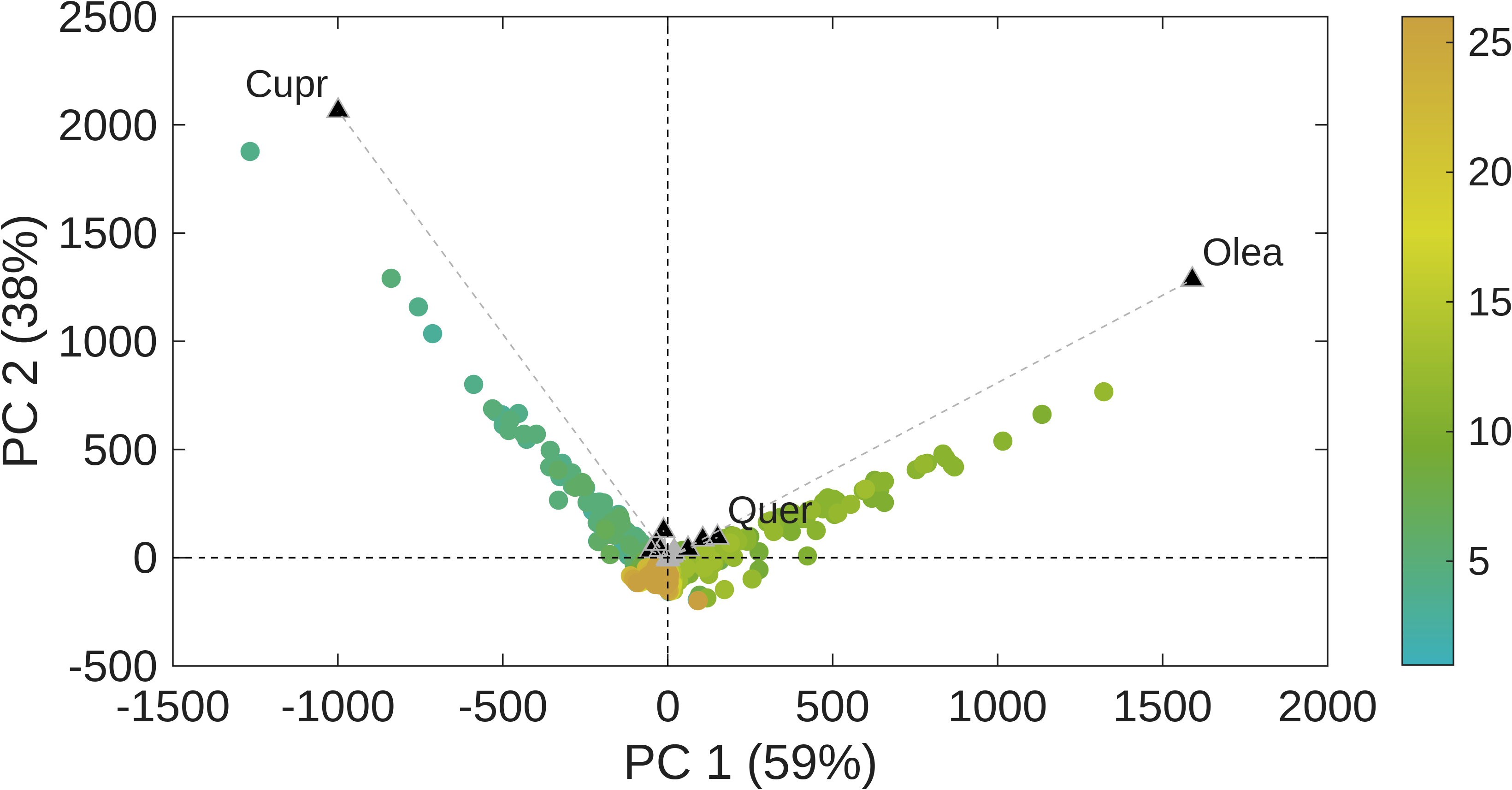}
        
    \end{minipage}
    \hfill
    \begin{minipage}{0.49\linewidth}
        \textbf{B} \par \vspace{1pt}
        \centering
        \includegraphics[height=6.5cm, width=\linewidth, keepaspectratio]{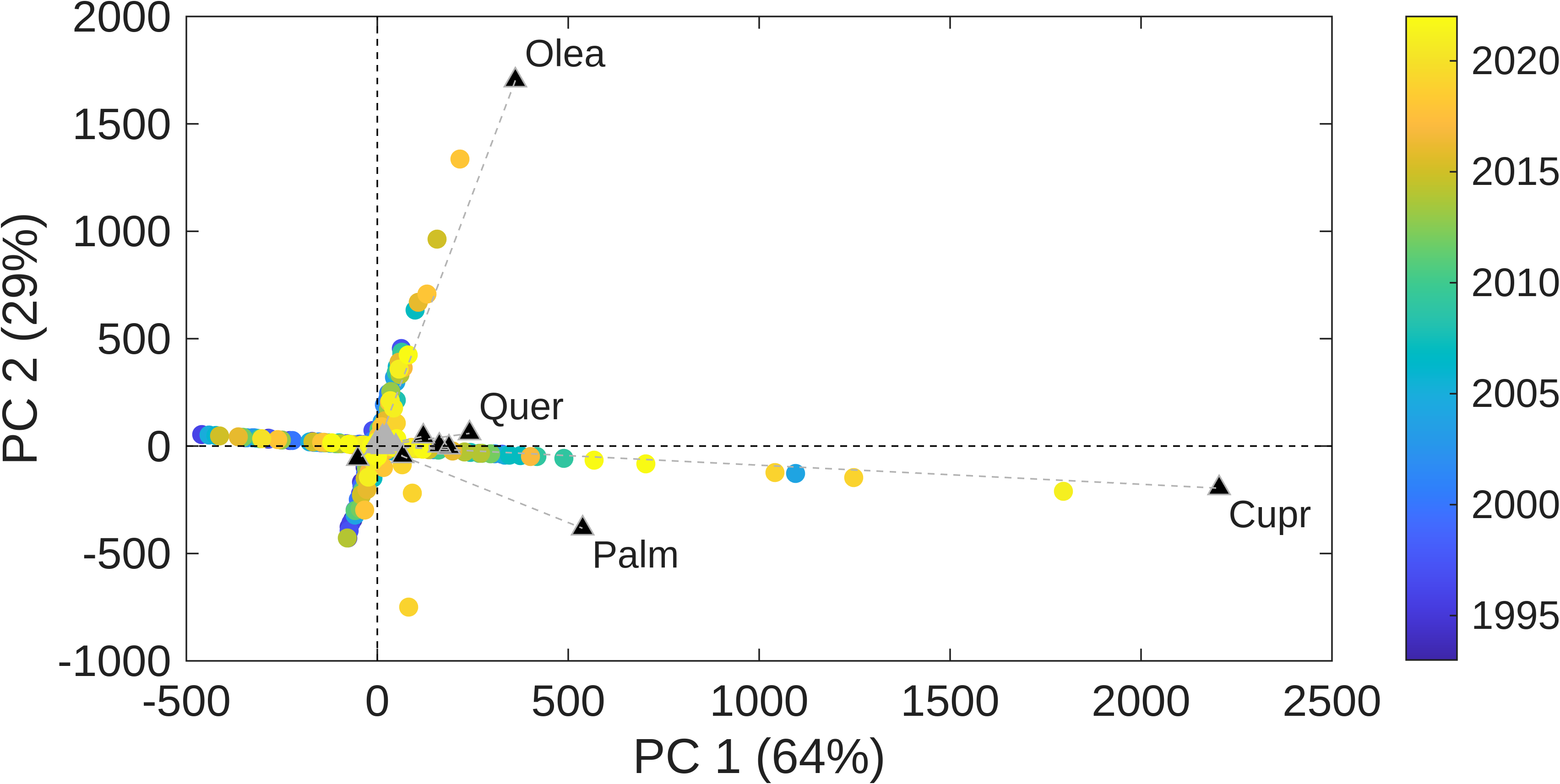}
    \end{minipage}

    \vspace{0.1cm}

    \begin{minipage}{0.49\linewidth}
        \textbf{C} \par \vspace{1pt}
        \centering
        \includegraphics[height=6.5cm, width=\linewidth, keepaspectratio]{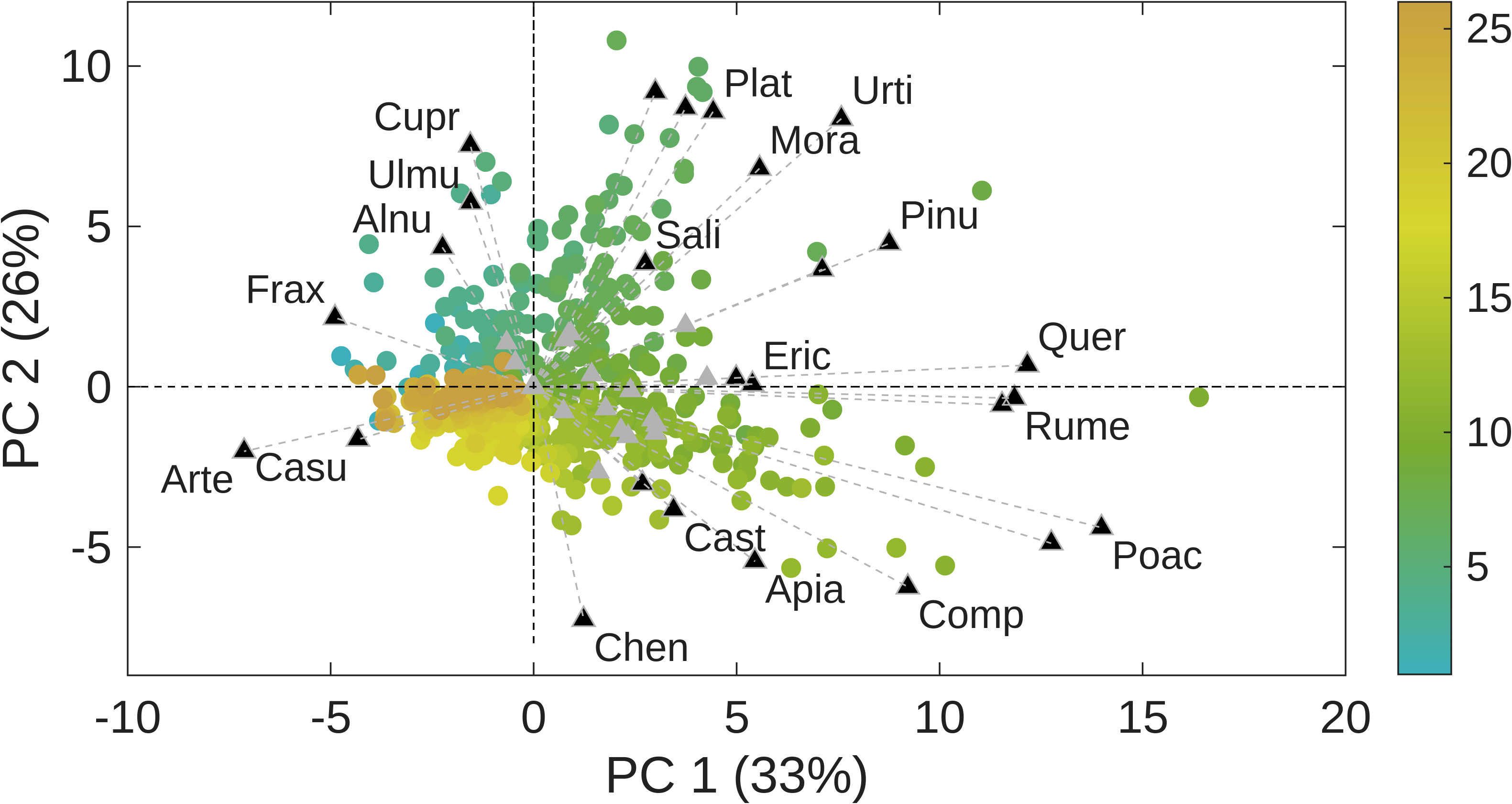}
        
    \end{minipage}
    \hfill
    \begin{minipage}{0.49\linewidth}
        \textbf{D} \par \vspace{1pt}
        \centering
        \includegraphics[height=6.5cm, width=\linewidth, keepaspectratio]{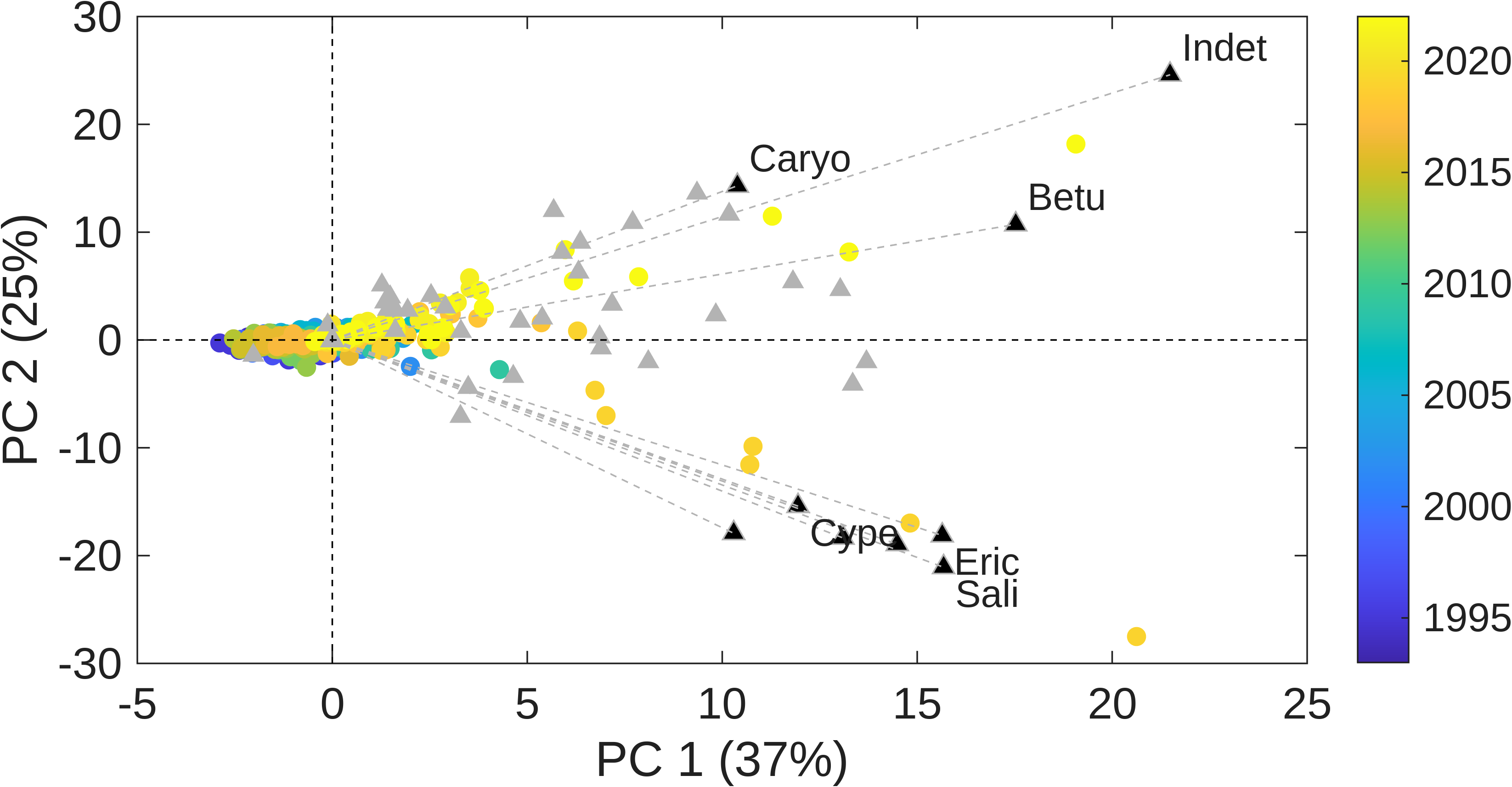}
    \end{minipage}

        \vspace{0.1cm}

    \begin{minipage}{0.49\linewidth}
        \textbf{E} \par \vspace{1pt}
        \centering
        \includegraphics[height=6.5cm, width=\linewidth, keepaspectratio]{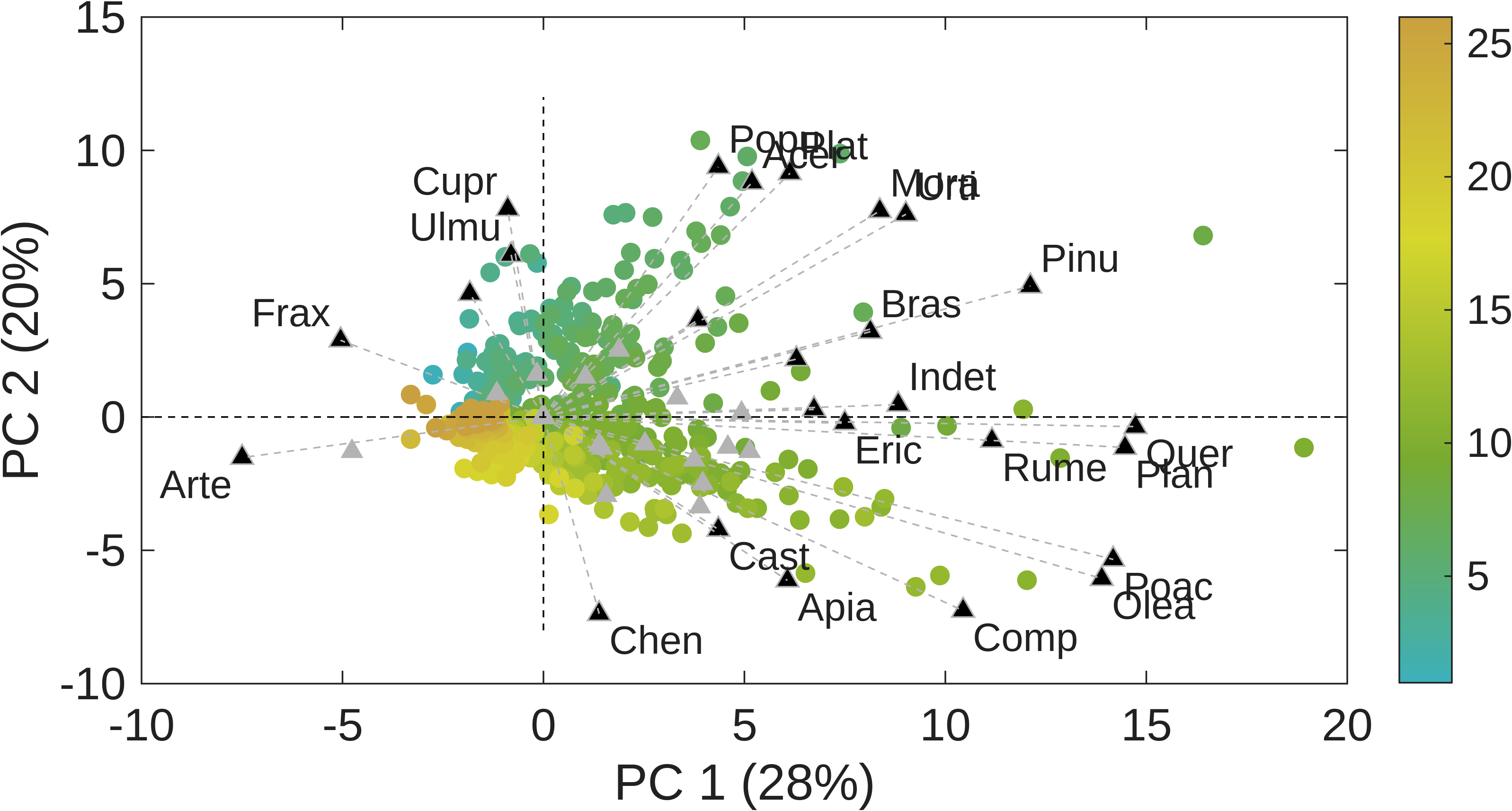}
        
    \end{minipage}
    \hfill
    \begin{minipage}{0.49\linewidth}
        \textbf{F} \par \vspace{1pt}
        \centering
        \includegraphics[height=6.5cm, width=\linewidth, keepaspectratio]{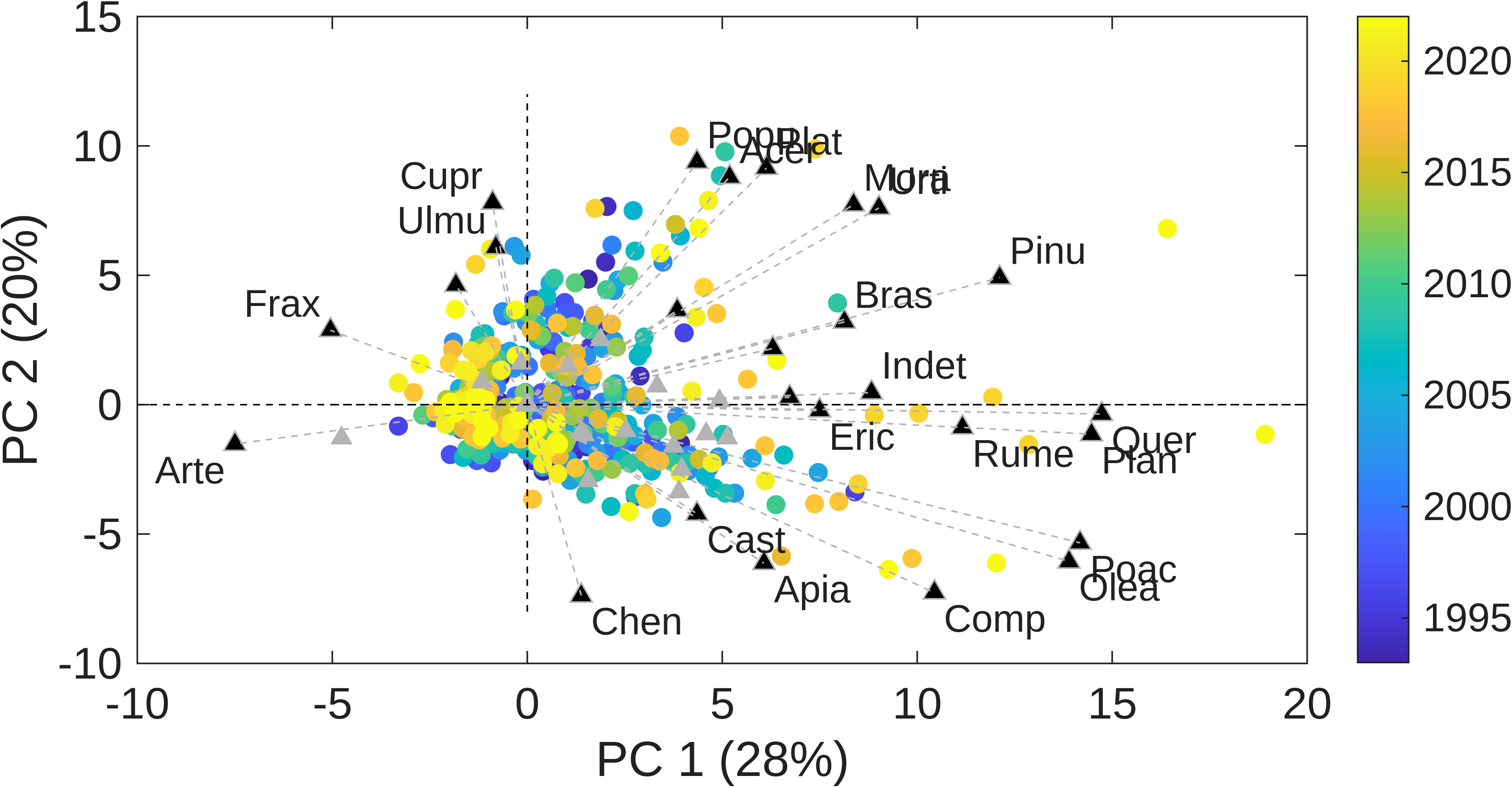}
    \end{minipage}

    \caption{\textbf{Supplementary ASCA models for the airborne pollen dataset}. \textbf{A-B} ASCA model with mean centering and factor year as nominal. \textbf{A} biplot for factor fortnight: the model is dominated by the pollen types with the highest concetrations: \textit{Olea} and \textit{Cupressaceae}. \textbf{B} biplot for factor year: this factor is also dominated by the same two pollen types. \textbf{C-D} ASCA model with autoscaling and factor year as nominal. \textbf{C} biplot for factor fortnight. \textbf{D} biplot for factor year. \textbf{E-F} interaction ASCA model with autoscaling and factor year as ordinal (main text model). \textbf{E} interaction biplot colored according to fortnight. \textbf{F} interaction biplot colored according to year. While more complete, these last two plots are more difficult to interpret, which is why we showcased just PC1 in the main text.\\
    \scriptsize The pollen types shown in the figures are named as acronyms: Arte, Artemisia; Casu, Casuarina; Frax, Fraxinus; Alnu, Alnus; Ulmu, Ulmus; Cupr, Cupressaceae; Popu, Populus; Acer, Acer; Plat, Platanus; Sali, Salix; Mora, Moraceae; Urti, Urticaceae; Pinu, Pinus; Bras, Brassicaceae; Indet, Indeterminate; Quer, Quercus; Plan, Plantago; Rume, Rumex; Poac, Poaceae; Cast, Castanea; Olea, Olea; Comp, Compositae; Apia, Apiaceae; Chen, Amaranthaceae.} 
    \label{fig:sup_ASCA_pollen}
\end{figure}

Figures \ref{fig:sup_ASCA_pollen}A and \ref{fig:sup_ASCA_pollen}B show the biplots of both factors. We can see that the model is completely dominated by just two pollen types: \textit{Olea} and \textit{Cupressaceae}. These two pollen types are the ones that feature the highest maximum concentrations. For our main model, we decided to use autoscaling as our preprocessing step. This, allowed us to build a richer model that let us see the different pollen types on an even field.

The second model uses autoscaling, but considers factor year to be nominal. What we want from this model is to further justify our decision to consider factor year as ordinal. Table \ref{tab:sup_pollen_tables}B showcases the ASCA table for this model, while figures \ref{fig:sup_ASCA_pollen}C and \ref{fig:sup_ASCA_pollen}D display the biplots of both factors. The fact factor fortnight's biplot is similar to the one featured in the model with factor year as ordinal suggests that the decision to analyze factor year as ordinal is adequate. Moreover, factor year's biplot displays the growth trend of the last years, making it the clearest reason to employ a model that considers this factor ordinal.

Lastly, we provide a more in-depth look at the interaction of the main text's model. Figures \ref{fig:sup_ASCA_pollen}E and \ref{fig:sup_ASCA_pollen}F display the biplot of the first two components of the interaction, color-coded according to fortnight and year, respectively. Compared to the first component presented in the main text, the second component appears to show a more nuanced differentiation of the evolution of pollen counts within the year, distinguishing between the first and second halves of the year (Figure~\ref{fig:sup_ASCA_pollen}E). However, this biplot is more difficult to interpret than the simplified visualization provided in the main text, so we decided to focus our interpretation on just the first component.

\begin{table}[ht]
\small
\centering

\raggedright \hspace{0.5cm} \textbf{A} \\
\centering
\resizebox{0.8\textwidth}{!}{%
\begin{tabular}{lrrrrrr}
\hline
 & \textbf{SS ($\times 10^7$)} & \textbf{\%SS} & \textbf{df} & \textbf{MS ($\times 10^4$)} & \textbf{F} & \textbf{$p$-value} \\
\hline
\hline
Year      & 0.18  & 3.1  & 29  & 6.15   & 1.71  & 0.831 \\ 
Fortnight & 2.96  & 52.2 & 25  & 118.0 & 32.85 & $<0.001$ \\
Residuals & 2.52  & 44.5 & 699 & 3.61   & --    & --       \\
\hline
Total     & 5.67  & 99.8 & 753 & 7.53   & --    & --       \\
\hline
\hline
\end{tabular}
}

\vspace{0.3cm}
\raggedright \hspace{0.5cm} \textbf{B} \\
\centering
\resizebox{0.8\textwidth}{!}{%
\begin{tabular}{lrrrrrr}
\hline
 & \textbf{SS ($\times 10^3$)} & \textbf{\%SS} & \textbf{df} & \textbf{MS} & \textbf{F} & \textbf{$p$-value} \\
\hline
\hline
Year      & 2.77  & 8.8  & 29  & 95.58  & 3.15  & $0.003$  \\
Fortnight & 7.59  & 24.0 & 25  & 303.61 & 10.01 & $<0.001$ \\
Residuals & 21.2 & 67.0 & 699 & 30.32  & --    & --       \\
\hline
Total     & 31.6 & 99.8 & 753 & 41.98  & --    & --       \\
\hline
\hline
\end{tabular}

}

\caption{ASCA tables of the supplementary models of the airborne pollen dataset. The columns indicate, respectively: the sum of squares (SS), its percentage relative to the total (\%SS), degrees of freedom (df), mean squares (MS), the F-statistic (F), and the associated $p$-value. The deviation of the total \%SS from 100\% is caused by a loss of orthogonality due to a slightly unbalanced design. \textbf{A} ASCA table of the mean centering model. \textbf{B} ASCA table of the model with year as nominal.}
\label{tab:sup_pollen_tables}
\end{table}

\newgeometry{left=2cm, top=1.5cm, right=2cm}

\restoregeometry

\bibliographystyle{model5-names}
\biboptions{authoryear}
\begin{small}
\bibliography{references}
\end{small}